\documentclass[useAMS,usenatbib]{mn2e}

\usepackage{graphicx}
\usepackage{subfigure}

\title[Radiation hydrodynamics of triggered star formation]{Radiation hydrodynamics of triggered star formation: the effect of the diffuse radiation field}
\author[Thomas J. Haworth and Tim J. Harries]{Thomas J. Haworth\thanks{E-mail:
haworth@astro.ex.ac.uk} and Tim J. Harries\\
School of Physics, University of Exeter, Stocker Road, Exeter EX4 4QL}
\begin{document}

\date{Accepted 2011 October 24. Received 2011 October 19; in original form 2011 September 20}

\pagerange{\pageref{firstpage}--\pageref{lastpage}} \pubyear{2011}

\maketitle

\label{firstpage}

\begin{abstract}
We investigate the effect of including diffuse field radiation when
modelling the radiatively driven implosion of a Bonnor-Ebert sphere
(BES). Radiation-hydrodynamical calculations are performed by
using operator splitting to combine Monte Carlo photoionization with grid-based Eulerian
hydrodynamics that includes self-gravity. It is found that the diffuse
field has a significant effect on the nature of radiatively driven
collapse which is strongly coupled to the strength of the driving
shock that is established before impacting the BES.  This can result
in either slower or more rapid star formation than expected using the
on-the-spot approximation depending on the distance of the BES from
the source object.  As well as directly compressing the BES, stronger
shocks increase the thickness and density in the shell of accumulated
material, which leads to short, strong, photo-evaporative ejections
that reinforce the compression whenever it slows. This happens
particularly effectively when the diffuse field is included as
rocket motion is induced over a larger area of the shell surface. The
formation and evolution of `elephant trunks' via instability is also
found to vary significantly when the diffuse field is included. Since
the perturbations that seed instabilities are smeared out elephant
trunks form less readily and, once formed, are exposed to enhanced
thermal compression.

\end{abstract}


\section{Introduction}
The majority of stars form in clusters, situated in molecular clouds that range in size from less than a single parsec to several hundred parsecs \citep{2003ARA&A..41...57L}.
In order for star formation to occur, gravitational collapse of material has to overcome internal thermal pressure, supersonic material motions (turbulence) and magnetic fields \citep[see e.g.][]{2007IAUS..237..270P, 2009apsf.book.....H}. The presence of OB stars in these systems has a dramatic impact on the surrounding material (and therefore star formation), as they emit large amounts of high energy radiation that photoionizes gas and gives rise to propagating ionization and shock fronts \citep{2011arXiv1101.3112E}. They also inject mechanical energy into the surroundings in the form of high-speed stellar winds and, eventually, supernova explosions. The net impact of radiative feedback from massive stars on star formation efficiency in a molecular cloud is currently unclear, though a number of individual processes that either inhibit or induce further star formation has been identified. 
\\

The two main radiative feedback mechanisms that inhibit star formation are the dispersal of material that might otherwise move towards the centre of the molecular clouds' gravitational potential \citep[e.g.][]{1962ApJ...135..736H} and the possibility of driving and maintaining turbulence that supports against collapse \citep[e.g.][]{2008PhST..132a4026P, 2009ApJ...694L..26G}. \\

The two primary established mechanisms for the induction of star formation are consequences of the expanding ionization and shock fronts about a massive star. The first is `collect and collapse' \citep{1977ApJ...214..725E,1995ApJ...451..675E, 2007MNRAS.375.1291D} in which material that is accumulated by the expanding ionization front of an HII-region becomes locally gravitationally unstable, fragmenting and collapsing to form stars. This mechanism is supported observationally by, for example, the identification of massive fragments located in a dust ring surrounding the HII region RCW 79 by \cite{2006A&A...446..171Z}.

The second mechanism is radiatively driven implosion (RDI), \citep{1989ApJ...346..735B} in which radiatively induced shocks drive into otherwise stable pre-existing density structures and cause them to collapse and form stars. RDI has been modelled by various groups, for example \cite{2003MNRAS.338..545K}, \cite{2006ApJS..165..283A}, \cite{2009MNRAS.393...21G}, \cite{2009MNRAS.398..157H}, \cite{2011ApJ...736..142B} and produces objects similar to observed bright rimmed clouds (BRCs) in HII regions \citep{2002AJ....123.2597O}. The rate at which collapse occurs and the associated star forming efficiency of these RDI models has unsurprisingly been found to be very sensitive to the incoming flux. 
\\

A variant of collect and collapse, in which the trigger for fragmentation of an ionization front is one of the many possible hydrodynamic or radiation hydrodynamic instabilities has also been explored \citep[e.g.][]{1983ApJ...274..152V, 1996ApJ...469..171G, 2002MNRAS.331..693W, 2006ApJ...647.1151M}. When sufficiently perturbed, faster moving components of the ionization front will funnel material transversely to the direction of I-front propagation, depositing it in the path of slower moving components \citep{1983ApJ...274..152V, 1996ApJ...469..171G}. This results in a collection of pillar-like objects with dense tips, much as is observed in e.g. the pillars of creation in the Eagle Nebula or the Dancing Queen's Trunk in NGC7822 \citep[see e.g.][]{1980ApJ...240...84S, 2004ApJS..154..385R, 2011PASJ...63..795C}.
\\

Recently, \cite{2009ApJ...694L..26G}  and \cite{2010ApJ...723..971G} used the ray-tracing, smooth particle hydrodynamics (SPH) code, iVINE \citep{2009MNRAS.393...21G} to consider the effects of a radiation front impinging upon a turbulent neutral medium. They found that radiation could support turbulence, which would prohibit large-scale star formation by supporting against collapse. They also found that ionizing radiation rapidly penetrated lower density regions, heating them up and compressing the remaining higher density structures. This again resulted in pillars with dense cores at the tips where stars might form. They termed this process `radiative round-up'. 
\\

Each of the aforementioned processes has been the subject of numerical modelling. Due to the intensive computational demand of these models, at least in three dimensions, a number of approximations have necessarily been developed. \cite{2011IAUS..270..301E} provides an overview and evaluation of some of the main approximations, summarised as follows.

a) Considering a monochromatic radiation field, usually Lyman 13.6eV photons, reduces calculation timescales (as with all of these approximations) compared with polychromatic models. This is because the entire source spectrum does not need to be resolved and a single value can be used for the gas opacity. However, this speed up is at the expense of being able to reliably calculate the resulting ionization and temperature structure of the system and neglects effects due to radiation hardening \citep[though radiation hardening effects can be estimated in monochromatic codes, e.g.][]{2006NewA...11..374M}.

b) Use of simplified thermal balance calculations, for example calculating the temperature as a simple function of the ionization fraction. This is more straightforward to implement and results in faster calculations than solving the thermal balance by comparing heating and cooling rates.

c) Assuming that the system is in photoionization equilibrium. This is
valid where recombination timescales are shorter than the dynamical
timescales of the gas.

d) The `on the spot' (OTS) approximation. Under this scheme diffuse field photons, those generated in recombination events, are not treated. This is justified in regimes where diffuse field photons will not propagate very far and therefore not modify the global ionization structure significantly. It is however, questionable in regions of low or rapidly varying density. Because diffuse field photons are emitted isotropically, it is possible that shadowed regions will not be exposed to a realistic amount of ionizing radiation when modelled under the OTS approximation.
\\

It is not yet clear what impact these approximations have on both the inhibiting and inducing mechanisms described above.
In an effort to understand the effect of the diffuse field, \cite{2011IAUS..270..301E} compared snapshots from the radiative round-up models run by iVINE, with full radiative transfer calculations using the Monte-Carlo radiative transfer code MOCASSIN \citep{2003MNRAS.340.1136E} and  noted significant differences between the ionization and temperature structures calculated by the two codes.  \cite{2011MNRAS.413..401E} subsequently attempted to account for the thermal effects of the diffuse field in iVINE by identifying shadowed regions and assigning them a parameterised temperature as a function of density based on comparisons of iVINE and MOCASSIN snapshots. Using this shadowing scheme a significant effect on the resulting pillar structures was observed. Far fewer pillars were formed and those that remained were narrower, denser and often cut off from the parent molecular cloud due to their paramaterised increased exposure to ionizing radiation from the diffuse field. This leads to earlier triggering of star formation and reduces the efficiency of radiation driven turbulence. 
This shadowing scheme is not without drawbacks, being susceptible to erroneous heating of true shadowed regions. The spread in temperatures at a given density calculated by MOCASSIN also results in quoted typical errors in parameterized temperature of approximately $50\%$.
It is certainly clear from this work that a more comprehensive knowledge of the effects of the diffuse field would be valuable. 
It will be necessary to establish just how different the result of a radiation hydrodynamics calculation can be when incorporating the diffuse field directly, to validate or reassess the use of simplified radiation handling in radiative feedback simulations.
\\

In this work we use the radiation hydrodynamics code TORUS \citep{2000MNRAS.315..722H, 2004MNRAS.351.1134K, 2004MNRAS.350..565H, 2010MNRAS.406.1460A, 2011MNRAS.416.1500H} to investigate the effects on radiatively driven implosion of a more sophisticated treatment of the diffuse field than previously applied in a radiation hydrodynamics calculation. Specifically, we will systematically deduce the relative effects of using a monochromatic OTS, polychromatic OTS and polychromatic-diffuse radiation field on the overall nature of collapse. 

\section{Numerical Method}
\label{numMethod}

\subsection{Hydrodynamics}
\label{hydroSection}
We use a flux conserving, finite volume hydrodynamics algorithm. It is
total variation diminishing (TVD) and makes use of the Superbee flux
limiter \citep{1985ams..conf..163R}.  We also incorporate a Rhie-Chow
interpolation scheme \citep{1983AIAAJ..21.1525R} to avoid
`checkerboard' effects that may otherwise appear in cell-centred
grid-based hydrodynamics codes. 

\subsection{Self-Gravity}
In order to include self-gravity we solve Poisson's equation 
\begin{equation}
	\nabla^2 \phi = 4 \pi G \rho
\label{poisson_eq}
\end{equation}
where $\phi$ and $\rho$ are the gravitational potential and matter density  respectively.

We use a multigrid method employing Gauss-Seidel sweeps with successive over-relaxation to
solve a linearized version of equation~\ref{poisson_eq} over the grid.
We impose Dirichlet boundary conditions with the potential at the boundary set to
a multipole expansion of the matter interior to the boundary (including terms up to
the quadrupole). The gravitational potential is subsequently  included as a source term
in the hydrodynamical equations.

\subsection{Photoionization}
\label{photoionSection}
Photoionization calculations are performed here using an iterative
Monte Carlo photon packet propagating routine, similar to that of
\cite{2003MNRAS.340.1136E} and \cite{2004MNRAS.348.1337W} which
in turn are based on the methods presented by \cite{1999A&A...344..282L}.

Photon packets are collections of photons for which the total energy
$\epsilon$ remains constant, but the number of photons contained
varies for different frequencies $\nu$. These are initiated at stars
in the model, with frequencies selected randomly based on the emission
spectrum of the star. The constant energy value $\epsilon$ for each photon packet
is simply the total energy emitted by stars (luminosity $L$) during the duration
$\Delta t$ of the iteration divided by the total number of photon
packets $N$:
\begin{equation}
\epsilon = \frac{L \Delta t}{N}
\end{equation}

Once initiated, a photon packet will propagate for a path length $p$
determined by a randomly selected optical depth as detailed in
\cite{1997A&AS..121...15H}. If the photon packet fails to escape a
cell after travelling $p$ then its propagation ceases and an
absorption event occurs. Under the OTS approximation, once a photon
packet is absorbed it is ignored, being assumed to either have been
re-emitted with a frequency lower than that required for
photoionization, or provide negligible further contribution to the
ionization structure by causing further photoionization on only small
scales. Using the principle of detailed balance, after an
absorption event a new photon packet is immediately emitted from the
same location with a new isotropically random direction and a new
random frequency based on the temperature dependent emission spectrum.
This process repeats until the photon packet escapes the grid, or its
propagation time matches the current simulation time. For each cell
in the grid the sum of the paths $l$ that photon packets traverse is
recorded.

Note that the energy density $dU$ of a radiation field
is given by
\begin{equation}
	dU = \frac{4\pi J_{\nu}}{c} d \nu
\label{energydensity}
\end{equation}
where $c$ is the speed of light. A photon packet having a path $l$ in
a particular cell contributes an energy $\epsilon (l/c) / \Delta t$
to the time-averaged energy density of that cell. Thus by summing
over all paths $l$ the energy density of a given cell (volume $V$) can be
determined, and by equation~\ref{energydensity} the mean intensity
may be estimated:
\begin{equation}
	\frac{4\pi J_{\nu}}{c} d \nu = \frac{\epsilon}{c\Delta t}
        \frac{1}{V} \sum_{d\nu} l.
\label{energydensityMC}
\end{equation}
This is then used to obtain ionization fractions by solving the ionization balance equation \citep{1989agna.book.....O}
\begin{equation}
	\frac{n(X^{i+1})}{n(X^i)} = \frac{1}{\alpha(X^i) n_e} \int_{\nu_1}^{\infty}\frac{4\pi J_{\nu} a_{\nu}(X^i) d\nu}{h\nu}
	\label{ionBalance}
\end{equation}
where $n(X^i)$, $\alpha(X^i)$, $a_{\nu}(X^i)$, $n_e$ and $\nu_1$ are the number density of the $i^{th}$ ionization state of species $X$, recombination coefficient, absorption cross section, electron number density and the threshold frequency for ionization of species $X^i$ respectively. In terms of Monte Carlo estimators (equation \ref{energydensityMC}), equation \ref{ionBalance} is given by
\begin{equation}
	\frac{n(X^{i+1})}{n(X^i)} = \frac{ \epsilon}{\Delta t V \alpha(X^i) n_e} \sum\frac{l a_{\nu}(X^i)}{  h\nu}
	\label{ionBalanceMC}
\end{equation}
This approach has the advantage that photon packets contribute to the estimate of the radiation field without having to undergo absorption events, thus even very optically thin regions are properly sampled. 
\\

TORUS performs photoionization calculations that incorporate a range of atomic species and in which thermal balance in each cell is calculated by iterating on the temperature until the heating and cooling rates match.

For the radiation hydrodynamics models in this paper a simplified thermal balance calculation is used and the only species considered are atomic and ionized hydrogen. This is to allow for comparison with previous works that use these schemes such as \cite{2009MNRAS.393...21G} and \cite{2009A&A...497..649B}. The temperature is calculated by interpolating between pre-determined temperatures, $T_{\rm{n}}$ and $T_{\rm{io}}$, ascribed to the state of fully neutral and fully ionized gas respectively as a function of the newly calculated fraction of ionized atomic hydrogen in the $i^{\rm{th}}$ cell $\eta_{i}$:
\begin{equation}
	T_{i} = T_{\rm{n}} + \eta_{i}(T_{\rm{io}} - T_{\rm{n}})
	\label{temperature}
\end{equation}
In this paper we use $T_{\rm{n}} = 10\,\rm{K}$ and $T_{\rm{io}} = 10000\,\rm{K}$ for all models 
that use this simplified thermal balance calculation. 

\subsection{Radiation Hydrodynamics}
\label{radHydro}
The hydrodynamics and photoionization schemes outlined in sections \ref{hydroSection} and \ref{photoionSection} are combined using operator splitting to perform radiation hydrodynamics calculations. A photoionization calculation is initially run to convergence, this generally allows subsequent calculations to run relatively quickly given that they are usually minor perturbations of the previous state. We then perform photoionization and hydrodynamics steps sequentially. In this work, the photoionization calculation for a time step is always calculated prior to the hydrodynamics calculation.

This operator splitting technique is flexible and relatively conceptually straightforward. It is however very computationally expensive, requiring a large number of Monte-Carlo photoionization calculations that render the gravitational and hydrodynamic components of the calculation negligible in comparative computational cost. Fortunately, this approach can be efficiently parallelized.

\subsection{Implementation}
\label{implement}
Despite the high computational cost of this radiation hydrodynamics scheme, it is extremely scalable. The computational domain is decomposed into subdomains over which the components of the radiation hydrodynamics calculation are computed by an individual processor (thread). In three dimensions subdomains take the form of cubes of equal volume, which at present can be either $1/8$, $1/64$ or $1/512$ of the total domain volume. An additional master thread performs governing and collating operations, giving a total of 65 threads for each of the three-dimensional models performed here. At lower dimensionality the grid can be decomposed in a similar manner into equally sized squares (2D) or lines (1D). 

In the photoionization component of a calculation, photon packets are
communicated between threads in stacks rather than individually to
reduce the communication latency overhead.  TORUS stores quantities using an
octree AMR grid however adaptive refinement and coarsening of the grid
is not yet fully tested in the hydrodynamics routine, a fixed grid is
therefore used for models in this work.

The RDI calculations presented in this paper were run on an SGI Altix
ICE system using 65 2.83GHz Intel Xeon cores across 9 dual
quad-core compute nodes. These typically completed within 600--1000
hours of wall time. The models that include the diffuse field do not
necessarily take the longest time to complete, as the additional
hydrodynamic and photoionization calculations required for models that
develop the highest velocity material motions outweigh the additional
time taken for each photoionization calculation when the diffuse field
is included.

\section{Numerical Tests}
\label{TESTS}
A number of tests have been conducted to confirm that TORUS is in agreement with accepted benchmarks. We treat the photoionization, hydrodynamics and self-gravity in isolation as well as the radiation hydrodynamics. Comprehensive tables of test parameters are provided to enable replication by other codes. 

\subsection{Photoionization: The HII40 Lexington Benchmark}
The HII40 Lexington benchmark is a one dimensional test in which the equilibrium temperature and ionization structure of an HII region heated by a star at 40000$\,$K is calculated and compared with the output of one of the many codes that reproduce the accepted result \citep[see][]{1995aelm.conf...83F}. Here we calculate a comparison set of results using the one dimensional semi-analytic code Cloudy \citep{1998PASP..110..761F}, one of the original contributors to the benchmark. The system modelled comprises a star at 40000$\,\rm{K}$ at the left hand edge of the grid, size 4.4$\times 10^{19}$cm comprising 1024 cells. This test incorporates more species than only hydrogen and does not rely on the simplified thermal balance calculation of equation \ref{temperature}, rather the temperature of the cell is determined through comparison of the heating and cooling rates. It also includes treatment of the diffuse field. 

\begin{table}
 \centering
 \begin{minipage}{180mm}
  \caption{Lexington benchmark parameters.}
  \label{LexingtonParams}
  \begin{tabular}{@{}l c l@{}}
  \hline
   Variable (Unit) & Value & Description\\
 \hline
   $\textrm{T}_{\rm{eff}} \textrm{(K)}$ & 40000 & Source effective temperature \\
   ${R}_*(R_\odot)$  & 18.67 & Source radius\\
   n$_{\rm{H}}$ (cm$^{-3}$) & 100 & Hydrogen number density\\
   log$_{10}$(He/H) & $-$1 & Helium abundance\\
   log$_{10}$(C/H) & $-$3.66 & Carbon abundance\\
   log$_{10}$(N/H) & $-$4.40 & Nitrogen abundance\\
   log$_{10}$(O/H) & $-$3.48 & Oxygen abundance\\
   log$_{10}$(Ne/H) & $-$4.30 & Neon abundance\\
   log$_{10}$(S/H) & $-$5.05 & Sulphur abundance\\
   L (cm) & 4.4$\times 10^{19}$ & Computational domain size\\
   $n_{\rm{cells}}$ & $1024$ & Number of grid cells\\
\hline
\end{tabular}
\end{minipage}
\end{table}

\begin{figure*}
		\includegraphics[scale = 0.365]{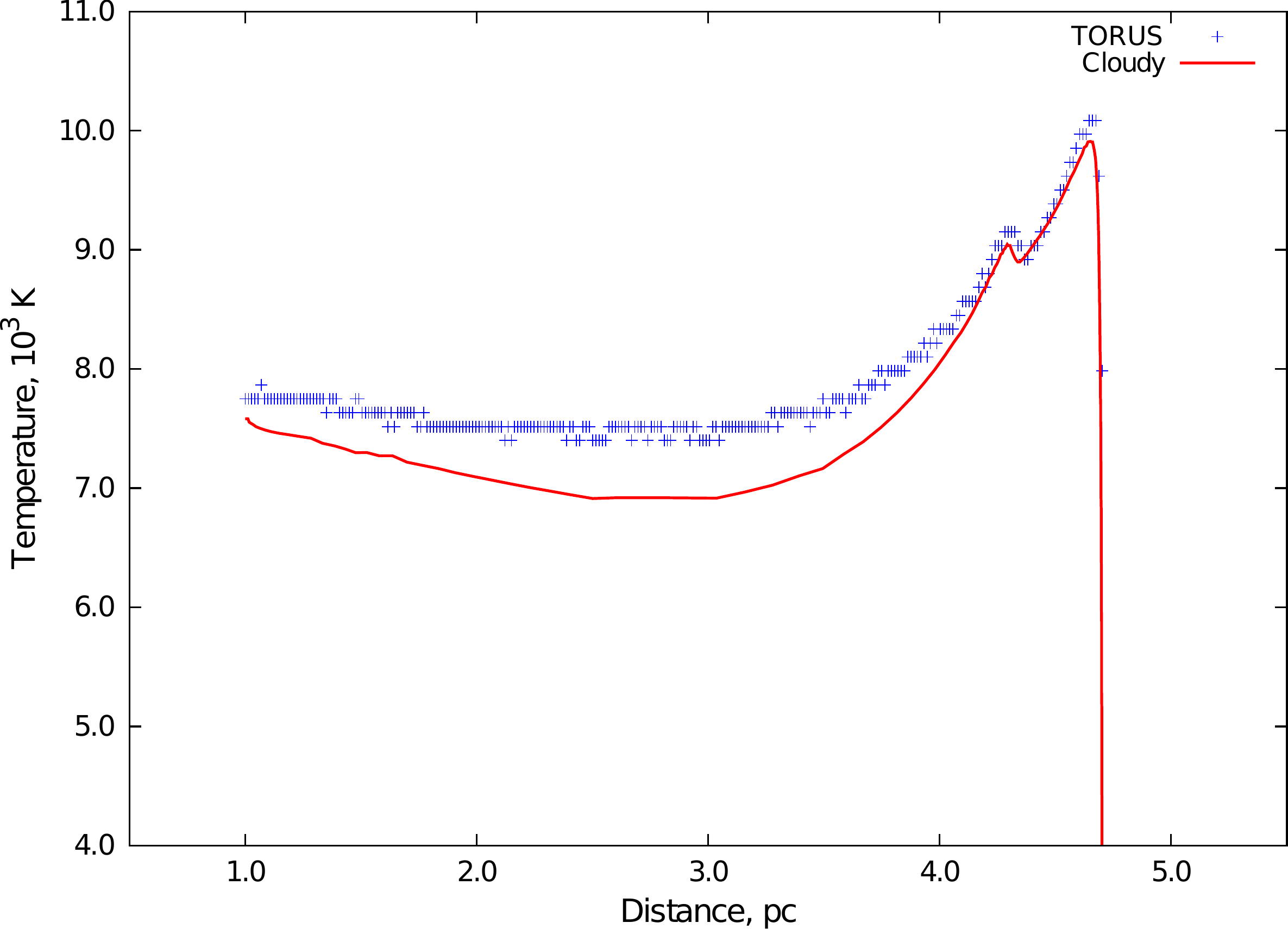}
		\includegraphics[scale = 0.365]{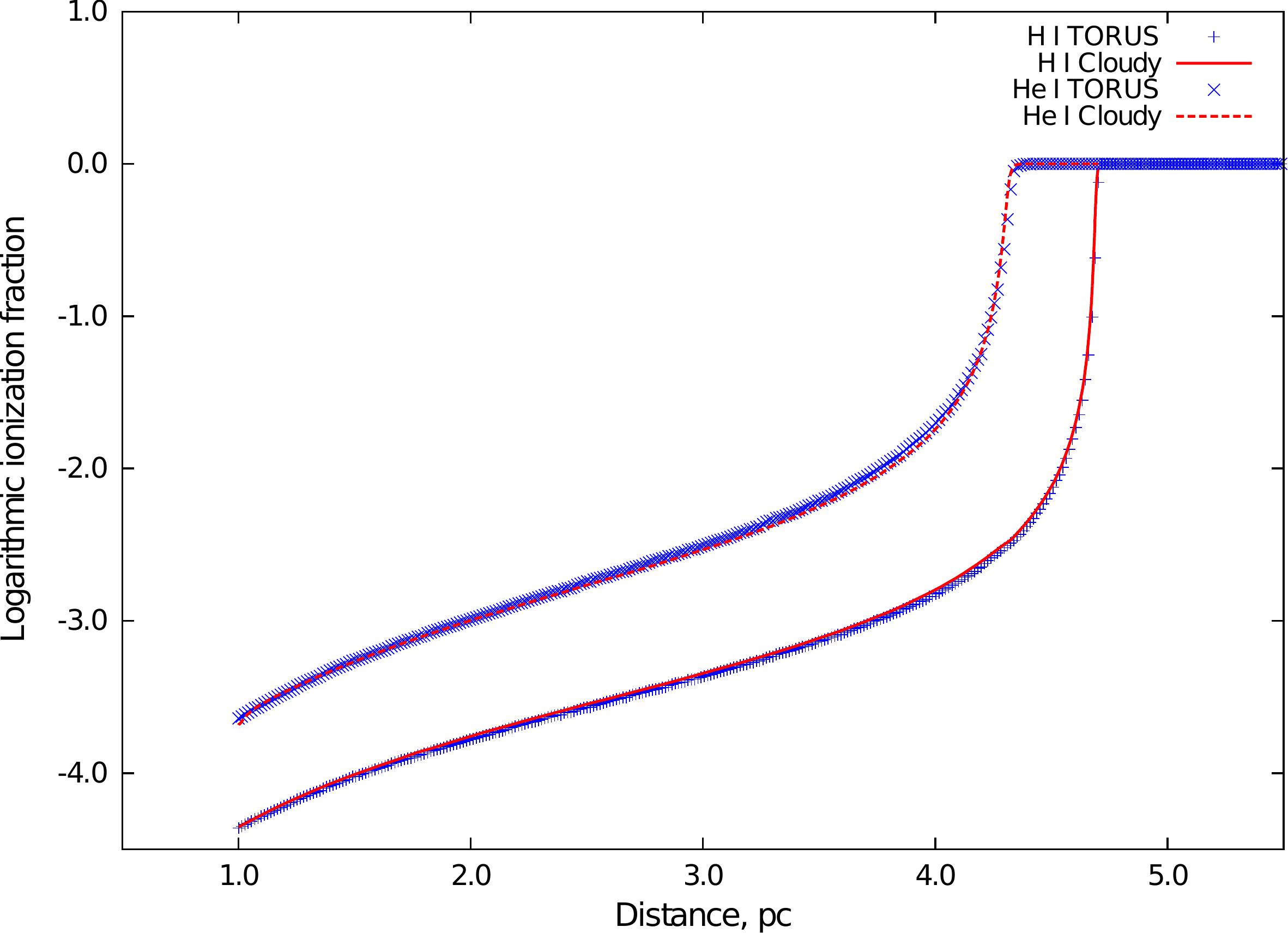}
		\includegraphics[scale = 0.365]{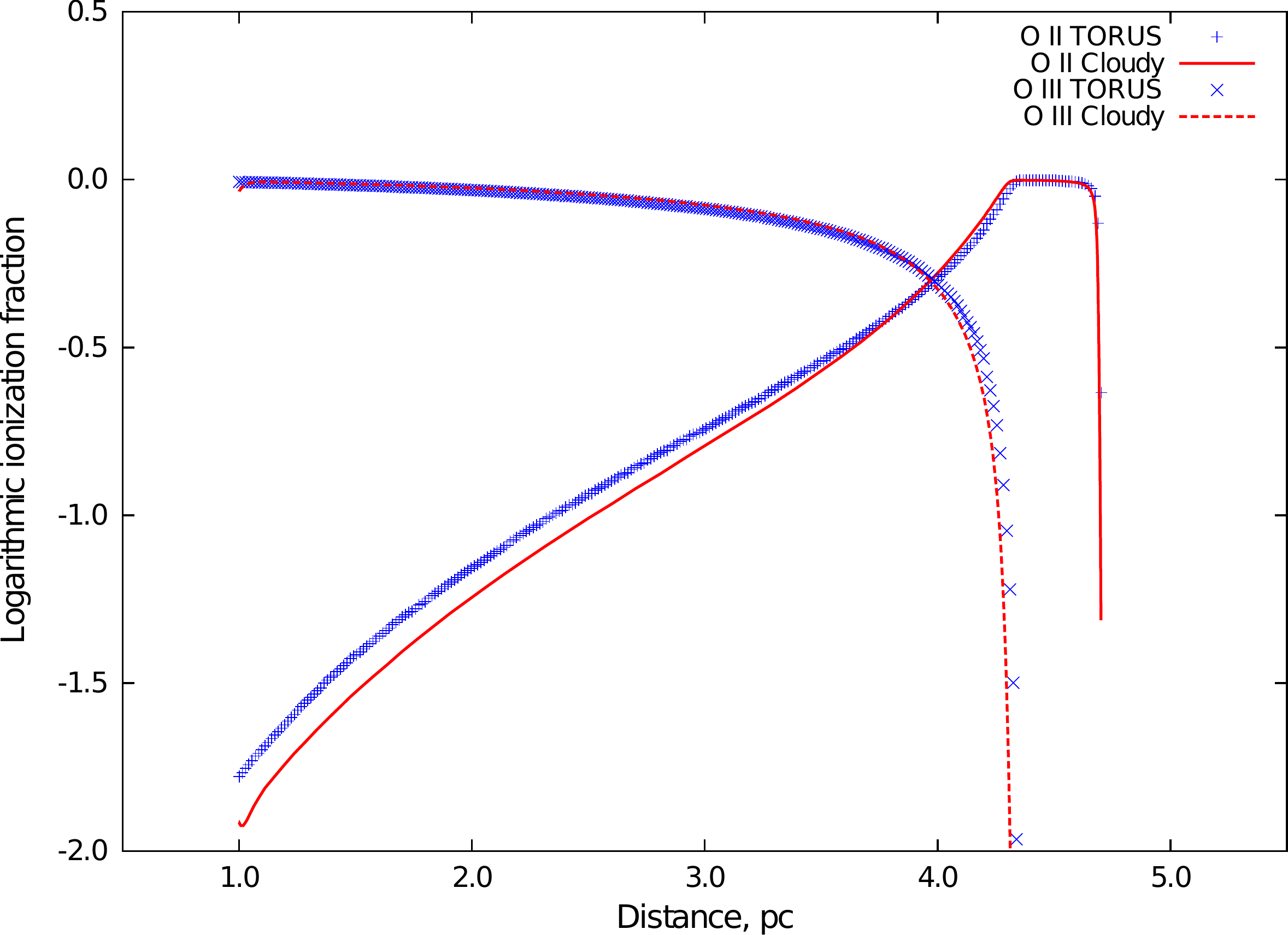}
		\includegraphics[scale = 0.365]{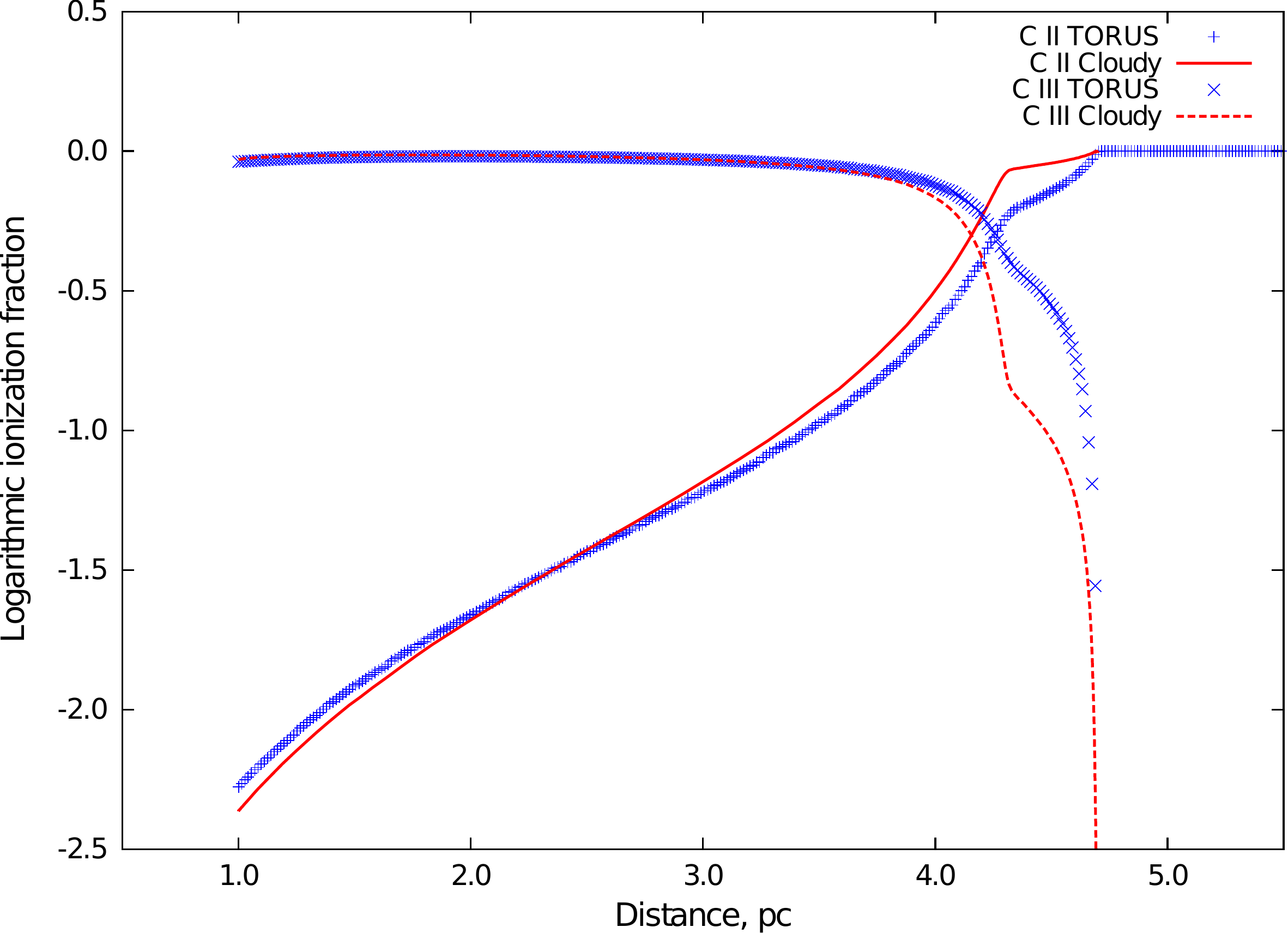}
		\includegraphics[scale = 0.365]{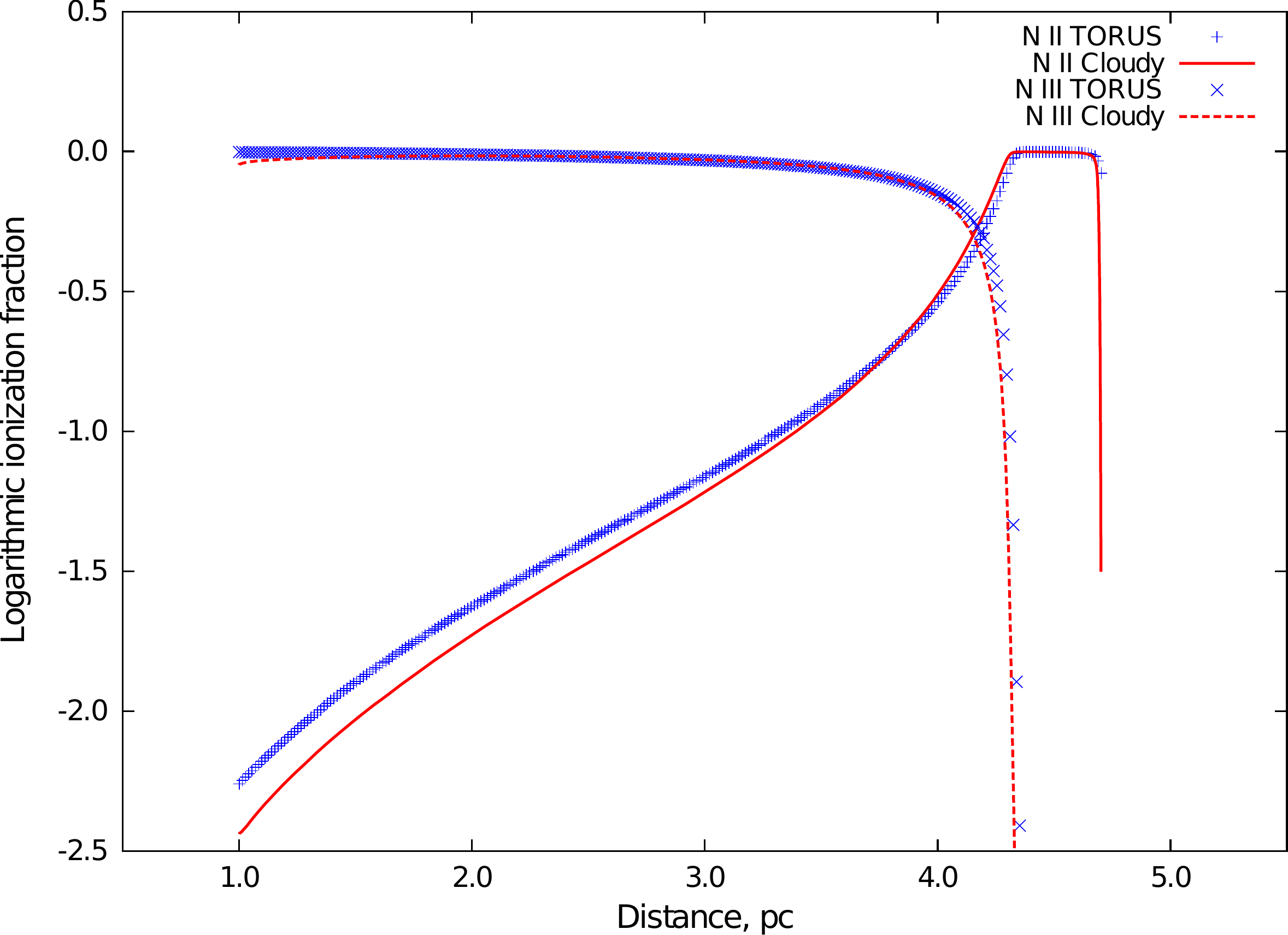}
		\includegraphics[scale = 0.365]{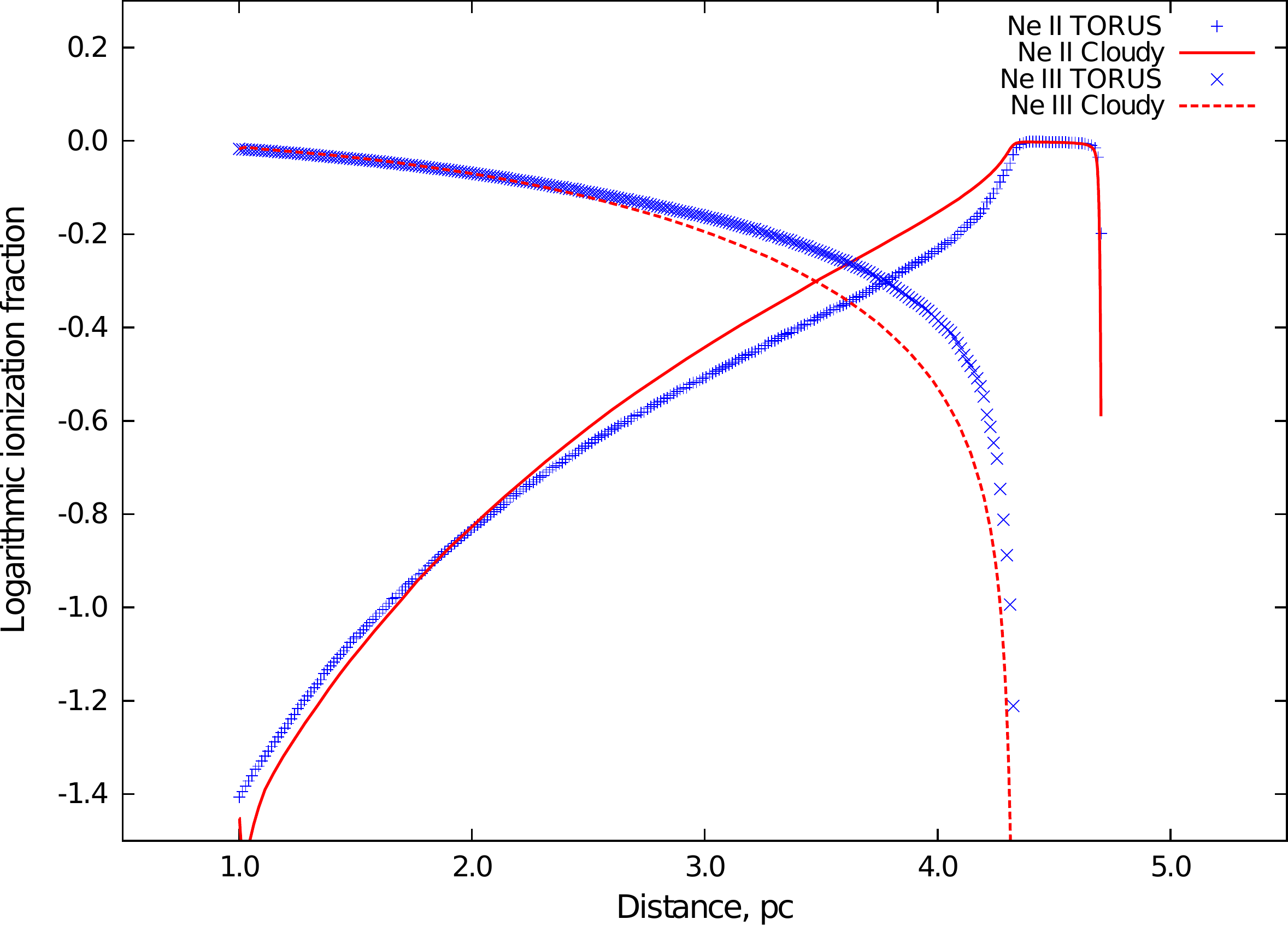}
  \caption{Top Left: The temperature distribution for the Lexington benchmark. Top Right: Hydrogen and helium ionization fractions. Middle Left: Oxygen ionization fractions. Middle Right: Carbon ionization fractions. Bottom Left: Nitrogen ionization fractions. Bottom Right: Neon ionization fractions.}
		\label{LexingtonFig}
\end{figure*}

A full list of parameters used for this benchmark is given in Table
\ref{LexingtonParams} and the resulting temperature and ionization
fractions as calculated using both TORUS and Cloudy are shown in
Figure \ref{LexingtonFig}. TORUS is consistent with
the Cloudy temperature distribution to within 10\% and is
generally much better than this. The higher temperature calculated by
TORUS in the inner regions is in agreement with the result obtained in
\cite{2004MNRAS.348.1337W}. The hydrogen and helium ionization
fractions agree extremely well, this is of particular importance with
regard to the hydrogen-only radiation hydrodynamics models in this
paper. The other ions match to within similar levels of agreement as
\cite{2003MNRAS.340.1136E} and
\cite{2004MNRAS.348.1337W}. Discrepancies in the result of this
benchmark are usually attributed to differences in the atomic data
used by the codes that are being compared.

\subsection{Hydrodynamics Tests}

\subsubsection{Sod Shock Tube}
The Sod shock tube test is a simple 1-dimensional model, initially comprising two equal volumes separated by a partition. Both partitions, the left hand partition (LHP) and right hand partition (RHP), contain ideal gases at zero velocity with different densities and hence different initial pressure and energy. 
\begin{table}
 \centering
 \begin{minipage}{180mm}
  \caption{Sod shock tube parameters.}
  \label{SodShockParams}
  \begin{tabular}{@{}l c l@{}}
  \hline
   Variable (Unit) & Value & Description\\
 \hline
	$\rho_{1}$ $(\mathrm{g\,cm}^{-3})$ & $1$ & Initial density in LHP\\
	$\rho_{2}$ $(\mathrm{g\,cm}^{-3})$ & $0.125$ & Initial density in RHP\\
	$\gamma$ & 7/5 & Adiabatic index\\
	$P_1$ $(\mathrm{dyn\,cm}^{-2})$ & $1$ & Initial pressure in LHP\\
	$P_2$ $(\mathrm{dyn\,cm}^{-2})$& $0.1$ & Initial pressure in RHP\\
	$E_1$ $(\mathrm{erg\,cm}^{-3})$& $2.5$ & Initial energy density in LHP\\
	$E_1$ $(\mathrm{erg\,cm}^{-3})$& $2.0$ & Initial energy density in RHP\\
	E.O.S. & Adiabatic & Equation of state\\
	L (cm)& $1$ & Computational domain size \\
        $n_{\rm{cells}}$ & $1024$ & Number of grid cells\\
\hline
\end{tabular}
\end{minipage}
\end{table}
At time $t=0$\,s, the partition is removed and the system allowed to evolve. The state after a time $t$ is detailed by \cite{Sod-1978}. 

We use an adiabatic equation of state, with adiabatic index $\gamma = 7/5$. Unless otherwise stated, a value of 0.3 is used for the Courant--Friedrichs--Lewy (CFL) parameter in all hydrodynamics models. In this model boundary conditions are reflective and 1024 cells are used for the grid.

\begin{figure}
	\hspace{-10pt}
	\includegraphics[scale = 0.365]{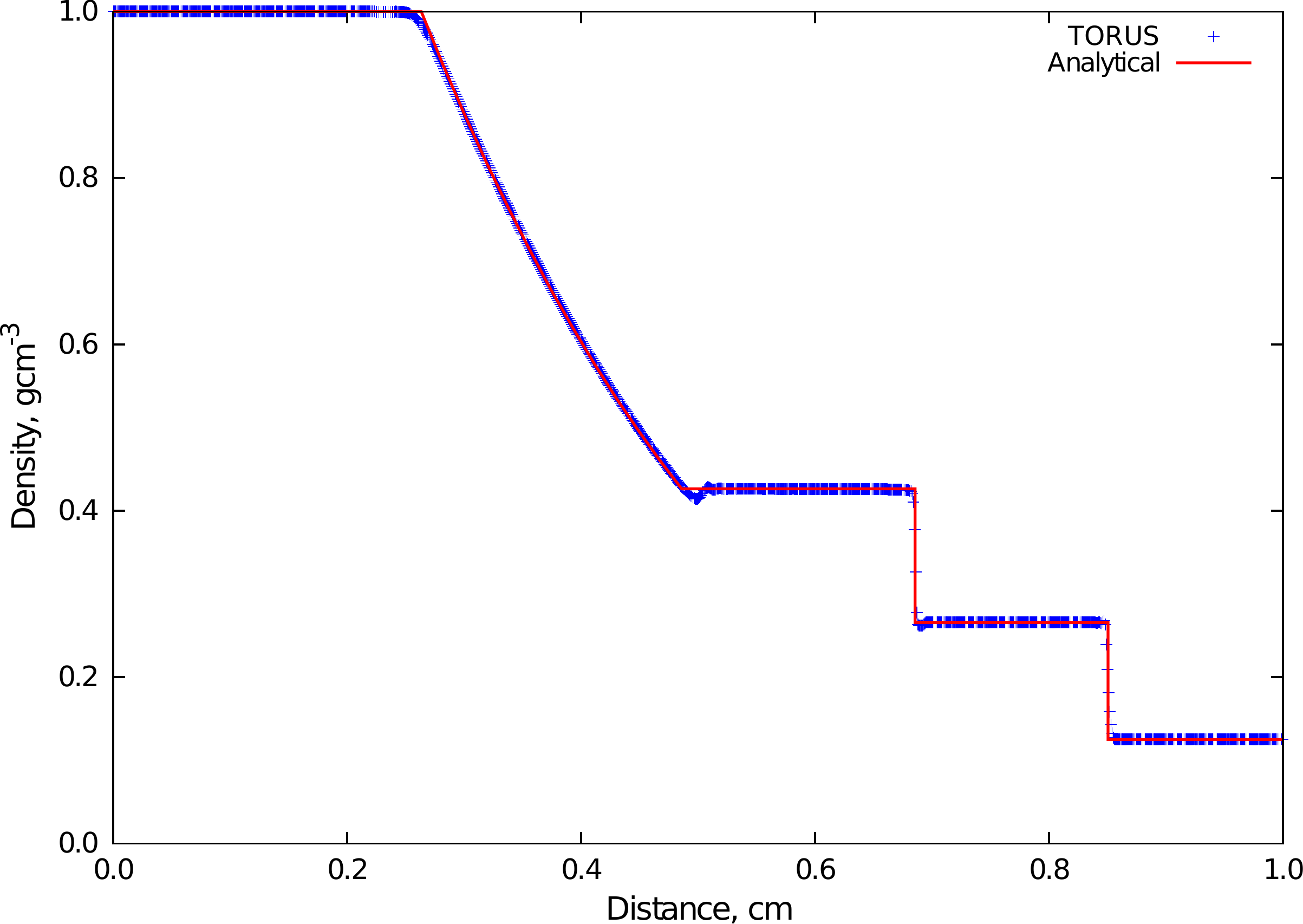}
	\caption{Density distribution of the Sod shock tube test at $t=0.2\rm{s}$, showing both the result given by TORUS (blue crosses) and the analytical solution (red line).}
		\label{SodResult}
\end{figure}

A summary of the model parameters is given in Table \ref{SodShockParams} and the result as computed by TORUS at $t=0.2$\,s is shown in Figure \ref{SodResult}.
The features in this result are, from right to left; the initial density in the RHP, a shock wave which forms as a result of the low density material recoiling away from the high density material, a contact discontinuity between the high density material of the LHP and the low density material of the RHP, a rarefaction wave formed because the contact discontinuity acts as a piston drawing left hand material to the right and the initial density of the LHP. 

TORUS shows excellent agreement with the analytical solution. The
slight density dip at the rarefaction wave-contact discontinuity
interface arises because TVD flux limited schemes only smooth out
oscillations near sharp shocks. This is a necessary compromise, so
that existing physical oscillations are not unphysically damped.

\subsubsection{Sedov-Taylor Blast Wave}
The Sedov-Taylor blast wave is an extreme model, which tests the
advection scheme beyond the demands that will be made of it in
star formation applications. In this 2-dimensional test a large amount
of energy is injected into a circular region, radius $0.01\,\rm{cm}$,
of a constant density ideal gas, causing a blast wave.  Self-similar
analytical solutions for the time evolution of such a blast wave were
found by \cite{1950RSPSA.201..159T} and \cite{Sedov-1946}.

\begin{figure}
	\hspace{-36pt}
		\includegraphics[scale = 0.365]{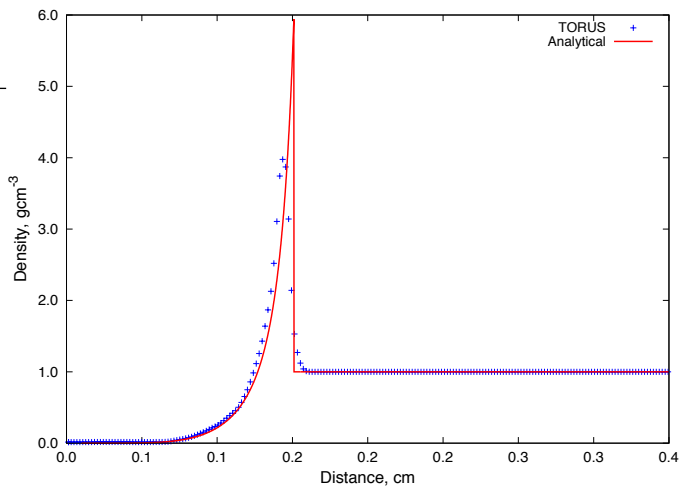}
		\caption{Density distribution of the Sedov-Taylor blast wave at $t=0.03\,\rm{s}$ showing both the result given by TORUS (blue crosses) and the analytical solution (red line).}
		\label{TaylorSedovResults}
\end{figure}

The initial ratio of thermal energy in the circular region to the rest
of the grid is $~3\times10^8:1$. Given the extreme nature of this
model, a CFL parameter value of 0.08 was required in order to capture
the early stages of evolution without numerical instability
arising. The boundary conditions used in this test are all reflective
and the grid comprises $512^2$ cells.  A full table of parameters is
given in Table \ref{TaylorSedovParams} and a comparison of the density
distribution at time $t=0.03$\,s, as calculated both by TORUS and
analytically, is given in Figure \ref{TaylorSedovResults}. TORUS
demonstrates a good level of agreement with the analytical solution
but, as with all numerical schemes, suffers from numerical
diffusion. This is responsible for the slight broadening of the shock
and reduction in the peak amplitude compared to the analytical
solution.
 \begin{table}
 \centering
 \begin{minipage}{180mm}
  \caption{Sedov-Taylor parameters.}
  \label{TaylorSedovParams}
  \begin{tabular}{@{}l c l@{}}
  \hline
   Variable (Unit)& Value & Description\\
 \hline
  $\gamma$ & 7/5 & Adiabatic index\\
  $v$ ($\textrm{cm\,s}^{-1}$) & 0 & Initial velocity\\
  $\rho$ $(\textrm{g\,cm}^{-2})$ & $1$ & Initial surface density\\
  $r_{\rm{i}}$ (cm) & 0.01 & Energy dump zone radius\\
  $\textrm{E}_{\rm{Blast}}$ (erg\,cm$^{-2}$) & 3183.1 & Dump zone surface energy density \\
  $\textrm{E}_{\rm{o}}$ (erg\,cm$^{-2}$)& $1\times 10^{-5} $ & Ambient surface energy density \\
  E.O.S. & Adiabatic & Equation of state\\
  L($\textrm{cm}$) & 1  & Computational domain size\\
  $n_{\rm{cells}}$ & $512^2$ & Number of grid cells\\
\hline
\end{tabular}
\end{minipage}
\end{table}

\subsubsection{Kelvin-Helmholtz Instability}
\label{KHsec}
Kelvin-Helmholtz (KH) instabilities are vortices that form at an
interface between two materials due to shear forces
\citep{Helmholtz-1871, Kelvin-1871}. Adaptations to SPH codes have
recently found to be required in order to form KH instabilities by
\cite{2007MNRAS.380..963A} and \cite{2008JCoPh.22710040P}, thus
reproducing these has proved to be an important test of hydrodynamical
algorithms.

\begin{figure}
	\begin{center}
		\hspace{-25pt}
		\includegraphics[scale=0.25]{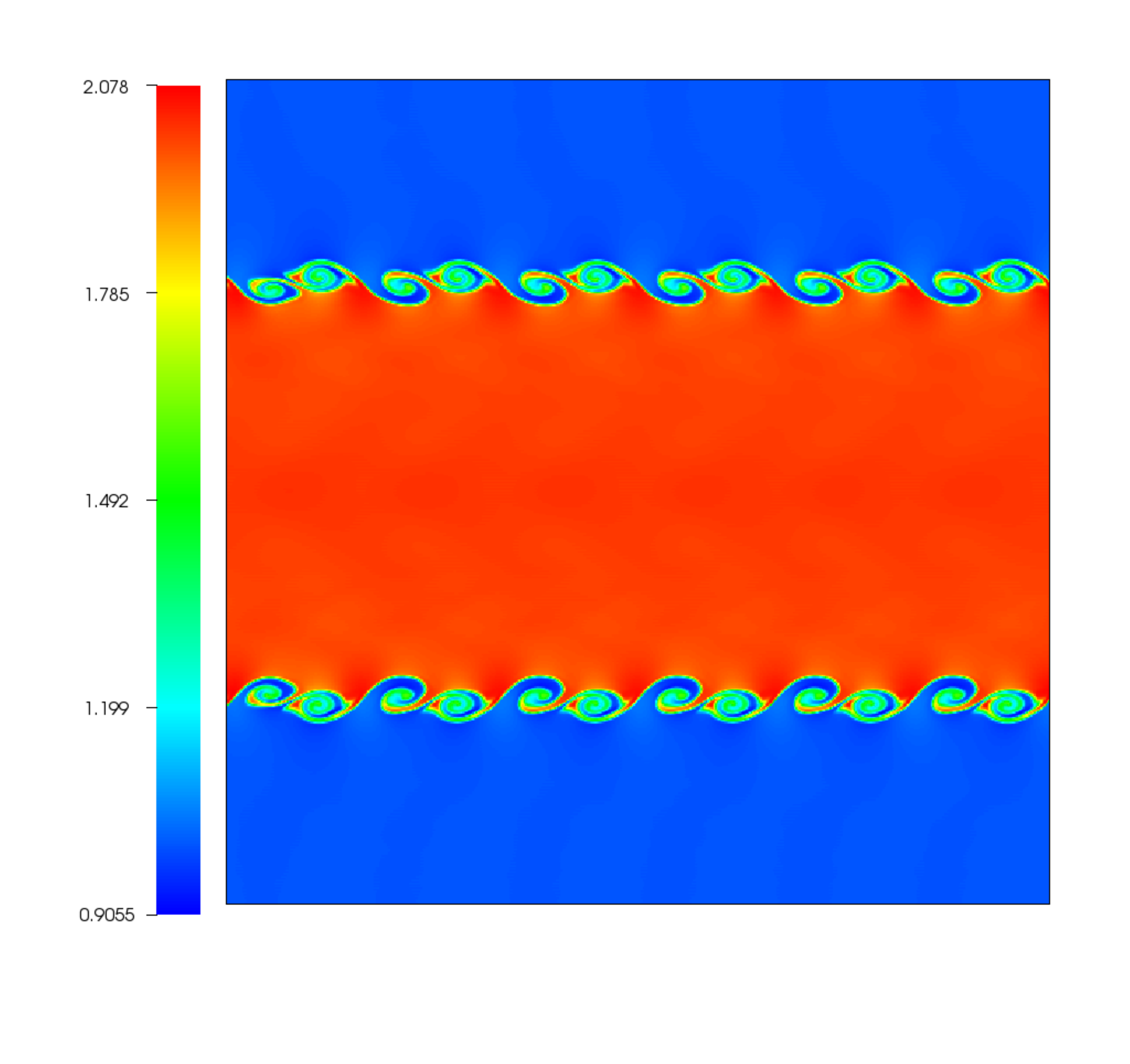}
		\vspace{-30pt}
		\caption{Surface density distribution in g\,cm$^{-2}$ at $\tau_{KH}$ across the $1\rm{cm}^2$ domain of the Kelvin-Helmholtz instability model.}
		\label{KHTest}
	\end{center}
\end{figure}

The 2-dimensonal system modelled here follows \cite{2008JCoPh.22710040P}, comprising two fluids in contact at different density and velocity, such that the ratio of their densities is 2:1 and their velocities are equal in magnitude but in opposite directions. Periodic boundary conditions are used at the $\pm x$ boundaries and reflective conditions at the $\pm z$ boundaries, corresponding to housing the system in a pipe that extends indefinitely in $\pm x$.
At time $t=0$\,s the interface between these fluids is subject to a perturbation of the form
\begin{center}
	\begin{equation}
		u = \left\{
    		\begin{array}{l l}
    			A\rm{sin}(-2\pi (x+1/2)(1/6)), & |z-0.25| < 0.025 \\
			 & \\
    			A\rm{sin}(2\pi (x+1/2)(1/6)), & |z+0.25| < 0.025. \\
    		\end{array} \right.
		\label{perturbation}
	\end{equation}
\end{center}
vortices should then form within a characteristic KH timescale given by
\begin{center}
	\begin{equation}
		\tau_{KH} = {2\pi \over \omega}.
	\label{tauKH}
	\end{equation}
\end{center}
Where, for materials in contact with density $\rho_1$ and $\rho_2$ and velocities $v_1$ and $v_2$ subject to a periodic perturbation of wavelength $ \lambda$
\begin{center}
	\begin{equation}
		\omega = {2\pi \over \lambda}{(\rho_1 \rho_2)^{1/2}|v_1 - v_2| \over (\rho_1 + \rho_2)}.
	\label{khFreq}
	\end{equation}
\end{center}
A full table of parameters used for this test is given in Table \ref{KHParams}.
 \begin{table}
 \centering
 \begin{minipage}{180mm}
  \caption{Kelvin-Helmholtz parameters.}
  \label{KHParams}
  \begin{tabular}{@{}l c l@{}}
  \hline
   Variable (Unit)& Value & Description\\
 \hline
	$\gamma$ & 5/3 & Adiabatic Index \\
	$\rho_{1}$ $(\textrm{g\,cm}^{-2})$& $1$ & Ambient fluid initial surface density \\
	$\rho_{2}$ $(\textrm{g\,cm}^{-2})$& $2$ & Central fluid initial surface density\\
	$u_1$ $(\textrm{cm\,s}^{-1})$& $-0.5$ & Ambient fluid initial velocity \\
	$u_2$ $(\textrm{cm\,s}^{-1})$& $+0.5$ & Central fluid initial velocity \\
	E.O.S & Adiabatic & Equation of State \\
	A & 0.025 & Constant in perturbation equation \\
	$\lambda$ (cm) & $1/6$ & Wavelength of perturbation \\
         L (cm)& $1$ & Computational domain size \\
	$n_{\rm{cells}}$ & $512^2$ & Number of grid cells\\

\hline
\end{tabular}
\end{minipage}
\end{table}
Using these parameters and equations \ref{tauKH} and \ref{khFreq} we obtain a KH timescale of approximately 0.35 seconds, the time within which vortices should form.  A plot of the density distribution as calculated by TORUS at $\tau_{KH}$ is given in Figure \ref{KHTest} and clearly demonstrates that primary and secondary vortices have formed within the KH timescale. This test has also been successfully performed using density ratios of 5:1 and 10:1.

\subsubsection{Rayleigh-Taylor Instability}
Rayleigh-Taylor instabilities can arise where a material is `on top'
of a second lower density material in a gravitational potential and
the interface between them is subject to a perturbation
\citep{Rayleigh1883, Taylor1950}. They are manifested as
`Rayleigh-Taylor fingers' that propagate into the low density material
along which Kelvin-Helmholtz instabilities may form. This gives rise
to a characteristic mushroom shape. In the following test we have
selected system parameters that should give rise to this
characteristic structure.

At time $t=0$\,s the interface between two
different density materials in a 1\,cm$^2$ box in the presence of a
gravitational field is subject to a small disturbance of magnitude
$-0.055\,\rm{cm}\,\rm{s}^{-1} $ across a finite range $0.45<x<0.55$ of
the interface, the system is then left to evolve. We use periodic
boundary conditions at the $\pm x$-direction bounds and reflective at
the $\pm z$-direction bounds. The gravitational potential $\phi$ at
height $z$ is given by
\begin{center}
	\begin{equation}
		\phi(z) = gz 
		\label{phi}
	\end{equation}
\end{center}
where $g$ is the acceleration due to gravity, here equal to $0.1\,\textrm{cm\,s}^{-2}$.

\begin{figure}
	\begin{center}
		\includegraphics[scale = 0.24]{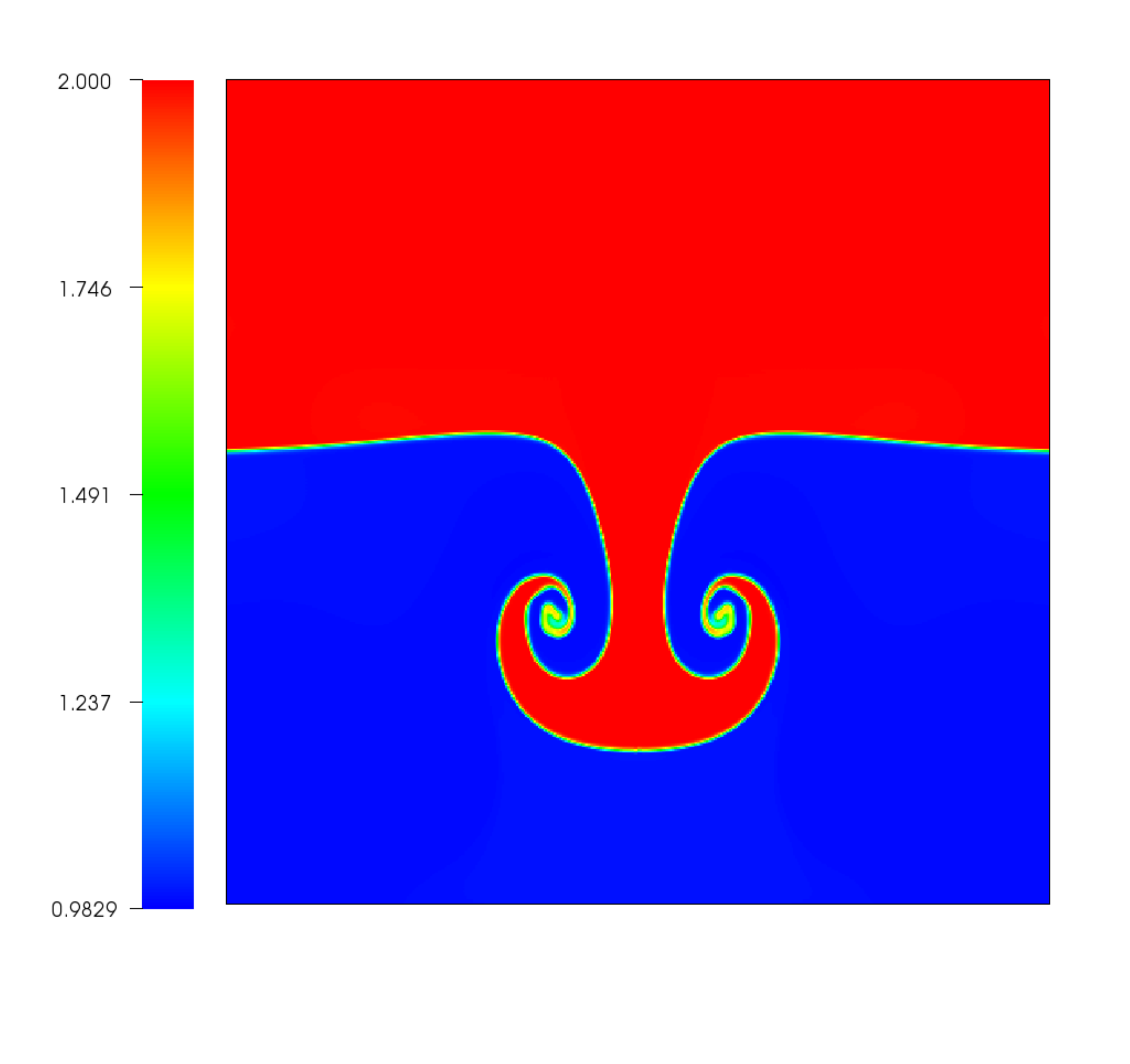}
		\vspace{-30pt}
		\caption{Surface density distribution in g\,cm$^{-2}$ showing the `mushroom' formed in our $1\,\rm{cm}^2$ domain size, Rayleigh-Taylor instability model. It comprises a penetrating column of material with Kelvin-Helmholtz instabilities forming where the materials flow past one another.}
		\label{RT}
	\end{center}
\end{figure}

A summary of parameters used in this test is given in Table
\ref{RTParams}. Figure \ref{RT} shows the distinctive mushroom-shape
formed via this method at time $t=5$\,s. The main body of the mushroom
is the Rayleigh-Taylor finger, at the tip of which Kelvin-Helmholtz
instabilities have formed.

\begin{table}
 \centering
 \begin{minipage}{180mm}
  \caption{Rayleigh-Taylor parameters.}
  \label{RTParams}
  \begin{tabular}{@{}l c l@{}}
  \hline
   Variable (Unit)& Value & Description\\
 \hline
  $\rho_{1}$ $(\textrm{g\,cm}^{-2})$ & $2$ & Surface density of upper material\\
  $\rho_{2}$ $(\textrm{g\,cm}^{-2})$& $1$ & Surface density of lower material\\
  g $(\textrm{cm\,s}^{-2})$& $0.1$ & gravitational acceleration\\
  E.O.S. & Adiabatic & Equation of state\\
  L (cm$^2$)& $1$ & Computational domain size \\
  $n_{\rm{cells}}$ & $512^2$ & Number of grid cells\\
\hline
\end{tabular}
\end{minipage}
\end{table}

\subsection{Self-Gravity}
This test follows, in three dimensions, the collapse of an initially
uniform sphere to form an $n=1$ polytrope. A polytropic cloud is one
in which the pressure $P$ varies according to the following relation:
\begin{center}
	\begin{equation}
		P = K\rho^{1+\frac{1}{n}}
		\label{polytropePressure}
	\end{equation}
\end{center}
where n is the index of the polytrope, $\rho$ is the density and $K$ is a constant.

The corresponding solution to Poisson's equation for a self-gravitating polytropic fluid is given by the Lane-Emden equation, which details the variation in pressure and density in terms of dimensionless variables $\zeta$ and $\theta$:
\begin{center}
	\begin{equation}
		\frac{1}{\zeta^2}\frac{d}{d\zeta}\left(\zeta^2 \frac{d\theta}{d\zeta} \right) + \theta^n = 0.
	\label{BESeqn}
	\end{equation}
\end{center}
$\theta$ and $\zeta$ are given by equations \ref{theta} and \ref{zeta}:
\begin{center}
	\begin{equation}
		\theta^n = \frac{\rho}{\rho_c}
	\label{theta}
	\end{equation}
\end{center}
and
\begin{center}
	\begin{equation}
		\zeta = r\left(\frac{4 \pi G \rho_c^2}{(n+1)P_c} \right)^{1/2}
		\label{zeta}
	\end{equation}
\end{center}
where $r$, $\rho_c$ and $P_c$ are the radial position, central density and pressure respectively.

\begin{figure}
		\hspace{-35pt}
		\includegraphics[scale = 0.365]{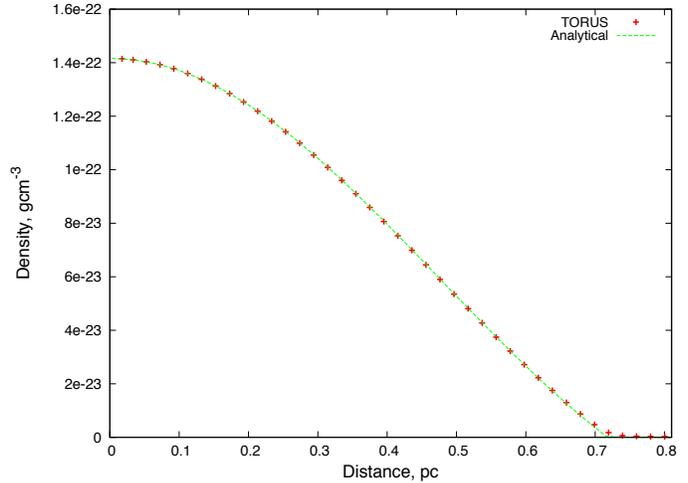}
		\caption{The analytical and TORUS-computed resulting density distribution for the collapse of a uniform density sphere to form an $n=1$ polytrope.}
		\label{selfGrav}
\end{figure}

The solution to the Lane-Emden equation for an $n=1$ polytrope is
simply a sinc function, which should be the form of the resulting
density distribution once collapse has occurred.  We employ reflecting
boundary conditions. A summary of the parameters used for this test is
given in Table \ref{selfGravParams}. We ran the model with significant
artifical viscosity in order to strongly damp the oscillations that
would otherwise occur.
\begin{table}
 \centering
 \begin{minipage}{180mm}
  \caption{Self-gravity test parameters.}
  \label{selfGravParams}
  \begin{tabular}{@{}l c l@{}}
  \hline
   Variable (Unit) & Value & Description\\
 \hline
  $M_{\rm{sphere}}$ ($M_\odot$) & 1 & Mass of initial sphere \\
  $\textrm{r}_{\rm{sphere}}$ (pc) & $1$ & Radius of initial sphere\\
  $\gamma$ & 2 & Adiabatic index\\
  E.O.S. & Polytropic & Equation of state\\
  $\textrm{K}$ & $4.1317\times 10^{29}$ & Equation \ref{polytropePressure} constant\\
  $\textrm{n}_{\rm{cells}}$ & $128^3$ & Number of grid cells\\
\hline
\end{tabular}
\end{minipage}
\end{table}
The resulting radial density distribution as calculated both
analytically and by TORUS is given in Figure \ref{selfGrav} and
demonstrates that TORUS is in excellent agreement with the expected
result.

\subsection{Radiation Hydrodynamics: Expansion of an HII region}
\label{hii_test}
The time evolution of the extent of an HII region around, for example, an O-type star can be divided into two regimes. 
First note that using a time-independent, photoionization equilibrium consideration, the HII region extends until the rates of ionization and recombination become equal. In a uniform density, hydrogen-only, isotropic medium this corresponds to a spherical HII region with extent $r_{\rm{I}}^{\rm{o}}$ known as the `Str\"{o}mgren radius', given by:
\begin{center}
	\begin{equation}
		r_{\rm{I}}^{\rm{o}} = \left( \frac{3 N_{\gamma}}{4 \pi n_{\rm{e}}^2 \alpha^{(2)}} \right)^{1/3}
	\label{stromgren}
	\end{equation}
\end{center}
Where $N_{\gamma}$ is the number of ionizing photons from the source per second, $n_{\rm{e}}$ is the electron number density and $\alpha^{(2)}$ is the recombination coefficient into all states except the ground state. 

In the time-dependent case the first phase of evolution is that in which $r_{\rm{I}} < r_{\rm{I}}^{\rm{o}}$. Here, the expansion is initially very rapid. Since the surrounding material near the star can be assumed to be ionized as soon as the stellar radiation reaches it, the ionization front propagates at the speed of light over a mean free path. Given that the ionizing front propagates much more rapidly than the speed of sound $c_{\rm{s}}$ little subsequent material motion occurs.

In the second phase of evolution, the hot ionized region expands into the surrounding, cool, material as a consequence of the pressure difference between them. Once the ionization front expansion velocity drops below $c_{\rm{s}}$, a shock moves ahead of the ionization front into the neutral material. \cite{1978ppim.book.....S} showed that the analytical radius of an HII region in the phase two expansion at time $t$ is given by 
\begin{center}
	\begin{equation}
		r_{\rm{I}} = r_{\rm{I}}^{\rm{o}}\left(1 + \frac{7}{4} \frac{c_{\rm{I}}t}{r_{\rm{I}}^{\rm{o}}} \right)^{4/7}
		\label{Ifront}
	\end{equation}
\end{center}
up until the point where pressure equilibrium is approached, where $c_{\rm{I}}$ is the speed of sound in the ionized gas. Equation \ref{Ifront} is constructed using the thin shell approximation.
\\

In this test the second phase expansion is modelled in three dimensions, using equation \ref{Ifront} as a comparison. The evolution of the first phase expansion, prior to $r_{\rm{I}}^{\rm{o}}$, is ignored because the calculations here assume photoionization equilibrium. The system consists of a star at 40000\,K with a blackbody spectrum at the centre of a $4\,$pc$^3$ box of neutral hydrogen with reflective boundary conditions. We perform a radiation hydrodynamics calculation as outlined in section \ref{numMethod} and follow the evolution of the ionization front position, defined as the point where the atomic hydrogen ionization fraction $X$(HI) = 0.5, from $r_{\rm{I}} = r_{\rm{I}}^{\rm{o}}$ with time. 

\begin{figure}
		\hspace{-20pt}
		\includegraphics[scale=0.365]{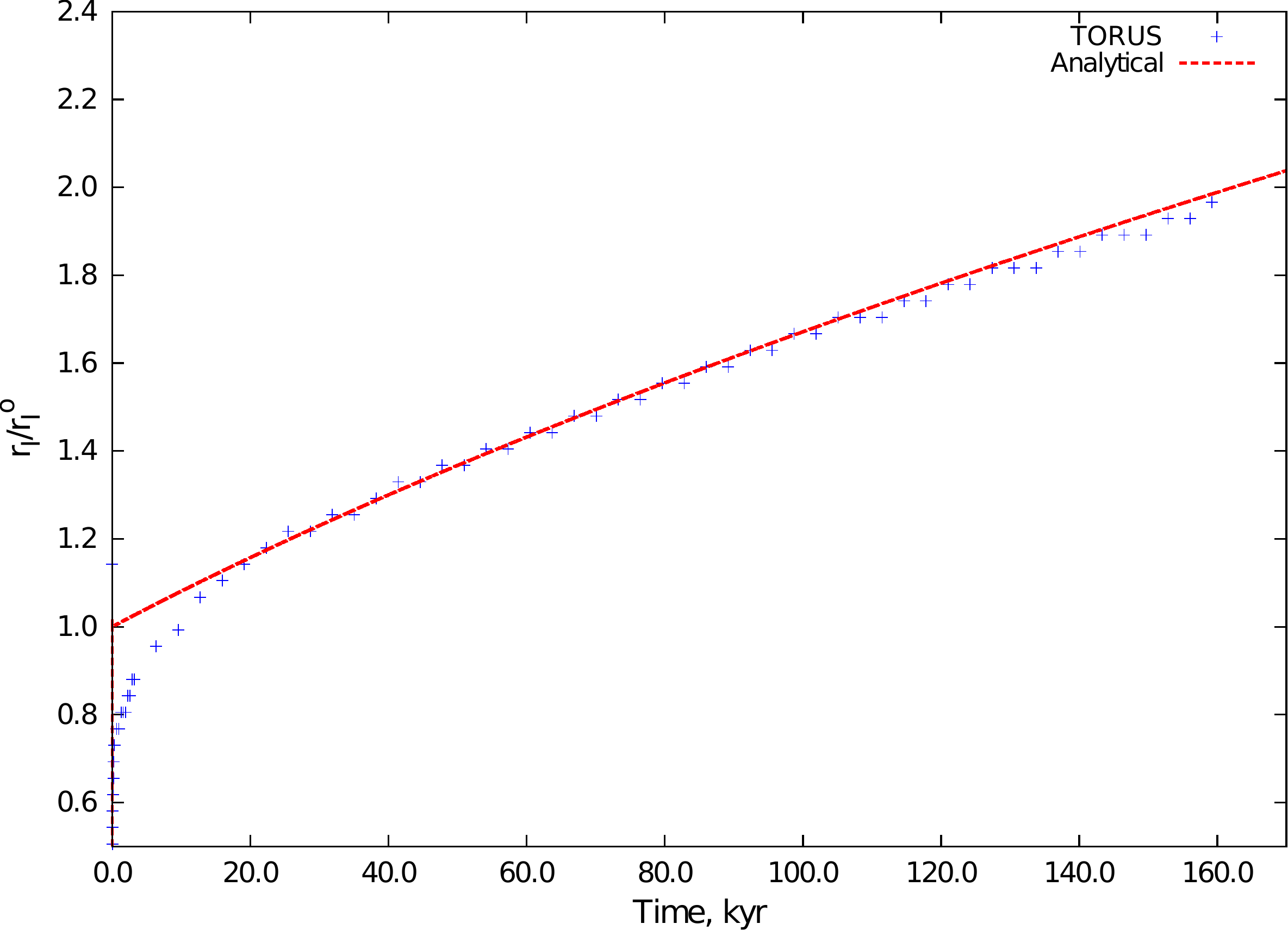}
		\caption{The position of the ionization front with time from $r_{\rm{I}} = r_{\rm{I}}^{\rm{o}}$}
		\label{IfrontTest}
\end{figure}

Table \ref{hiiExpansionParams} lists the parameters used in this test
and the results are shown in Figure \ref{IfrontTest}.  TORUS shows
excellent agreement with equation \ref{Ifront} shortly after reaching
the Str\"{o}mgren radius. (The discrepancies at early times occur
because TORUS evolves from a neutral starting point whereas the
analytical solution starts with the ionization front at the
Str\"omgren radius). At late times the
evolution starts to deviate from the analytical solution as sufficient
material is accumulated for the thin shell approximation to no longer
apply.

\begin{table}
 \centering
 \begin{minipage}{180mm}
  \caption{Parameters used for the HII expansion model.}
  \label{hiiExpansionParams}
  \begin{tabular}{@{}l c l@{}}
  \hline
   Variable (Unit) & Value & Description\\
 \hline
 $\rho_{\rm{o}}$ ($\textrm{m}_{\rm{H}}\,\textrm{cm}^{-3}$) & $100 $ & Initial density\\
 $T_{\rm{o}}$ (K) & 10& Initial temperature of grid\\
 $u_{\rm{o}}$ ($\textrm{cm\,s}^{-1}$) & $0$ & Initial velocity throughout grid\\
 $\gamma $ & 5/3 & Adiabatic index\\
 E.O.S & Isothermal & Equation of state\\
 $T_{*} $ (K) & $40000$ & Effective source temperature\\
 $R_* $ $(\textrm{R}_\odot)$ & $10 $ & Source radius \\
 L(pc$^3$) & 11.36 & Grid size\\
 $n_{\rm{cells}}$ & $128^3$ & Number of grid cells\\
\hline
\end{tabular}
\end{minipage}
\end{table}

\subsection{Testing Summary}
We have shown that TORUS satisfies a number of benchmark tests. The
radiative transfer scheme, including treatment of the diffuse field,
is in good agreement with the Lexington benchmark as calculated by
Cloudy \citep{1998PASP..110..761F}. The hydrodynamics algorithm
satisfies the Sod shock tube test and Sedov-Taylor blast wave density
distributions at a given point in time and has also been shown to
produce Kelvin-Helmholtz and Rayleigh-Taylor
instabilities. Furthermore our self-gravity calculation reproduces the
same $n=1$ polytropic density distribution as given by the Lane-Emden
equation following the collapse of a uniform density sphere. Finally,
the radiation hydrodynamics scheme has been shown to agree with the
analytical work of \cite{1978ppim.book.....S} for the rate of
expansion of an HII region. These results verify that TORUS
hydrodynamics and photoionization modules reproduce the standard
radiation and hydrodynamical benchmarks and is therefore ready for
application to new problems.

\section{Radiatively Driven Implosion Of A Bonnor-Ebert Sphere}
\label{MAIN}

Here we model the radiatively driven implosion of a Bonnor-Ebert sphere (a sphere in which the density varies according to equation \ref{BESeqn}) using three different treatments of the radiation field so as to distinguish their relative contributions to the evolution of the system. The three different radiation fields used are:

a) a monochromatic radiation field with the OTS approximation

b) a polychromatic radiation field with the OTS approximation

c) a polychromatic radiation field with the diffuse field, as outlined in section \ref{photoionSection}. 

The details of the system are very similar to that of \cite{2009MNRAS.393...21G}. A Bonnor-Ebert sphere of radius $1.6$\,pc resides at the centre of a $1.5\times10^{19}\,\textrm{cm}^3$ ($4.87$\,pc$^3$) grid of $128^3$ cells. Our model domain is slightly larger than that used by \cite{2009MNRAS.393...21G} so that the evolution of larger extent of any shadowed region can be studied. The BES has a core number density of $10^3$cm$^{-3}$, with the material surrounding the BES having a number density equal to that at the BES edge (such a BES is known as `incomplete'). The resolution of these models is currently limited by computational cost, however the number of grid cells here is equivalent to that used in the successful HII-region expansion test of section \ref{hii_test}. This model is also on a smaller length scale than that of the HII-region expansion test so the resolution will be sufficiently high. As mentioned in section \ref{implement}, we will be implementing an AMR grid that will enable the calculation of models at higher resolutions in future work.

The primary photon source is a star that lies outside the grid in the $-x$ direction. The radiation field is assumed to enter the grid plane parallel at the $-x$ boundary, at which photon packets are initiated at random locations and the flux is modified to account for geometric dilution. 


As well as considering the three different radiation schemes mentioned above, we also treat the three different flux regimes considered in \cite{2009MNRAS.393...21G}. These are denoted high, medium and low flux and correspond to the BES being located just within, on the edge of and just beyond the Str\"{o}mgren radius respectively. The fluxes and corresponding stellar properties that were used to generate these different flux regimes are given in Table \ref{rdiParams}, along with the other parameters used for this model.  

In all of the models presented here, the hydrodynamic boundary conditions are periodic at $\pm y$ and $\pm z$ and free outflow/no inflow at $\pm x$. Dirichlet boundary conditions are used for the self-gravity calculation and the radiation field boundaries are free outflow/no inflow. Using this boundary condition for the radiation field can lead to reduced sampling at the domain boundaries, where material will be subject to non-symmetric diffuse flux.
\\ 

The free fall time for this cloud is approximately $3$\,Myr,
estimated using $1/(\sqrt{G\rho_{\rm{max}}})$ where $\rho_{\rm{max}}$
is the central density and G is the gravitational constant. This is a
factor of $15$ longer than the total simulation time of
$200$\,kyr. Radiation hydrodynamics will thus dominate gravitational
effects in the evolution of the system. We dump the state of the
computational grid every 5\,kyr.

\begin{table}
 \centering
  \caption{Parameters used for the RDI of a  BES model.}
  \label{rdiParams}
  \begin{tabular}{@{}l c l@{}}
  \hline
   Variable (Unit) & Value & Description\\
 \hline
 $R_{\rm{c}}$ (pc) & 1.6 & Cutoff radius\\
 $n_{\rm{max}}$ ($\textrm{cm}^{-3} $)& $1000$ & Peak BES number density\\
 $T_{\rm{o}}$ (K) & 10 & Initial temperature of grid\\
 $\gamma $ & 1 & Adiabatic index\\
 E.O.S & Isothermal & Equation of state\\
 $\Phi_{\rm{lo}}$ ($\textrm{cm}^{-2}$) & $9.0\times10^8$& Low ionizing flux\\
 $D_{\rm{lo}}$ (pc) & $(-10.679,0,0)$ & Source position (low flux) \\
 $ \Phi_{\rm{med}}$ ($\textrm{cm}^{-2}$) & $4.5\times10^9$ & Intermediate ionizing flux\\
 $D_{\rm{med}}$ (pc) & $(-4.782,0,0)$ & Source position (medium flux) \\ 
 $\Phi_{\rm{hi}}$ ($\textrm{cm}^{-2}$) & $9.0\times10^9$ & High ionizing flux\\ 
 $D_{\rm{hi}}$ (pc) & $(-3.377,0,0)$ & Source position (high flux)\\
 L(pc$^3$) & 4.87 & Grid size\\
 $n_{\rm{cells}}$ & $128^3$ & Number of grid cells\\
CFL & 0.3 & CFL parameter\\
\hline
\end{tabular}
\end{table}

\subsection{Results and Discussion}

\subsubsection{Initial properties}
\begin{figure*}	
		\hspace{10pt}
		\includegraphics[width=51mm, height=51mm]{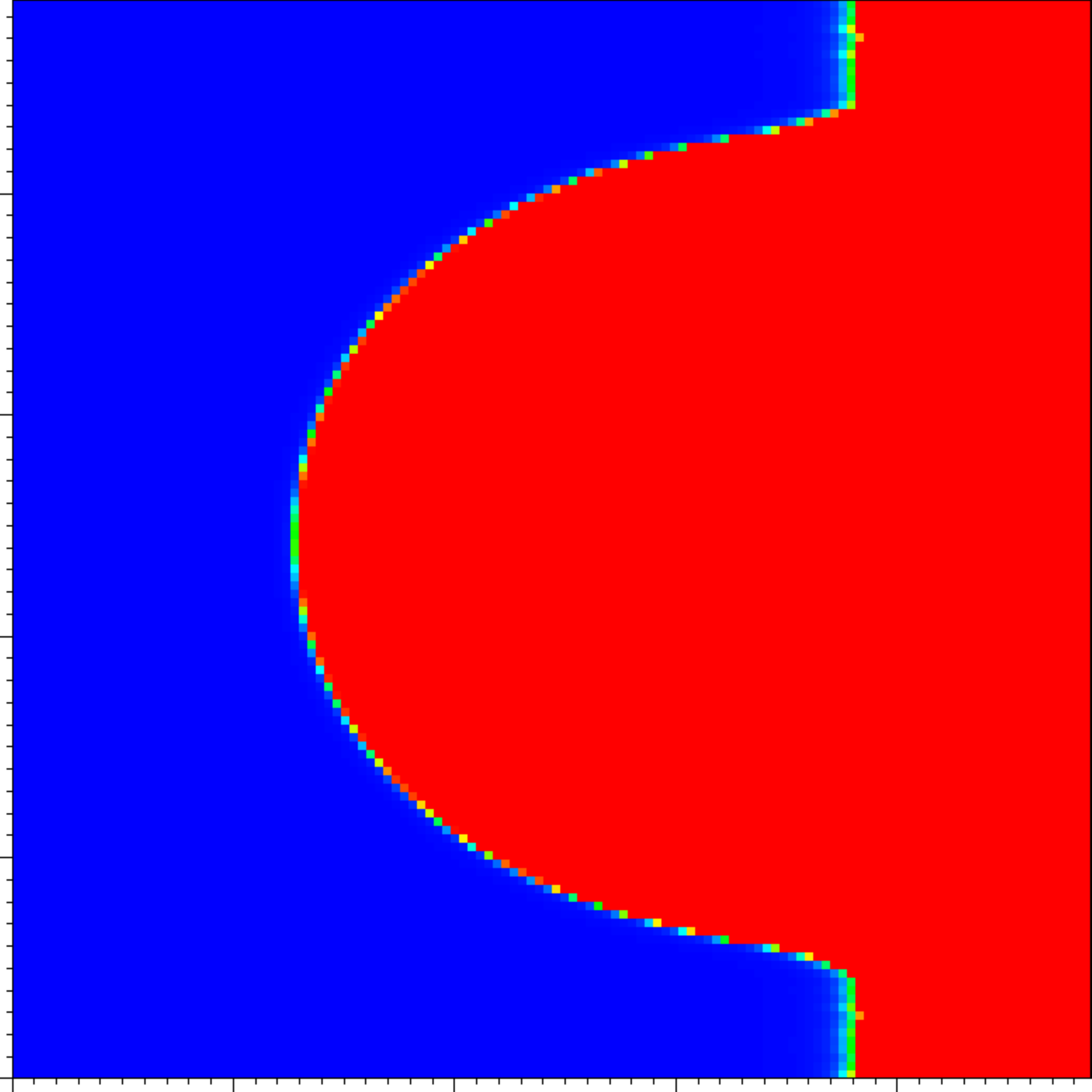}
		\hspace{3pt}
		\vspace{10pt}
		\includegraphics[width=51mm, height=51mm]{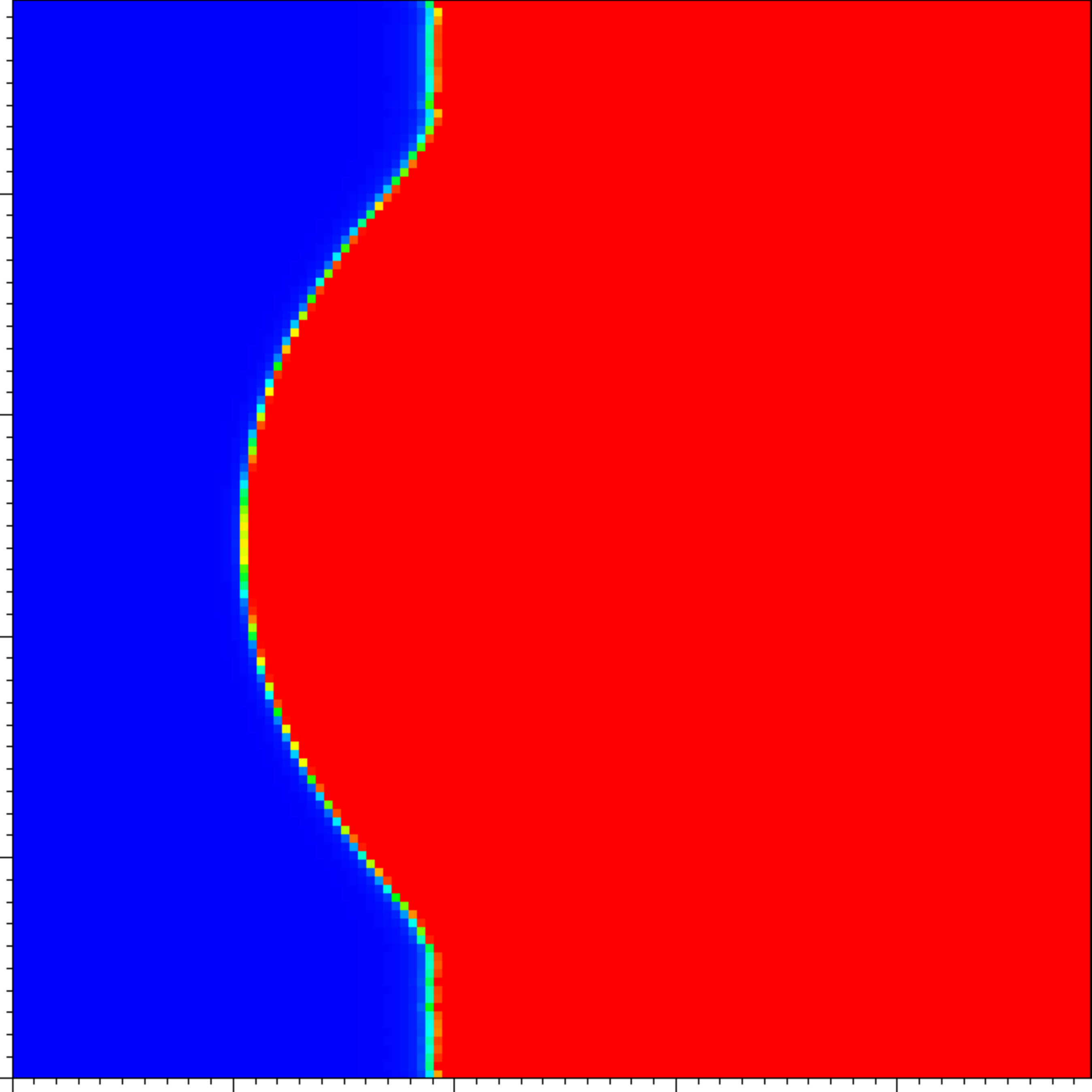}
		\hspace{3pt}
		\includegraphics[width=61.5mm, height=51mm]{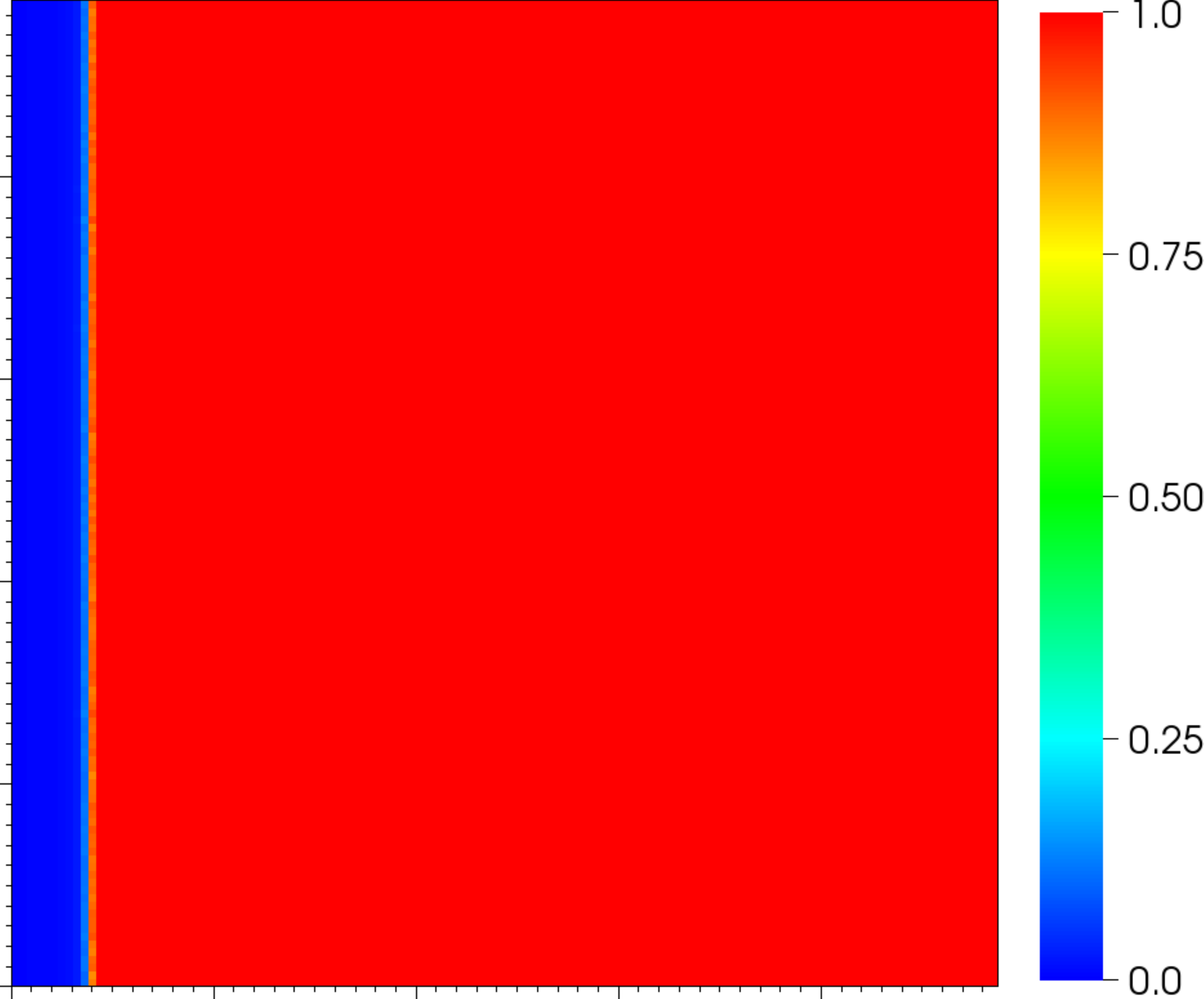}		

		\hspace{-20pt}
		\vspace{10pt}
		\includegraphics[width=51mm, height=51mm]{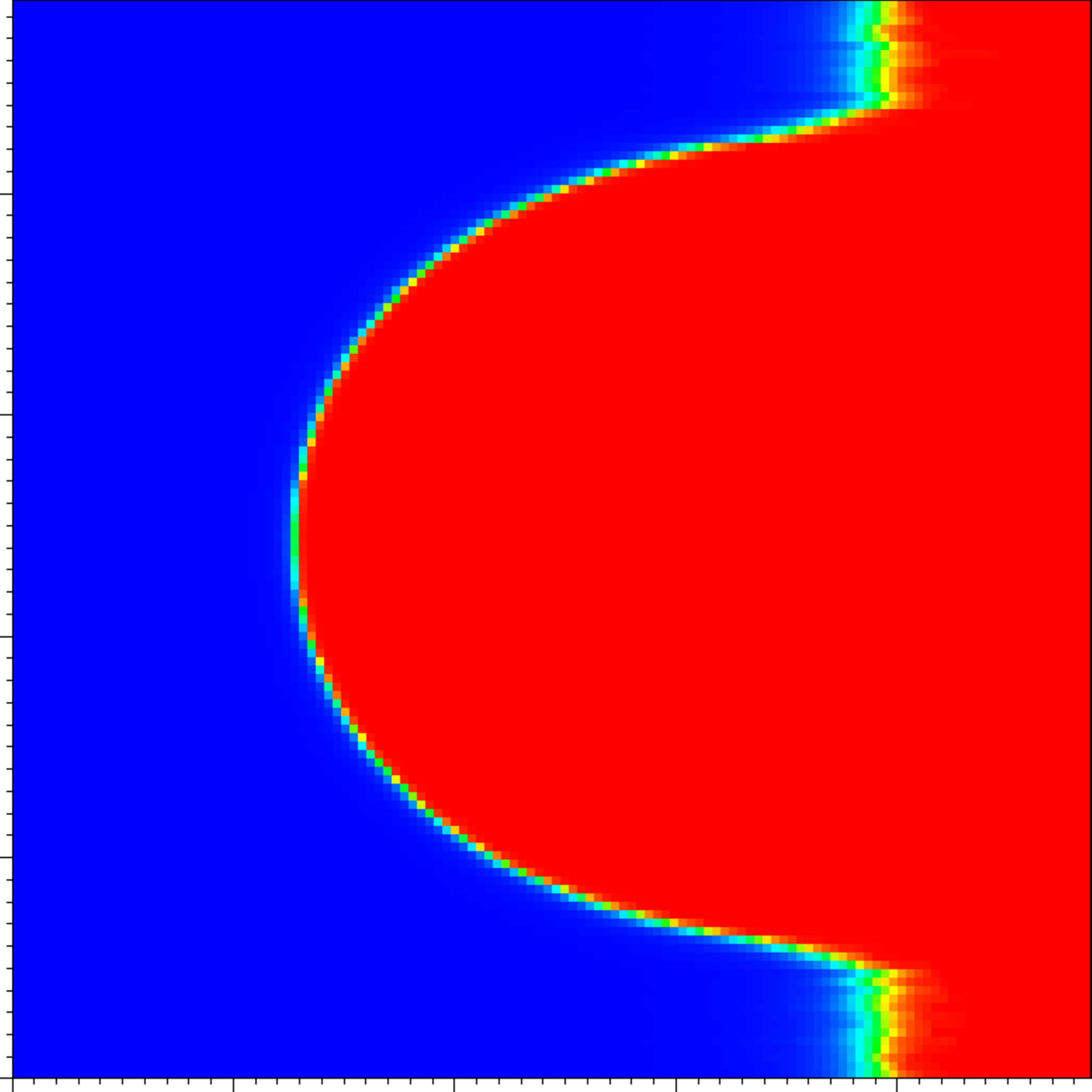}
		\hspace{3pt}
		\includegraphics[width=51mm, height=51mm]{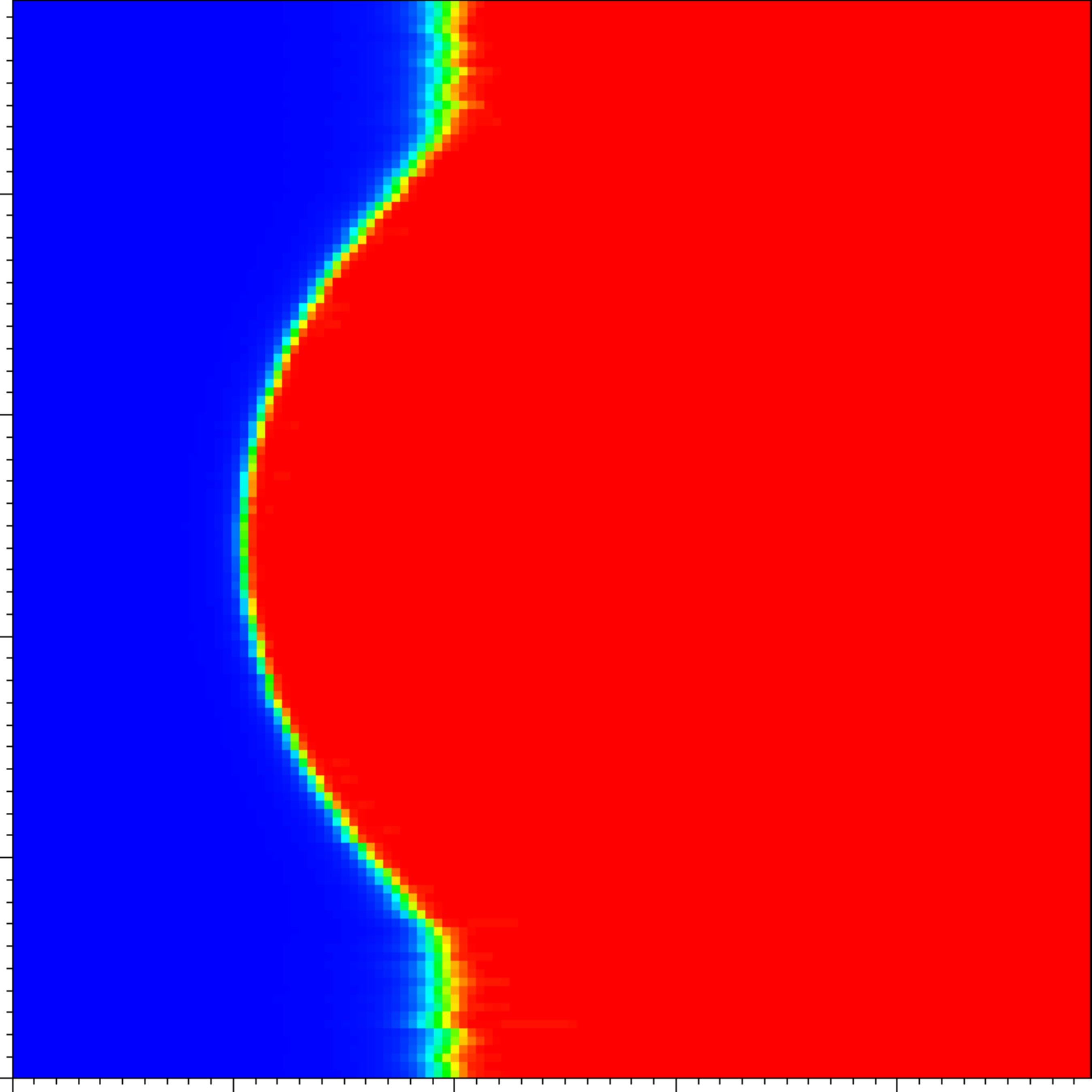}
		\hspace{3pt}
		\includegraphics[width=51mm, height=51mm]{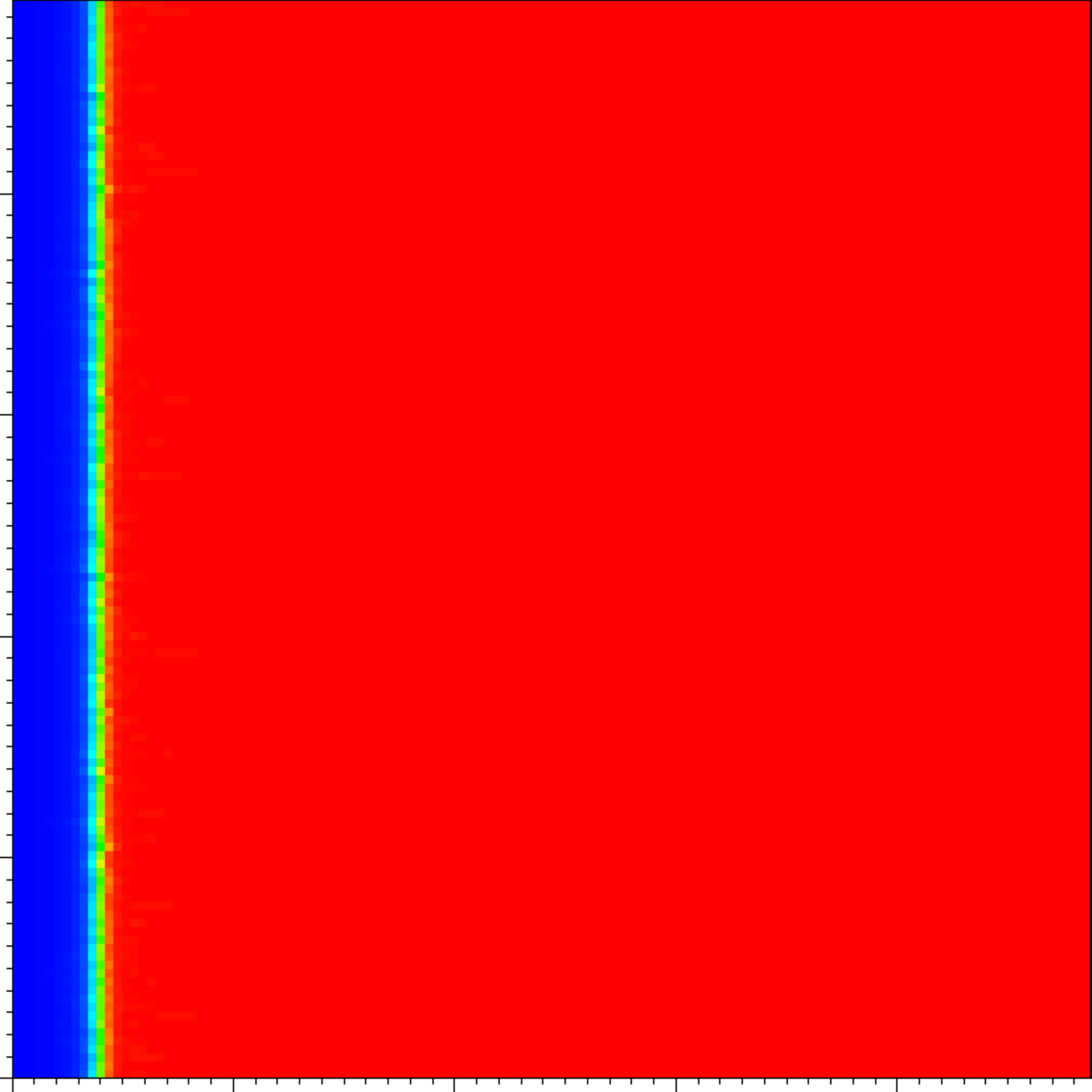}
	
		\hspace{-20pt}
		\vspace{10pt}
		\includegraphics[width=51mm, height=51mm]{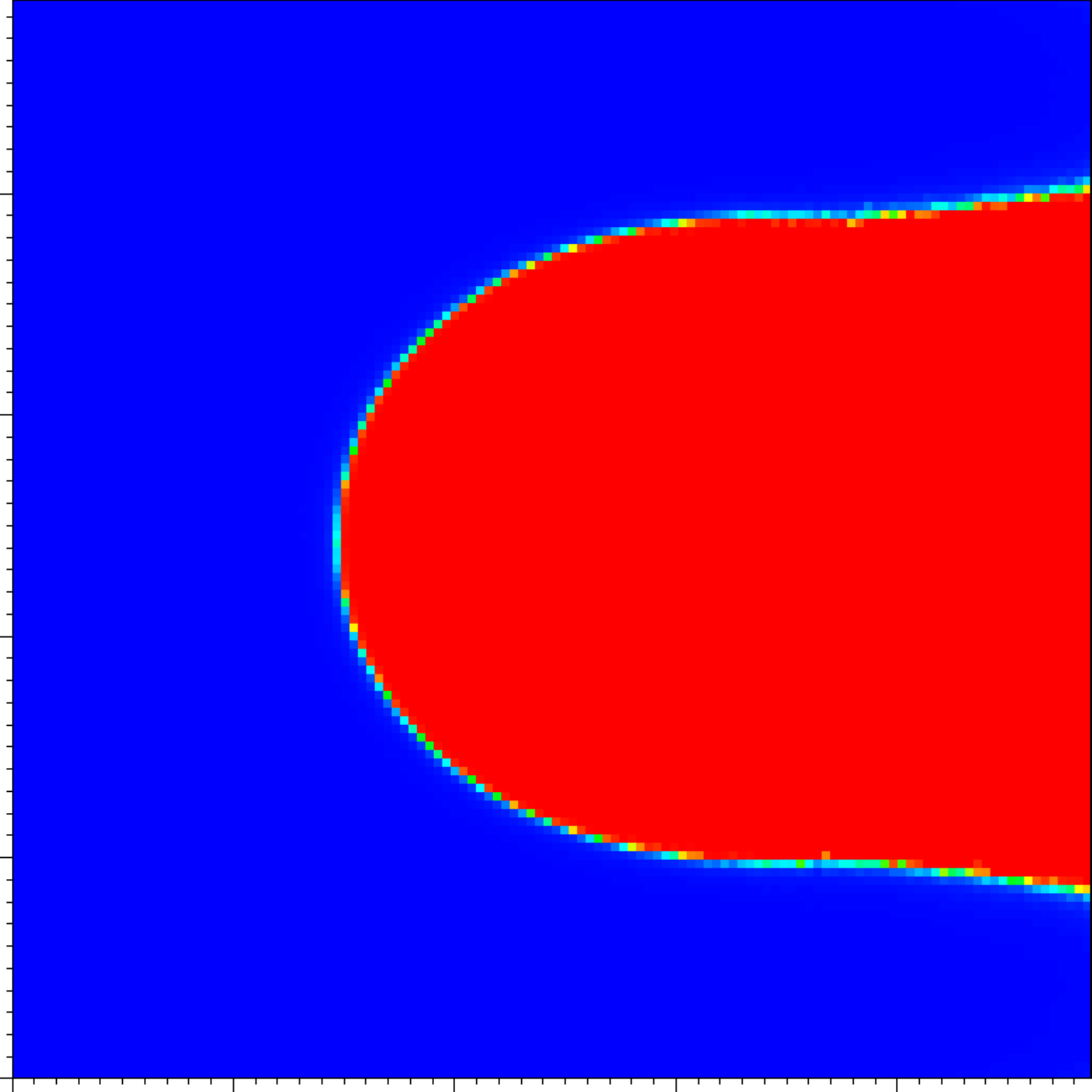}
		\hspace{3pt}
		\includegraphics[width=51mm, height=51mm]{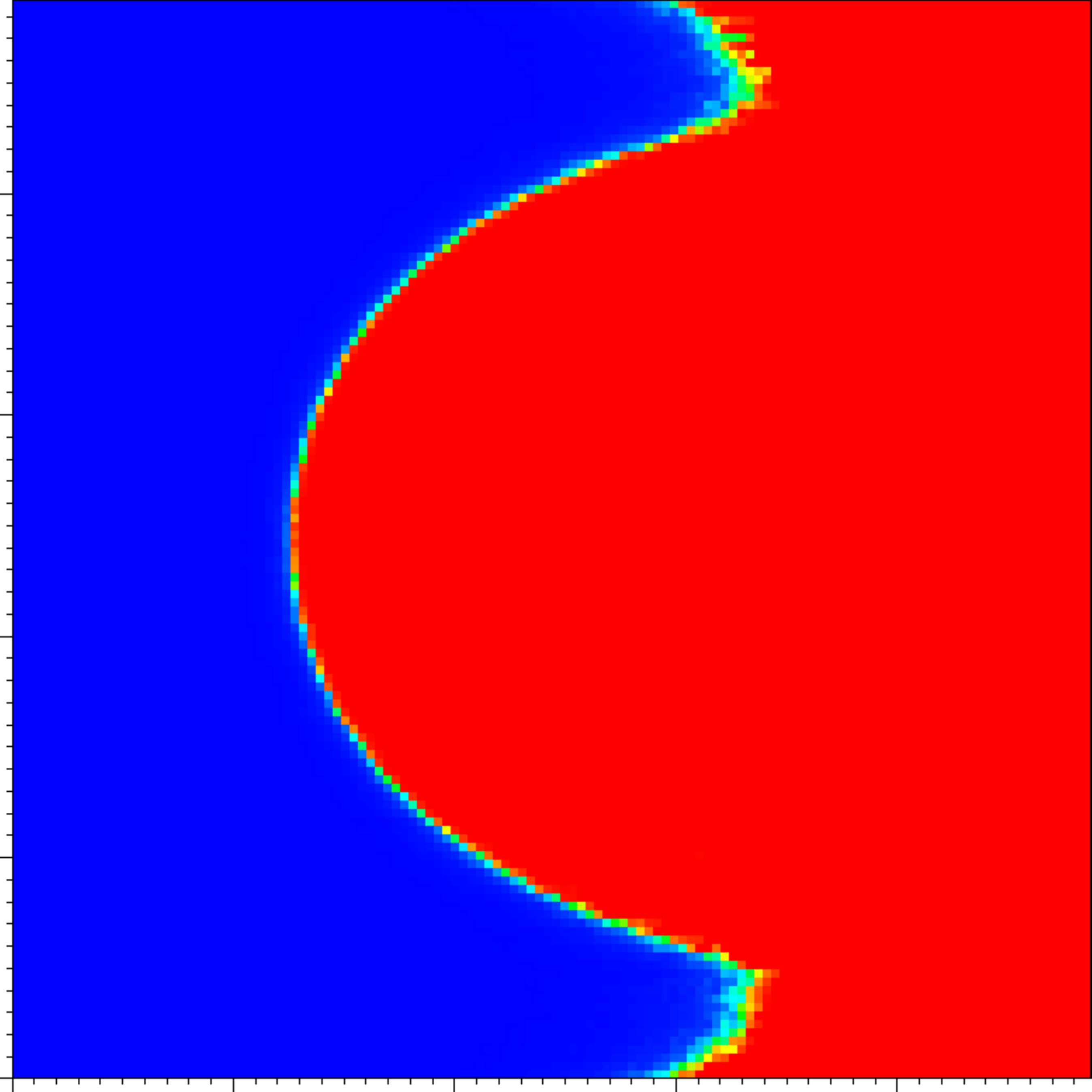}
		\hspace{3pt}
		\includegraphics[width=51mm, height=51mm]{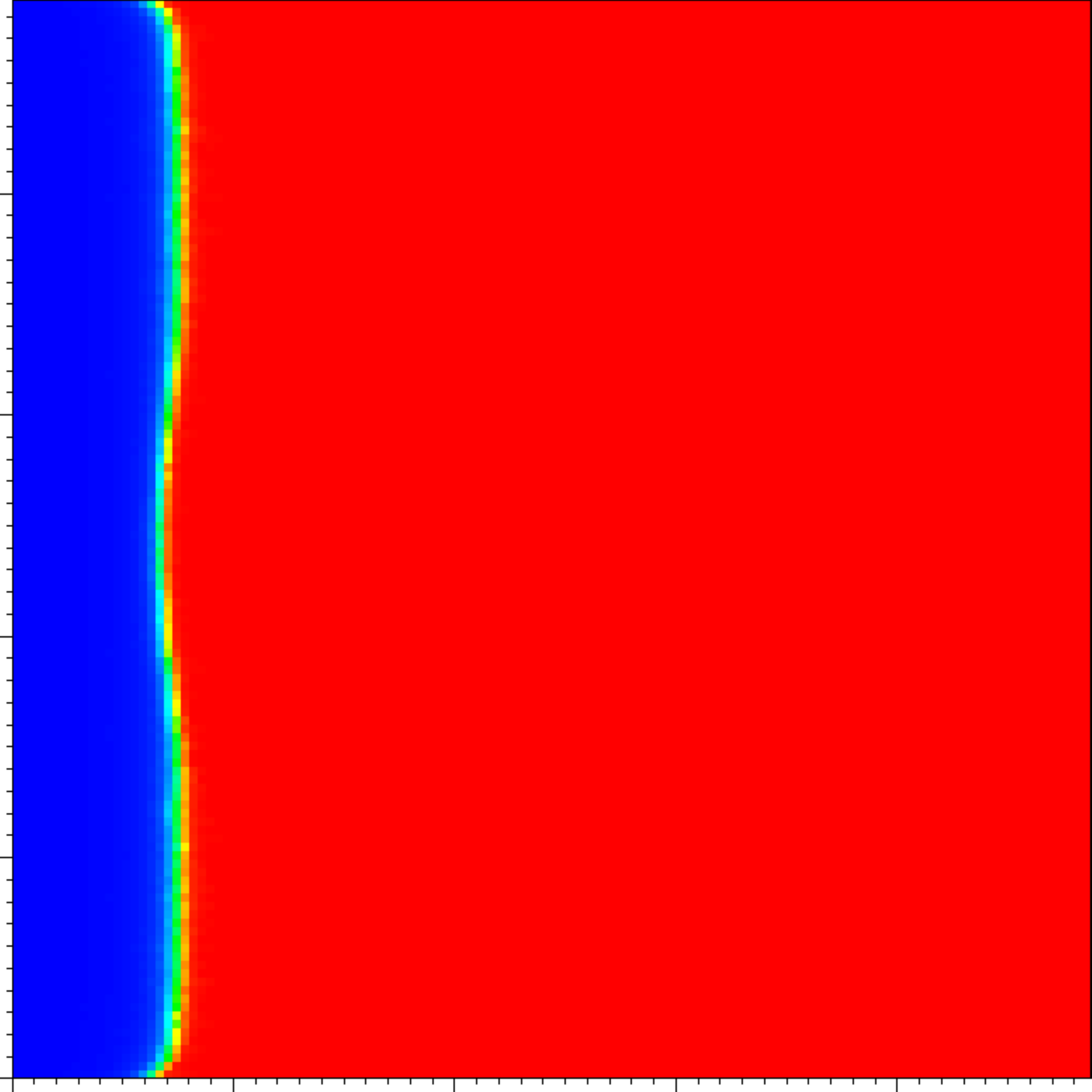}
		\caption{Top row) The hydrogen ionization fraction for the OTS monochromatic model. Middle row) The hydrogen ionization fraction for the OTS polychromatic model. Bottom row) The hydrogen ionization fraction for the model which includes the diffuse field. Columns are, from left to right, high, medium and low flux regimes. Each frame is a slice through the computational grid, which is a cube with sides 4.87\,pc long. Major ticks are separated by 1\,pc.}
		\label{inis}
\end{figure*}

The initial ionization state of the system for all three flux regimes and photoionization schemes is shown in Figure \ref{inis}. At this stage there is already a noticeable difference between them in the extent of their un-ionized regions.
The top row from Figure \ref{inis} represents the starting point that would be obtained by most pre-existing models \citep[specifically, it is a zoomed out version of][]{2009MNRAS.393...21G}.
The middle row is the start point for the models in which a polychromatic radiation field is considered, but the OTS approximation is still applied. In comparison with the top row, it is clear that the extent of the ionized region has increased slightly and the transition region has been smoothed out. This is due to hard radiation, which penetrates more deeply since the photoionization cross section approximately decreases in proportion to the inverse cube of the photon frequency \citep{1989agna.book.....O}.
The bottom row is the starting H{\sc i} fraction for the models that include the diffuse field. In all three flux regimes the extent of the ionized region is significantly different to that of the other two sets of models. At high flux material in the wings of the model is completely ionized, the medium flux model I-front has significantly wrapped itself around the BES and the low flux model I-front now grazes the BES.
Note that the curved I-front wings towards the edge of the low and medium flux models arise because these models do not include periodic photon packet boundary conditions and so these regions are subject to a non-symmetric diffuse ionizing flux.
\\

\begin{figure*}		
%
%
%
		\hspace{10pt}
		\includegraphics[width=51mm, height=51mm]{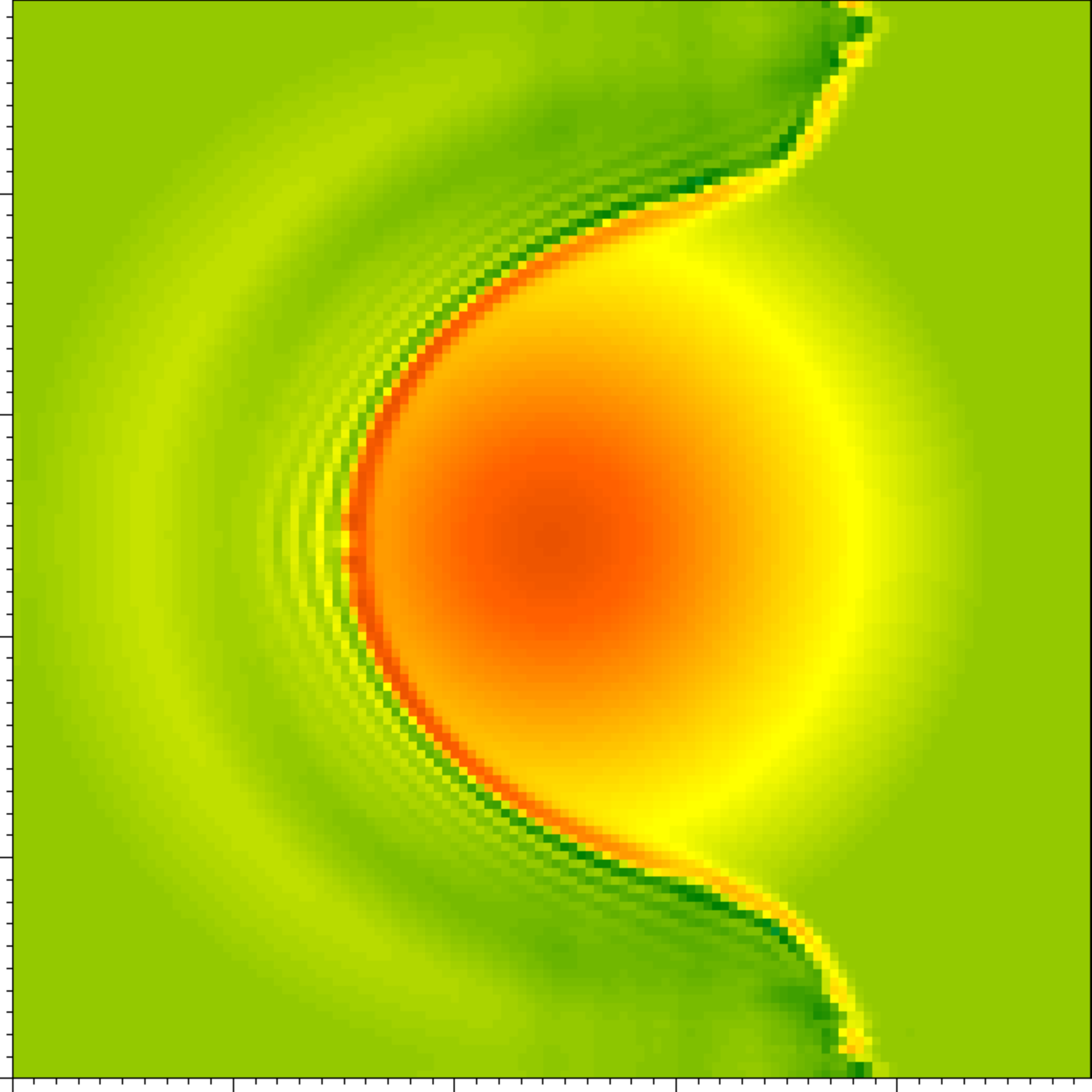}
		\hspace{3pt}
		\vspace{10pt}
		\includegraphics[width=51mm, height=51mm]{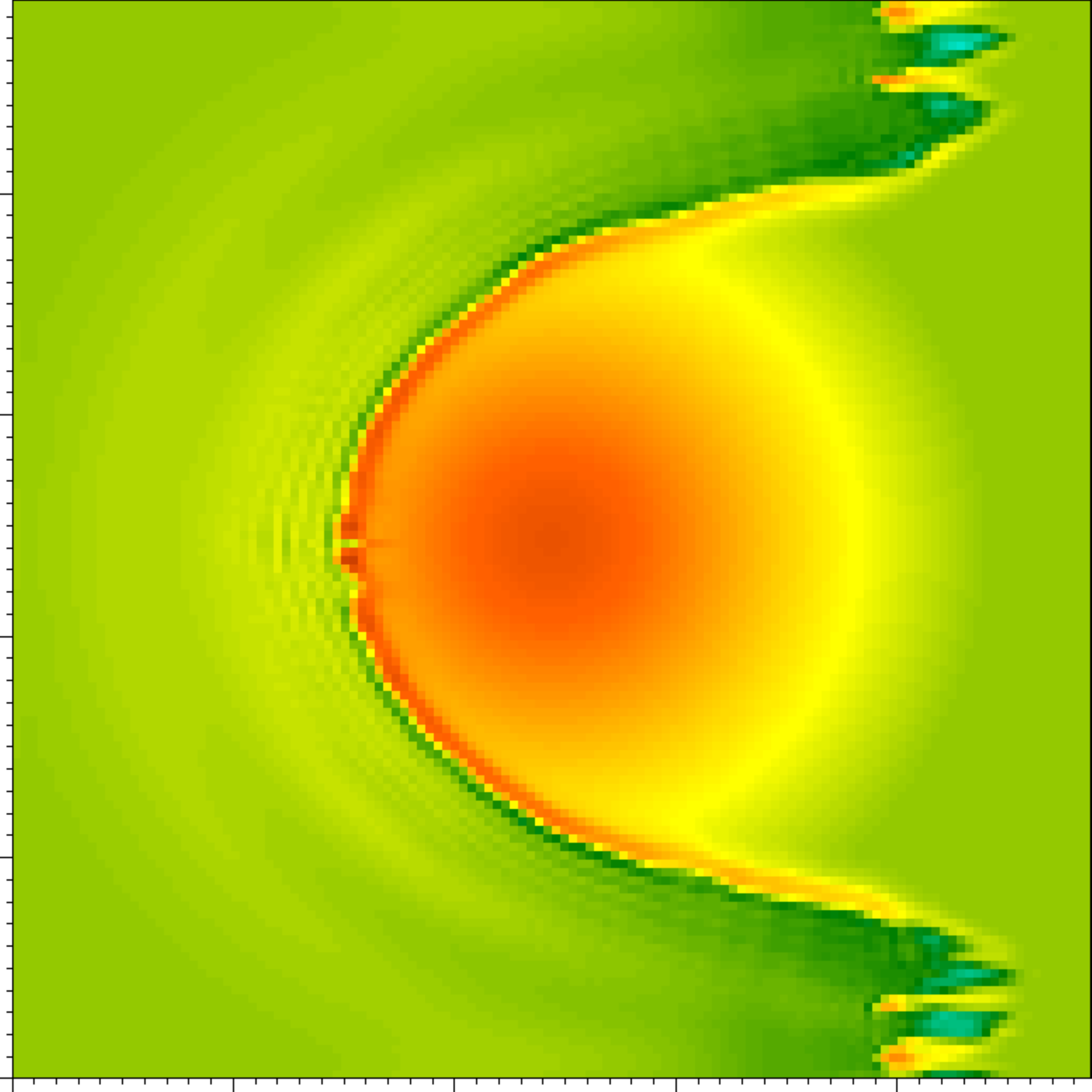} 
		\hspace{3pt}
		\includegraphics[width=61.7mm, height=51.2mm]{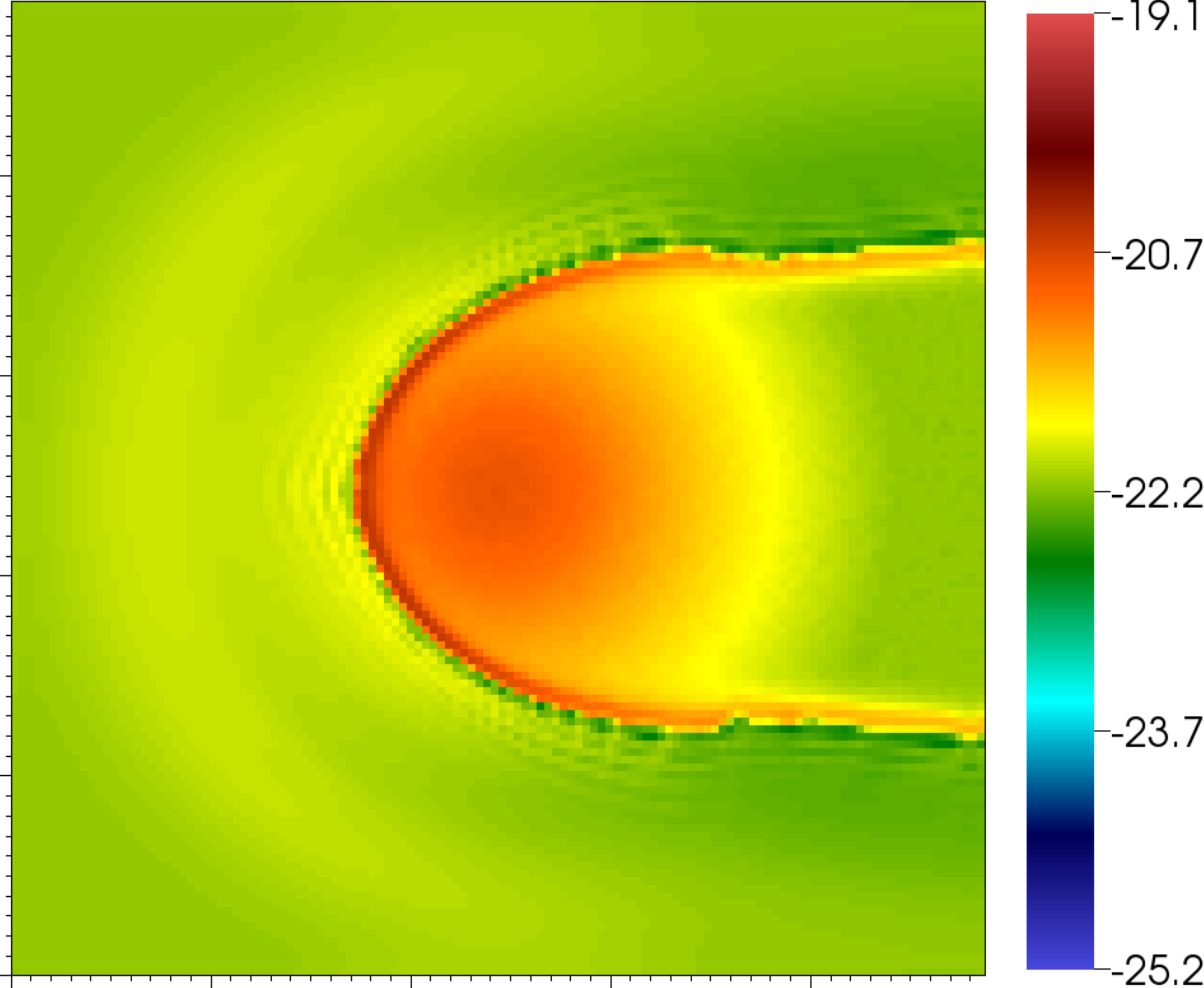}

		\hspace{-20pt}
		\vspace{10pt}
		\includegraphics[width=51mm, height=51mm]{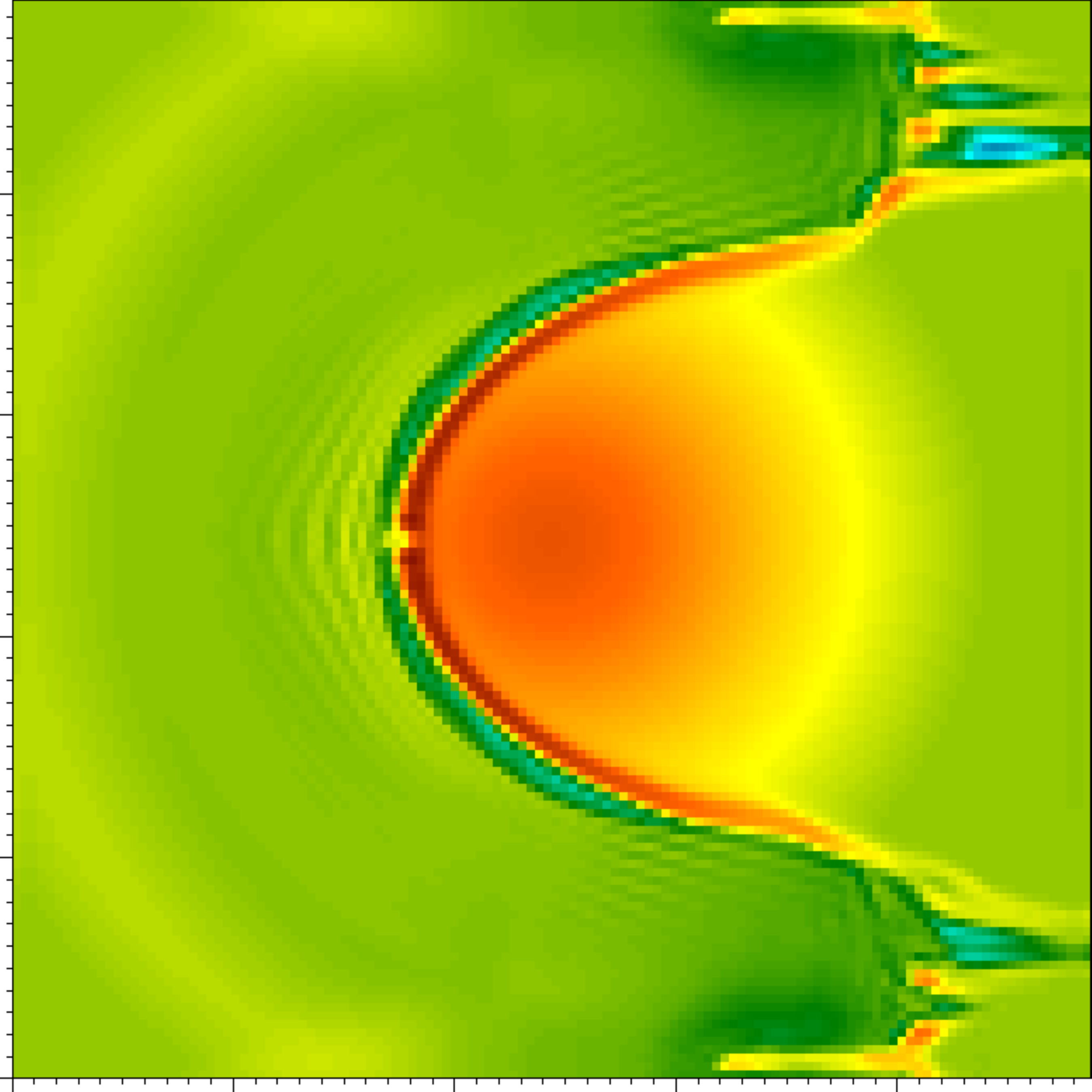}
		\hspace{3pt}
		\includegraphics[width=51mm, height=51mm]{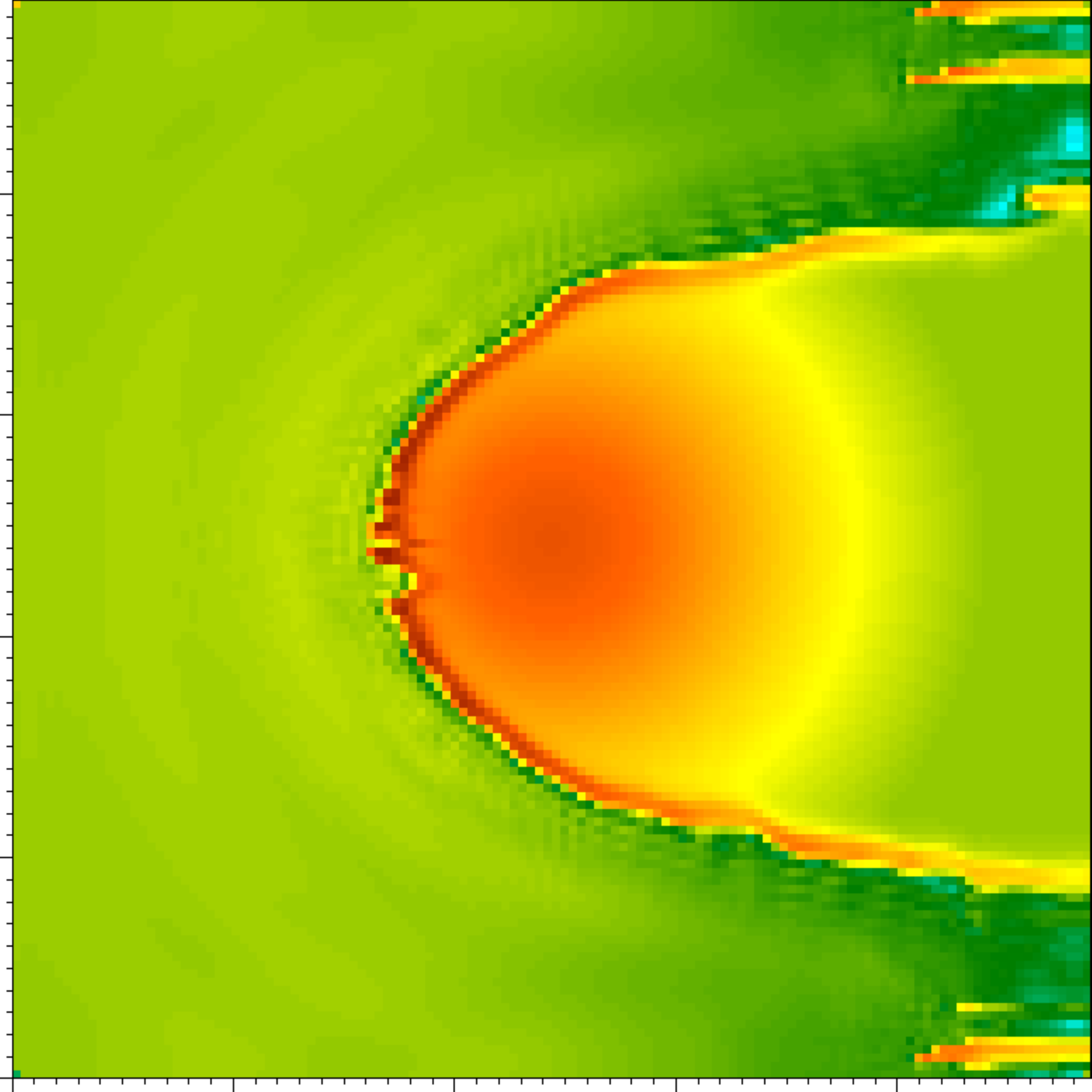}
		\hspace{3pt}
		\includegraphics[width=51mm, height=51mm]{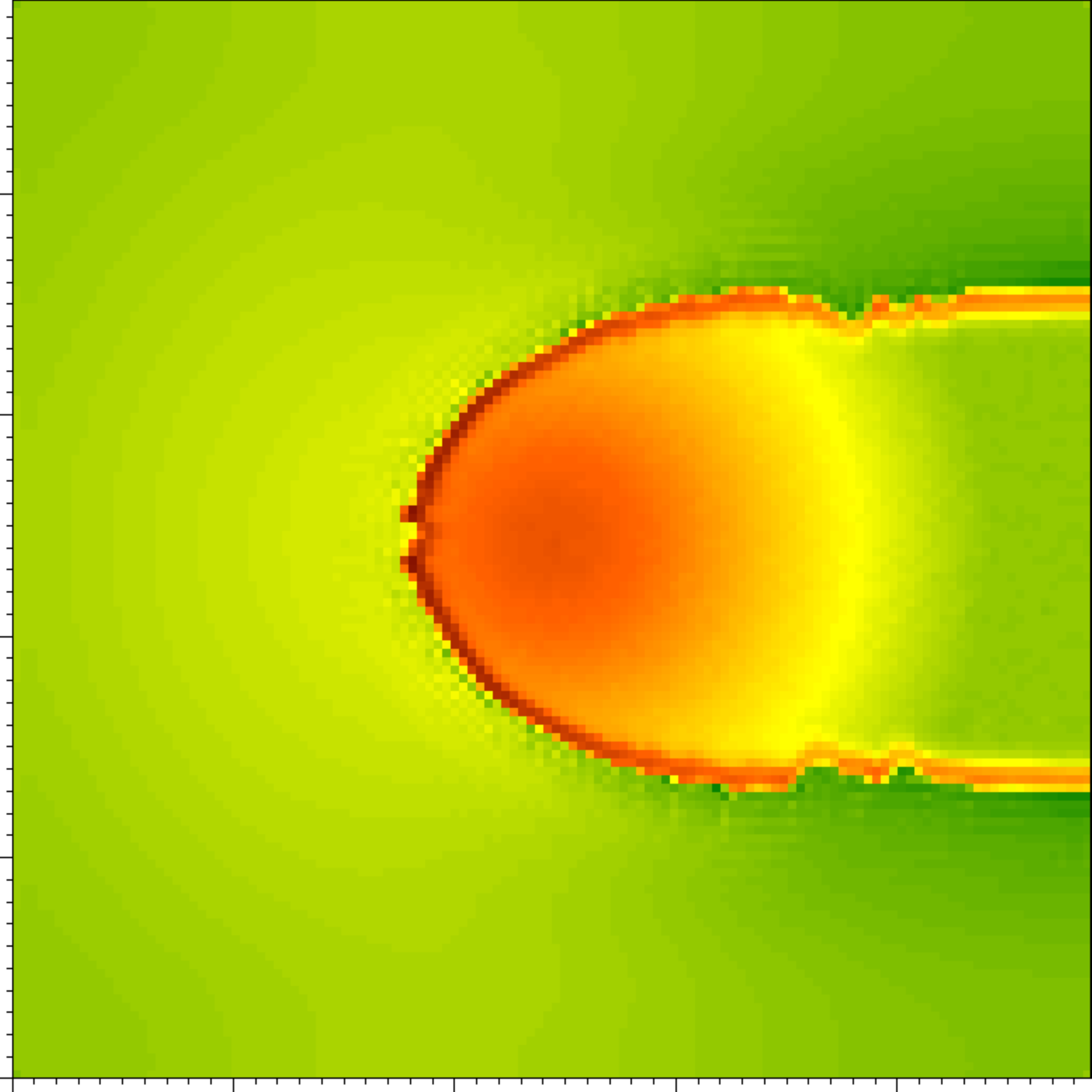}

		\hspace{-20pt}
		\vspace{10pt}
		\includegraphics[width=51mm, height=51mm]{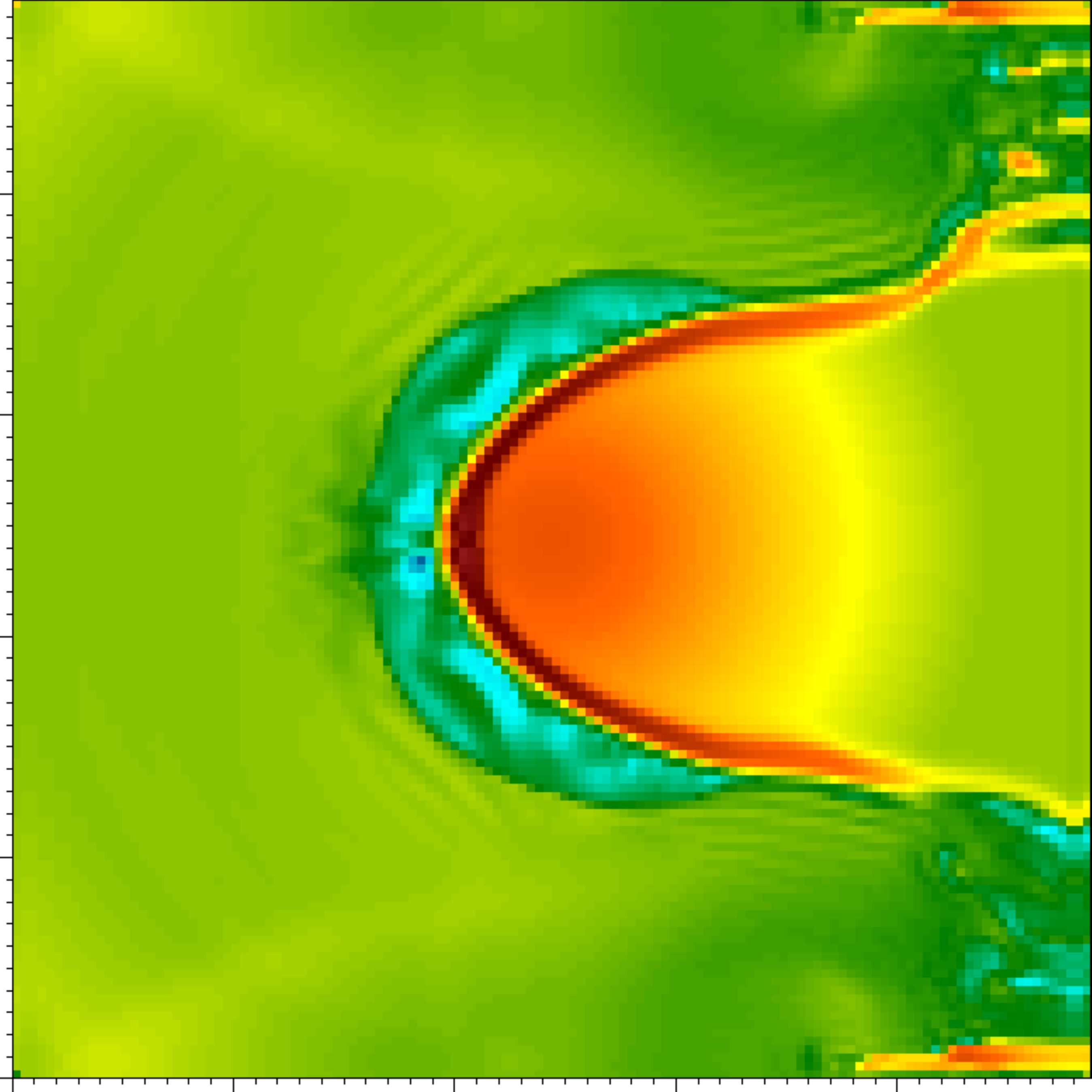}
		\hspace{3pt}
		\includegraphics[width=51mm, height=51mm]{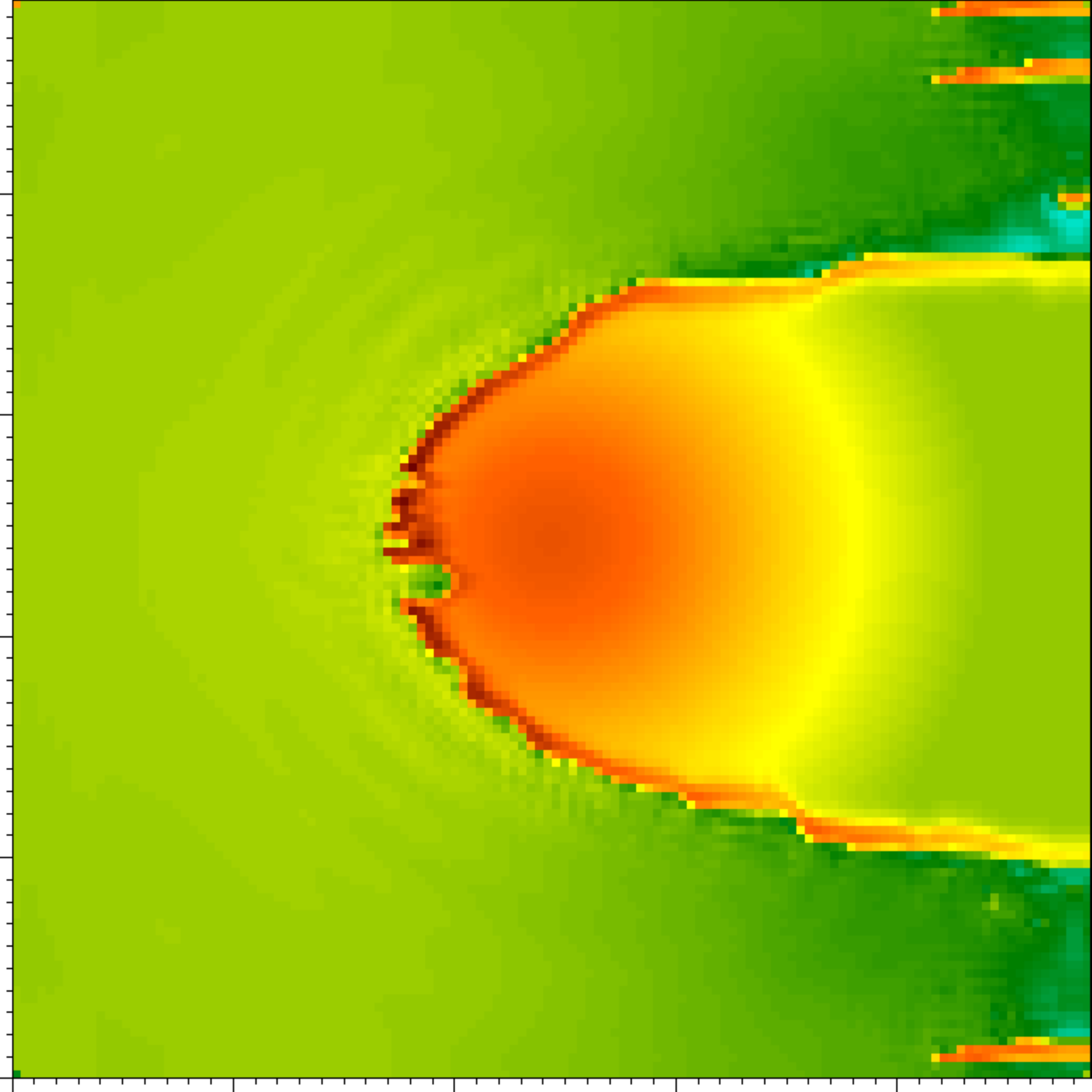}
		\hspace{3pt}
		\includegraphics[width=51mm, height=51mm]{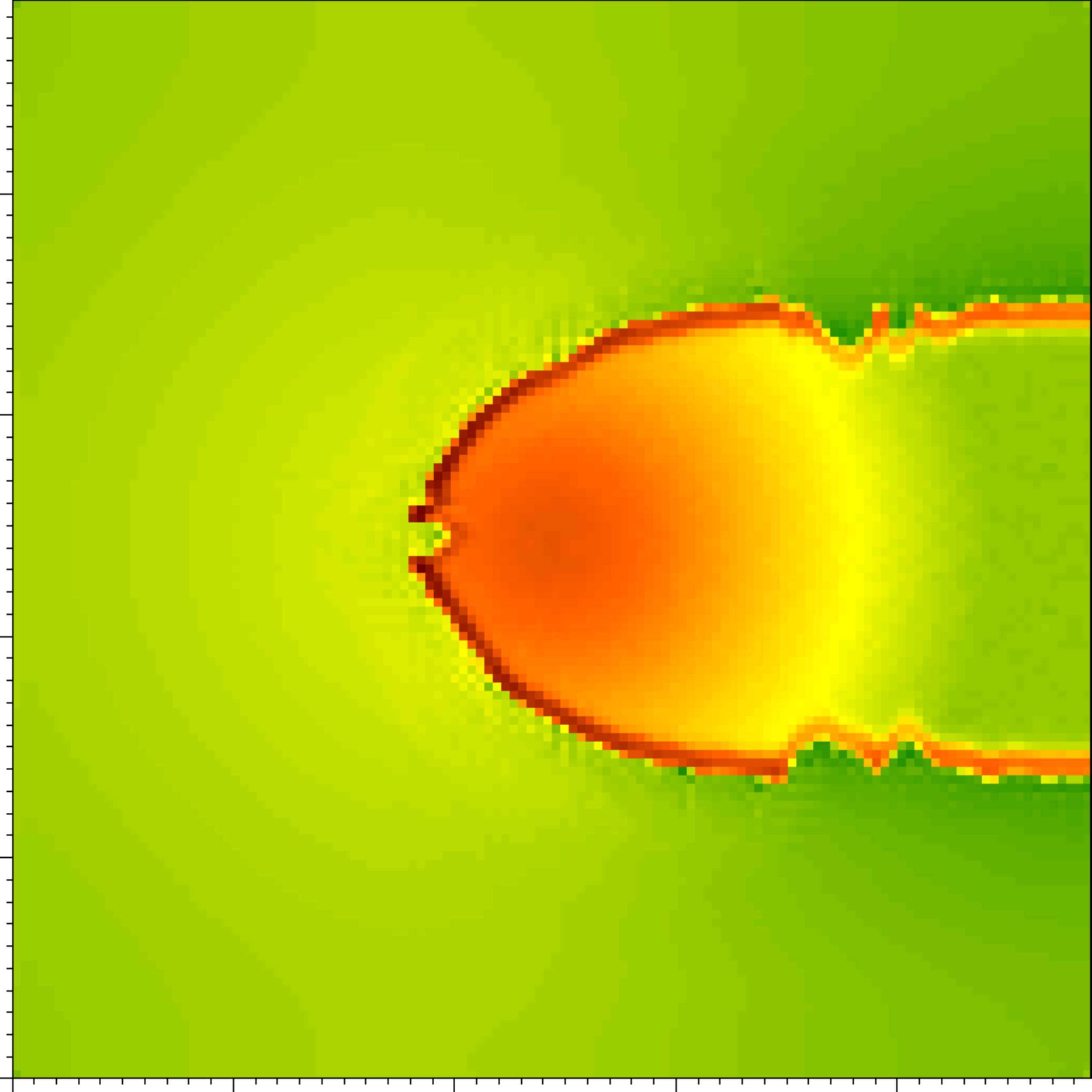}

		\hspace{-20pt}
		\vspace{10pt}
		\includegraphics[width=51mm, height=51mm]{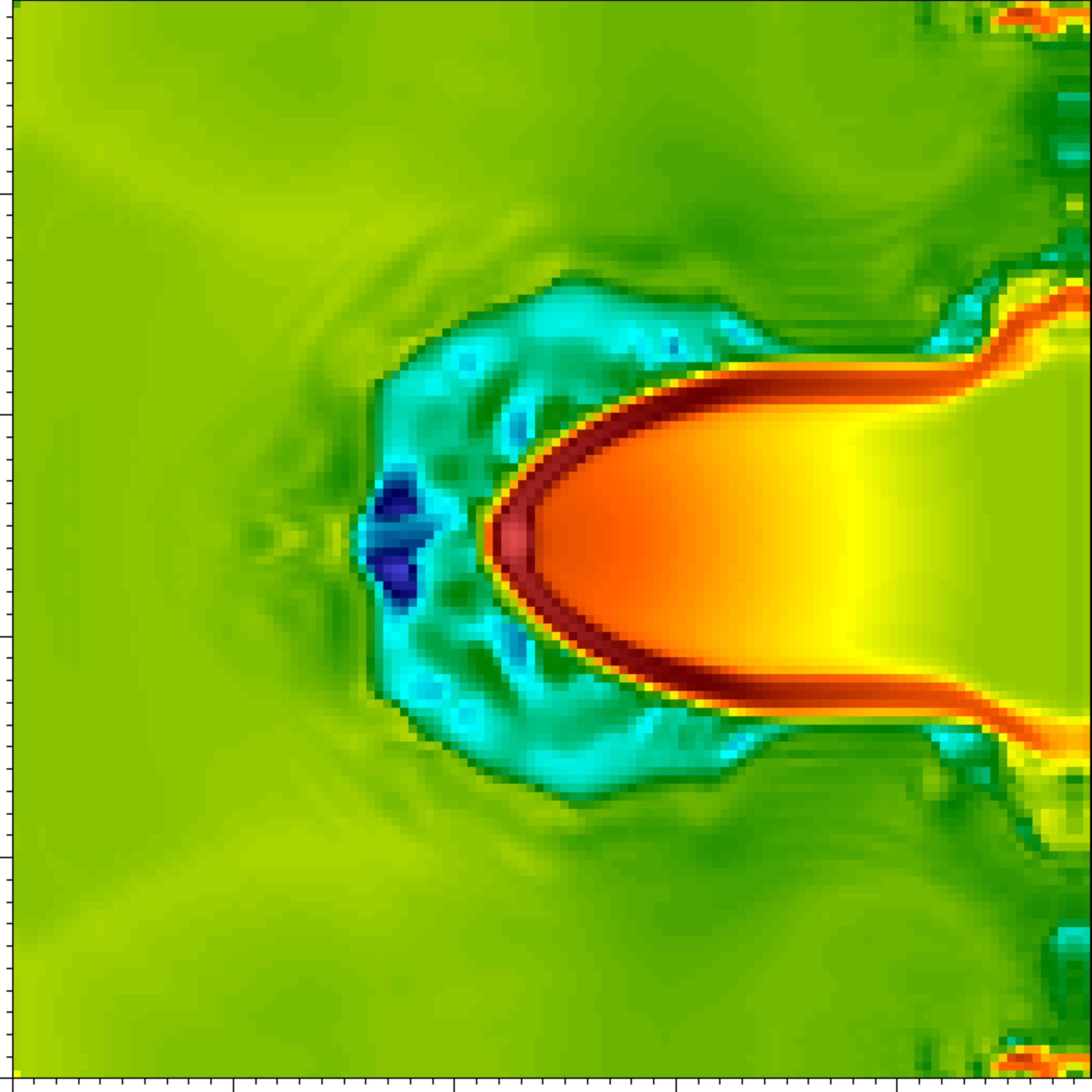}
		\hspace{3pt}
		\includegraphics[width=51mm, height=51mm]{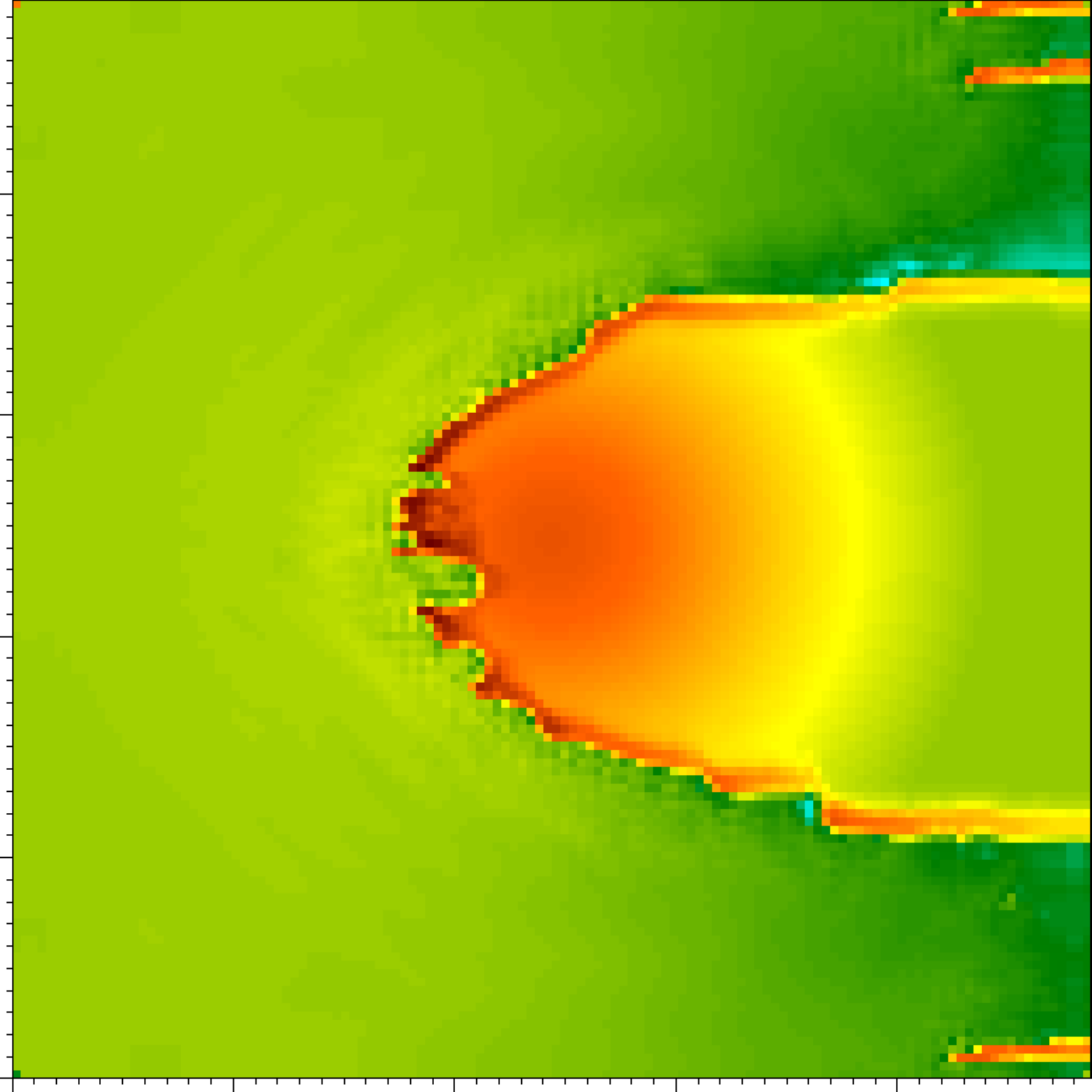}
		\hspace{3pt}
		\includegraphics[width=51mm, height=51mm]{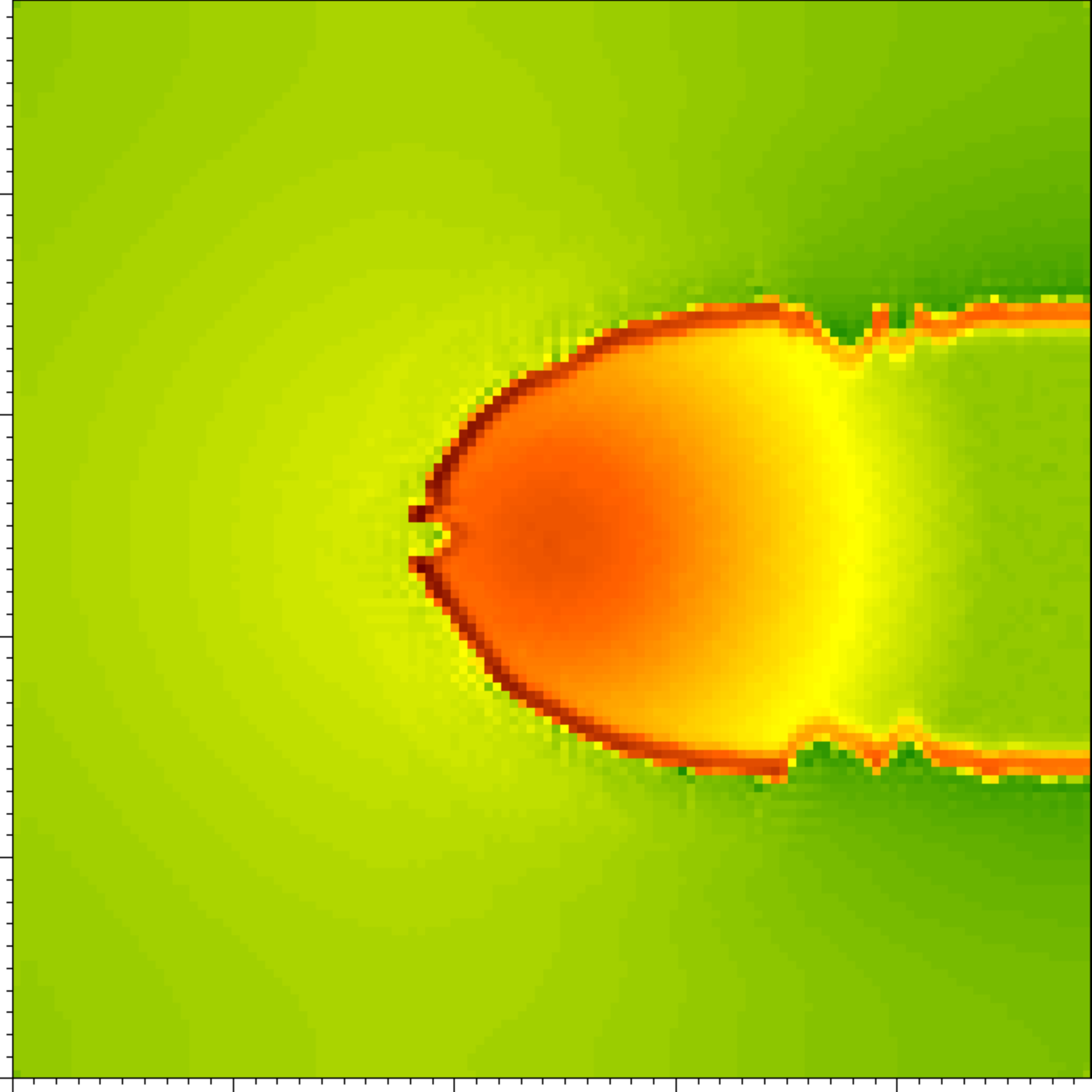}
  \caption{The high flux model logarithmic density distributions (cgs). Left column) Monochromatic models. Middle column) Polychromatic models. Right column) Polychromatic-diffuse models. Time is increasing from top to bottom, with snapshots at 50, 100, 150 and 200\,kyr. Each frame is a slice through the computational grid, which is a cube with sides 4.87\,pc long. Major ticks are separated by 1\, pc.}
		\label{RDI_HI}
\end{figure*}

\begin{figure*}
%
%
%
%

		\hspace{10pt}
 		\includegraphics[width=51mm, height=51mm]{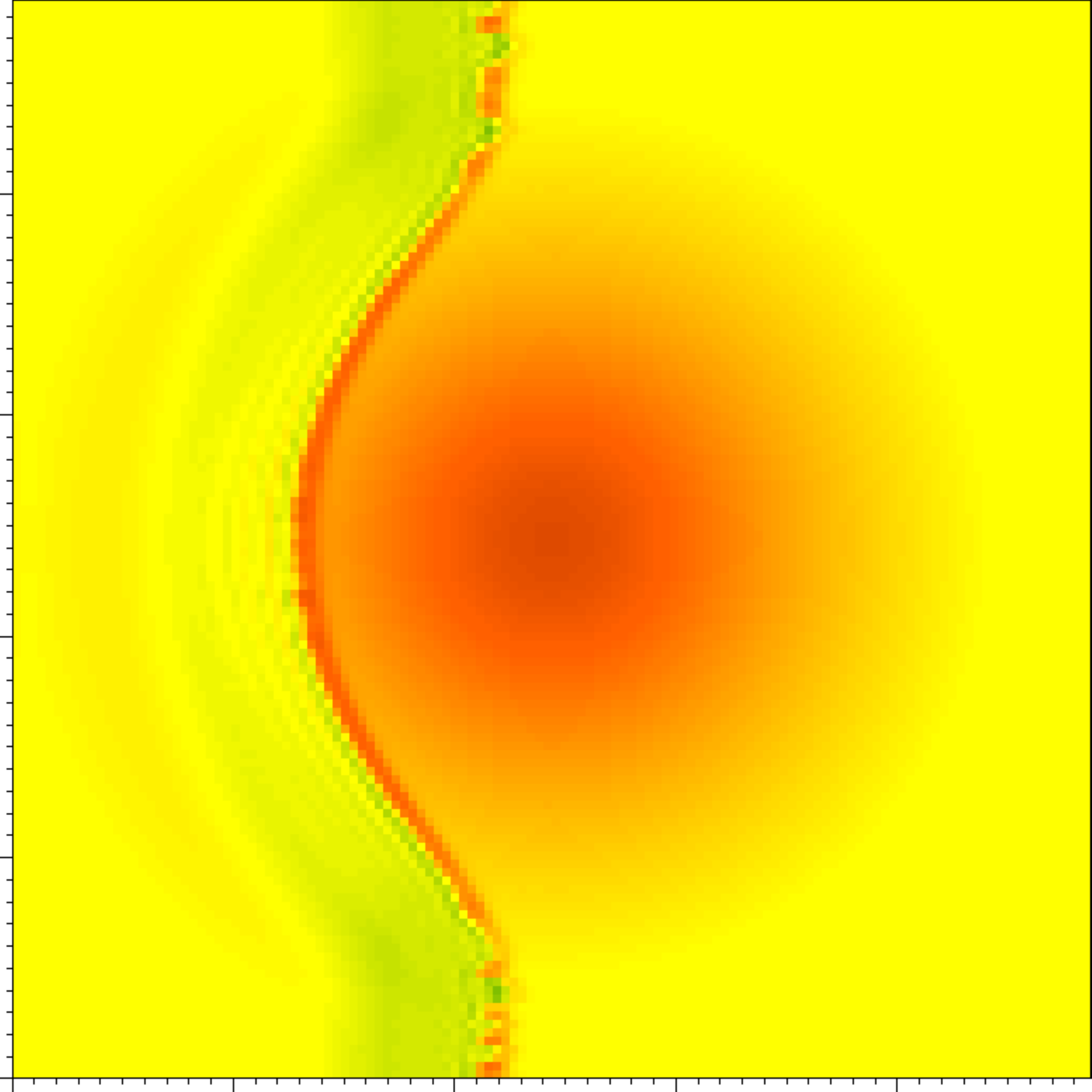}
		\hspace{3pt}
		\vspace{10pt}
		\includegraphics[width=51mm, height=51.1mm]{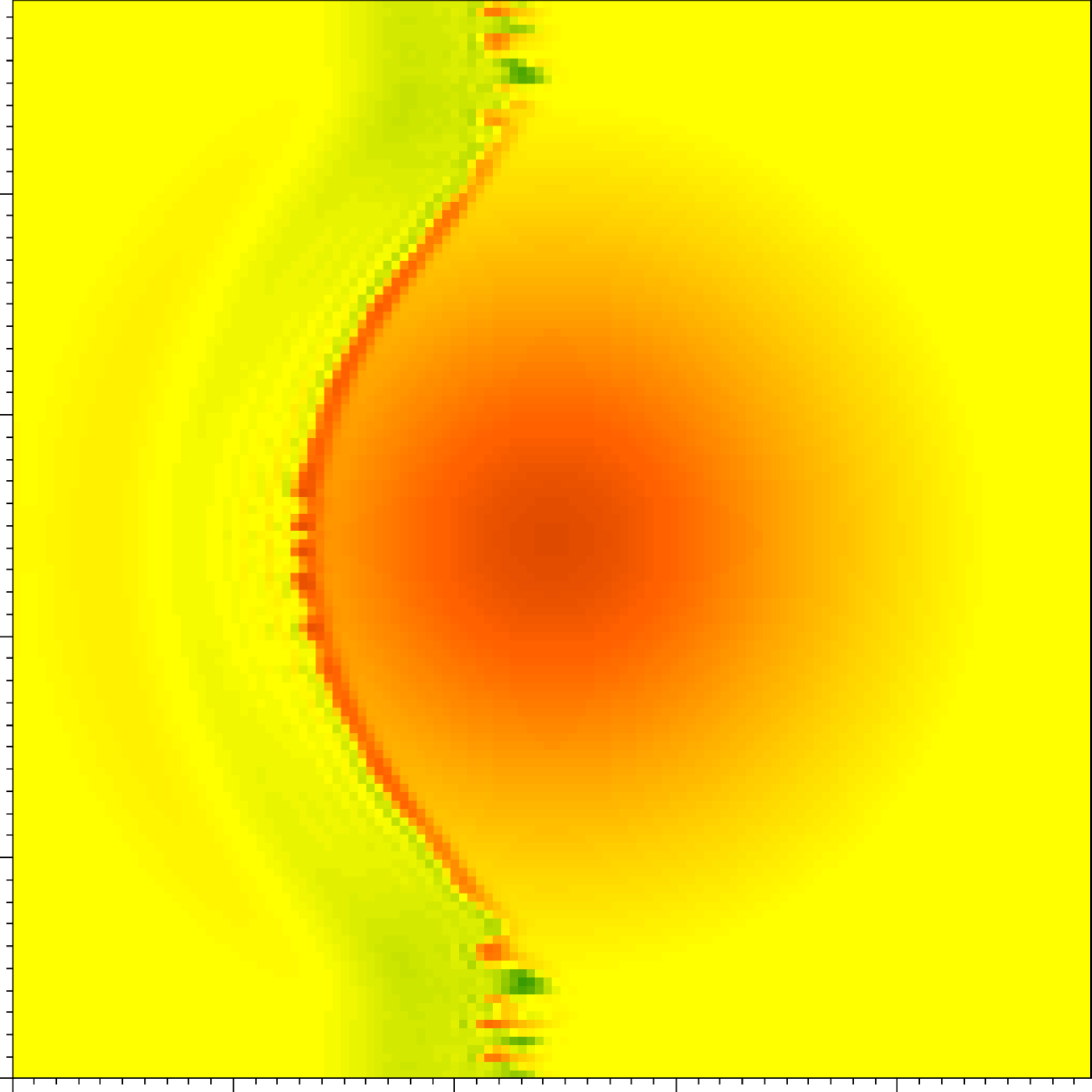}
		\hspace{3pt}
		\includegraphics[width=61.7mm, height=51.2mm]{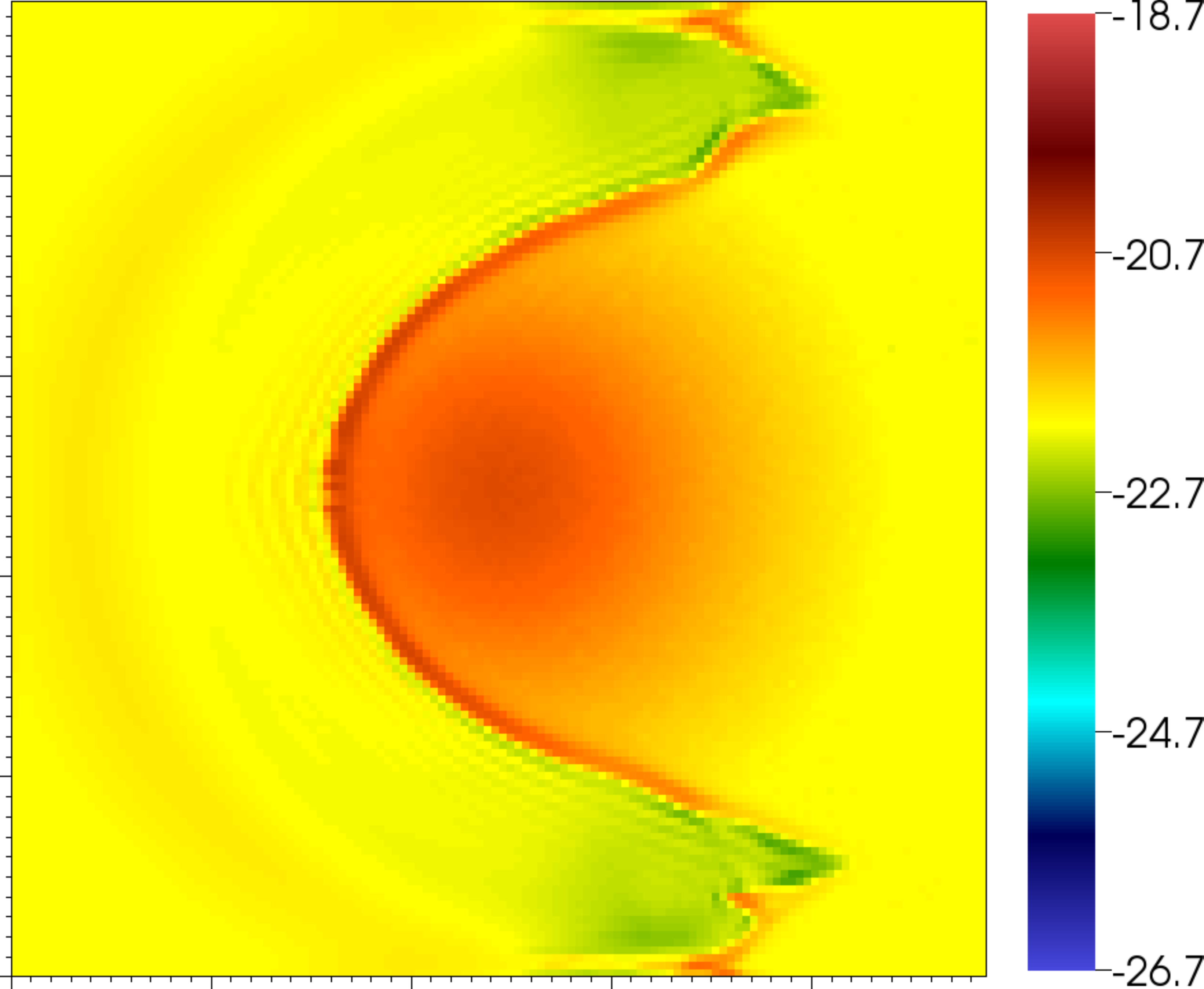}

		\hspace{-20pt}
		\vspace{10pt}
		\includegraphics[width=51mm, height=51mm]{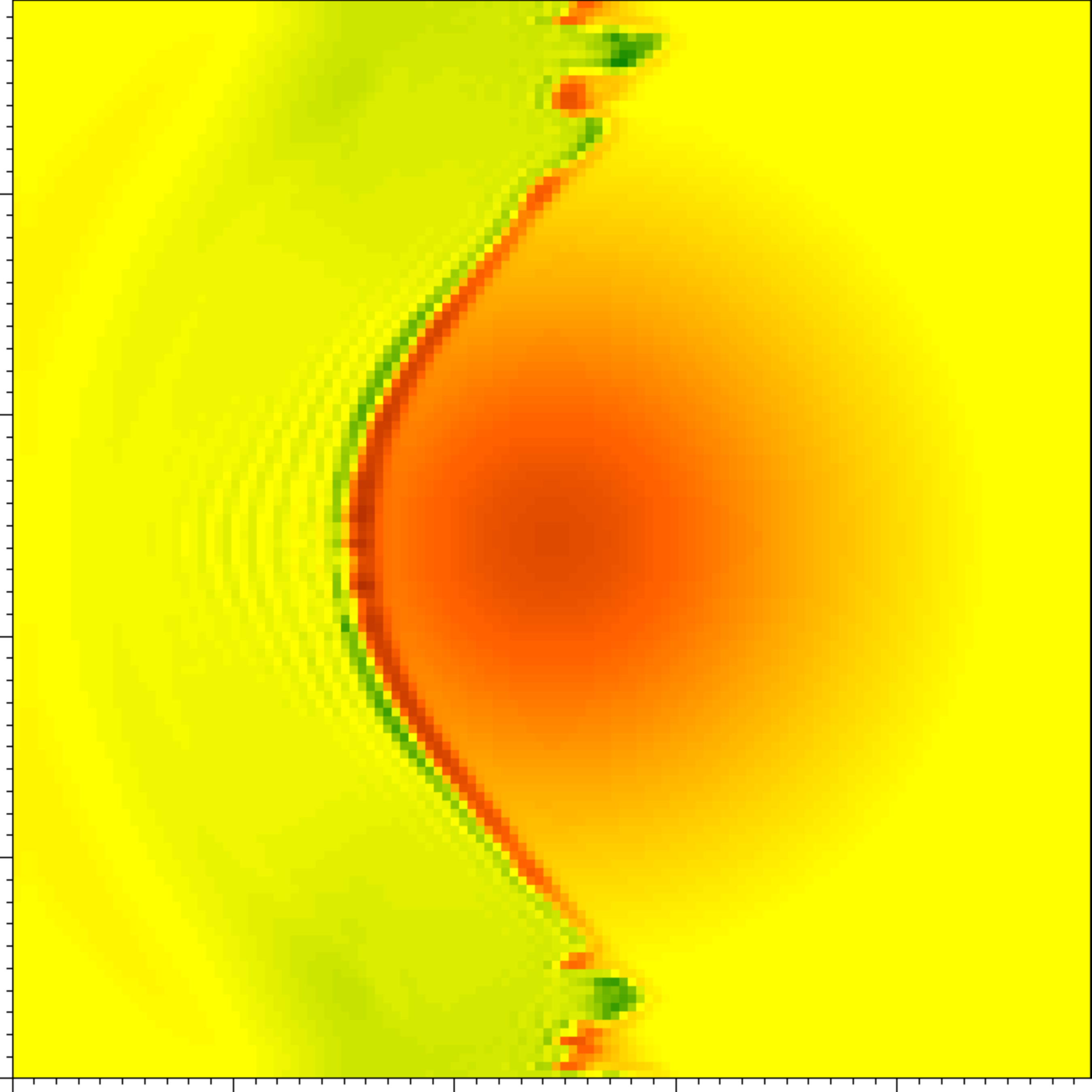}
		\hspace{3pt}
		\includegraphics[width=51mm, height=51mm]{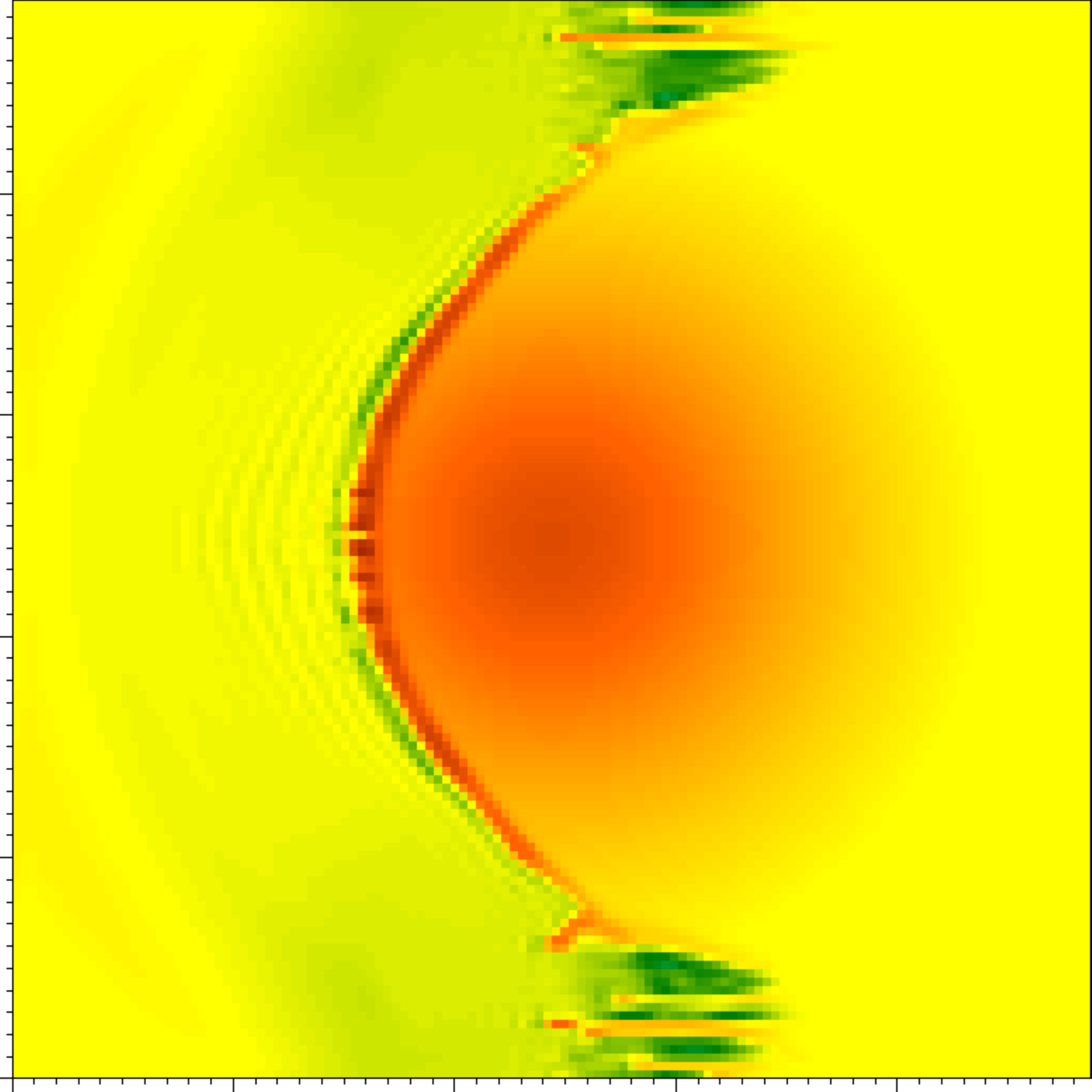}
		\hspace{3pt}
		\includegraphics[width=51mm, height=51mm]{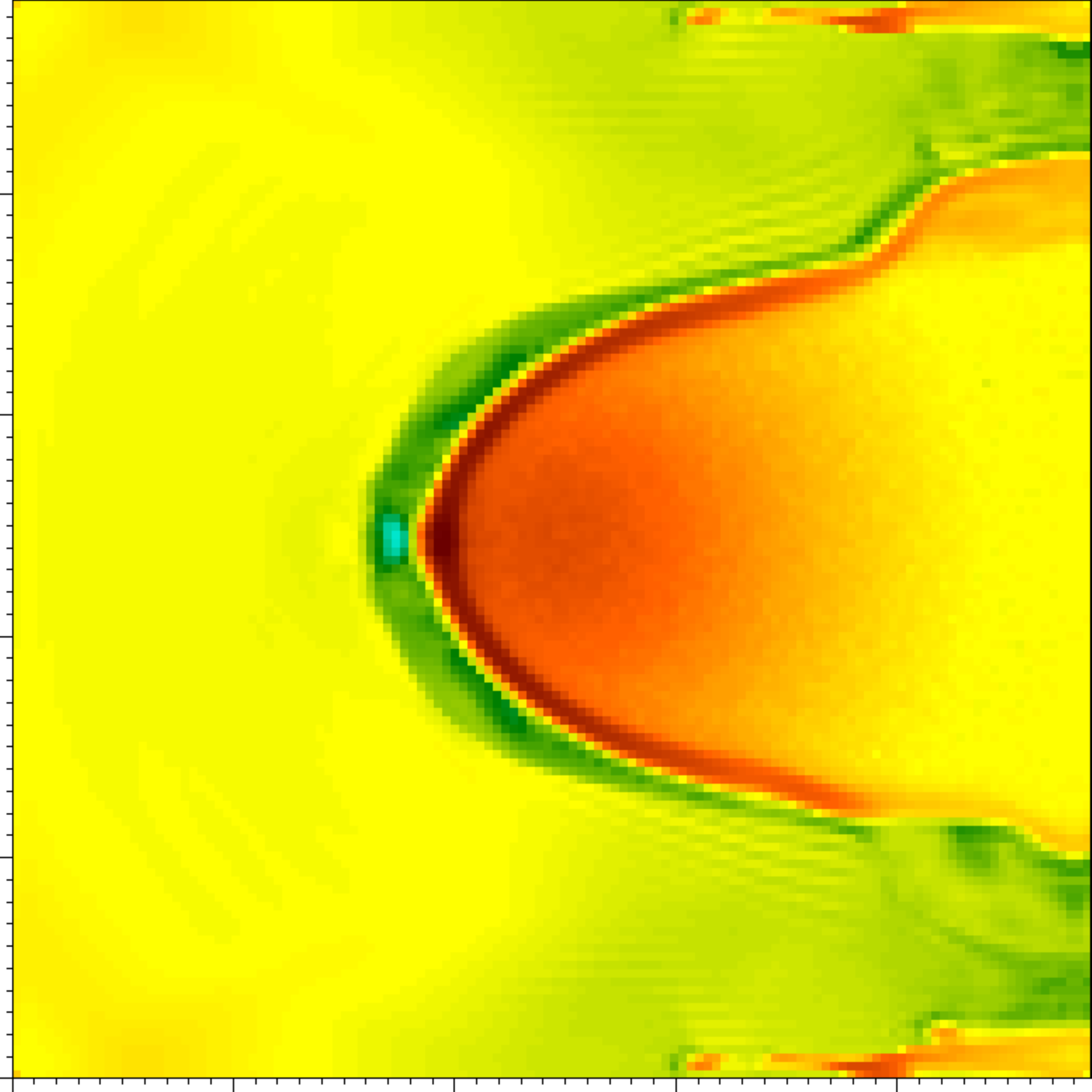}

		\hspace{-20pt}
		\vspace{10pt}
		\includegraphics[width=51mm, height=51mm]{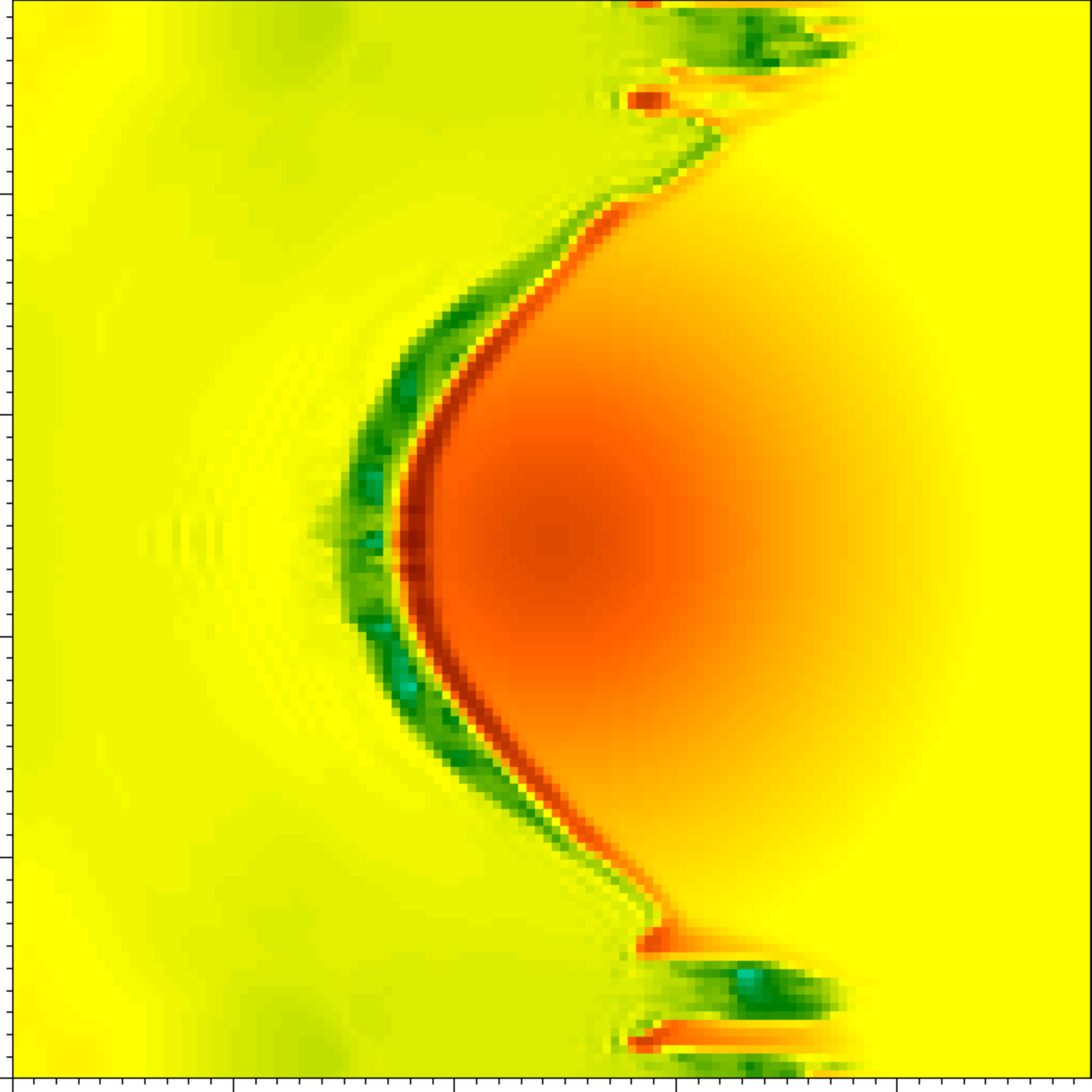}
		\hspace{3pt}
		\includegraphics[width=51mm, height=51mm]{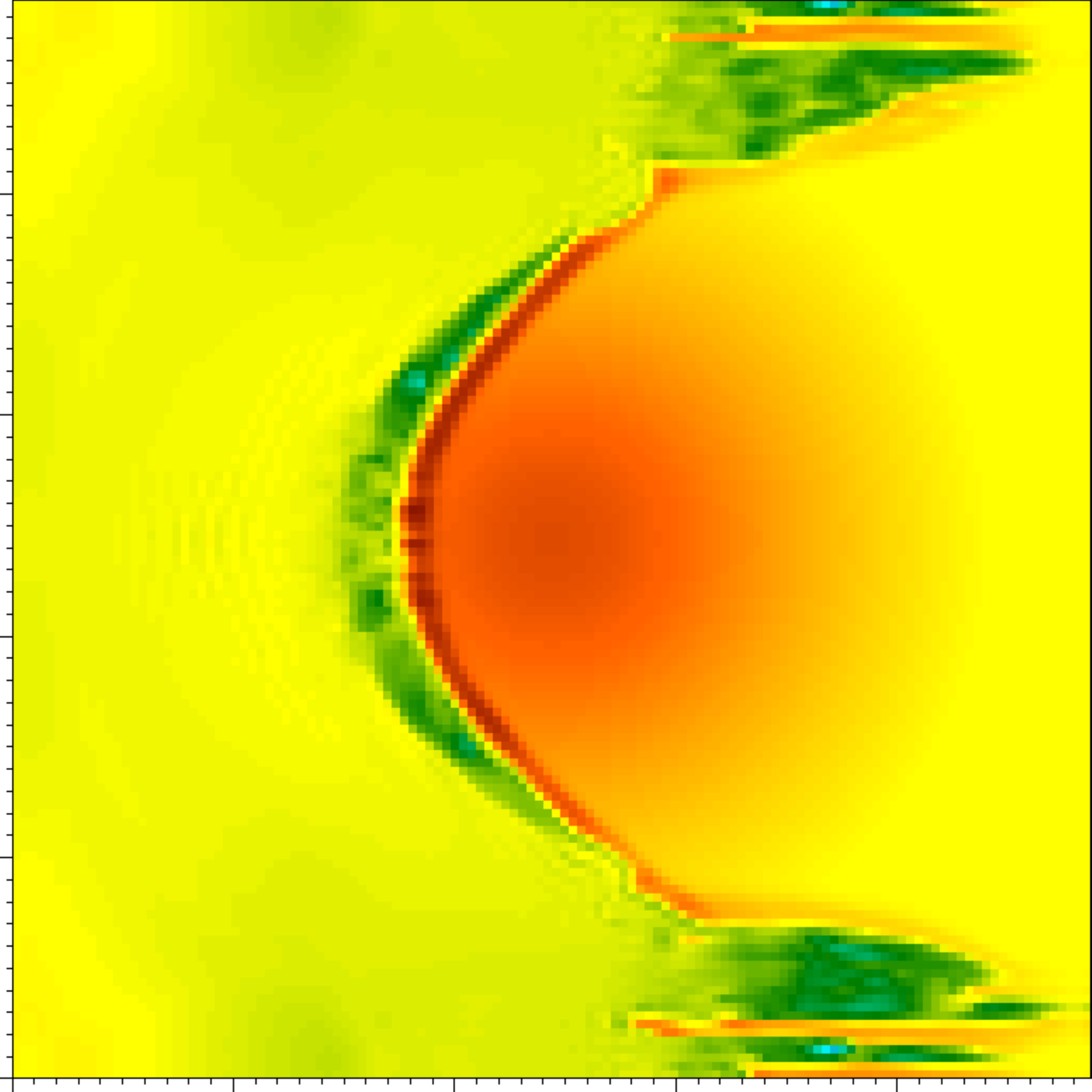}
		\hspace{3pt}
		\includegraphics[width=51mm, height=51mm]{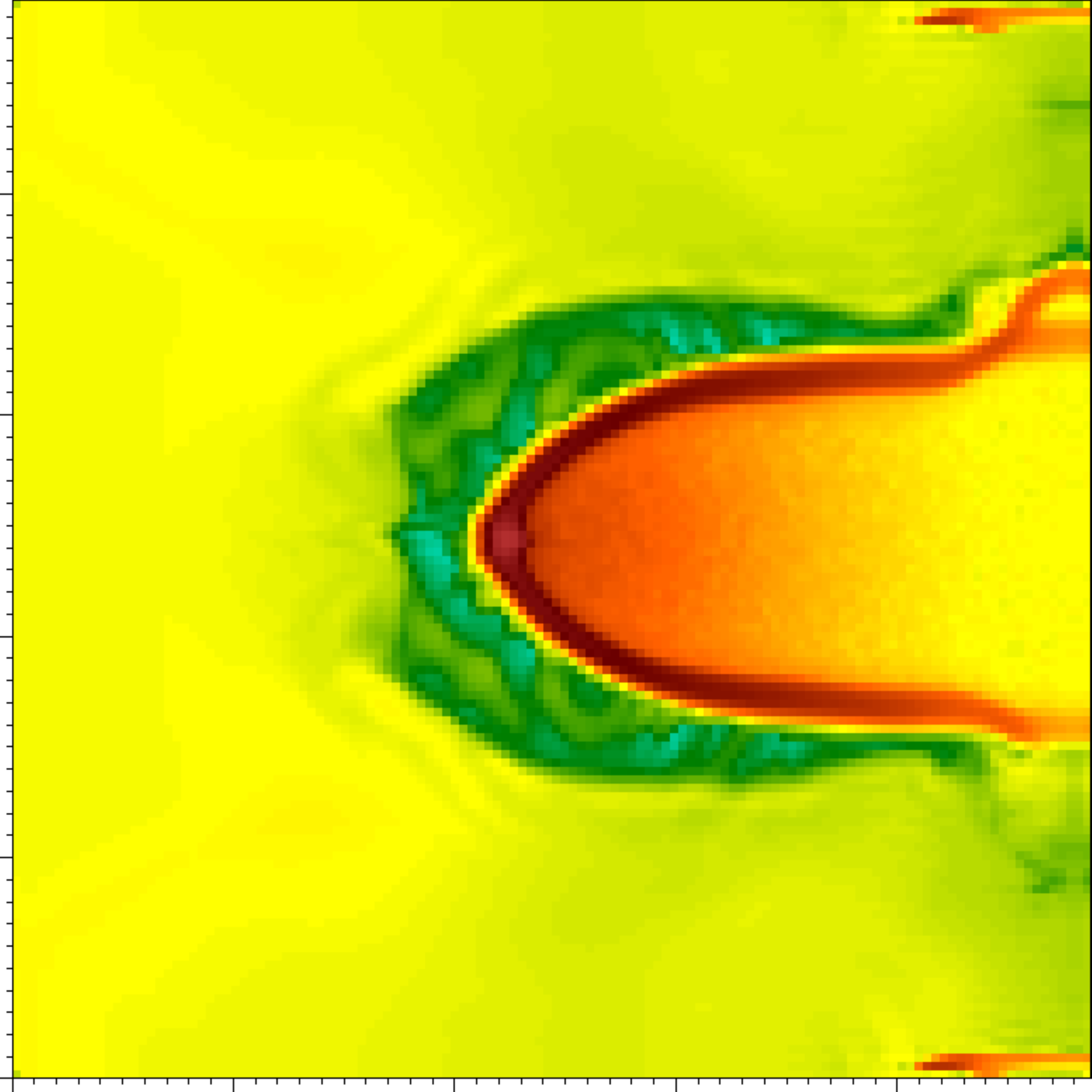}

		\hspace{-20pt}
		\vspace{10pt}
		\includegraphics[width=51mm, height=51mm]{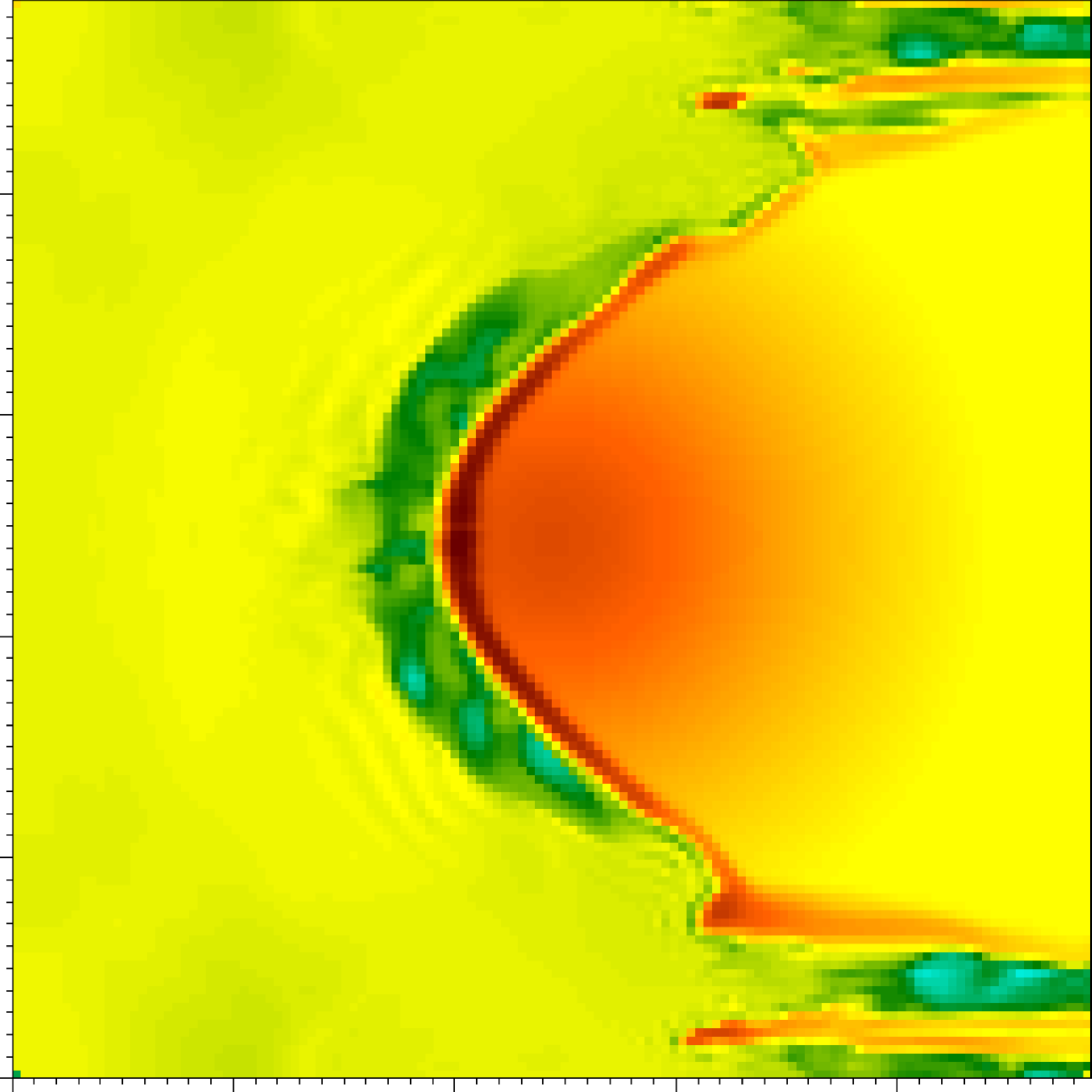}
		\hspace{3pt}
		\includegraphics[width=51mm, height=51mm]{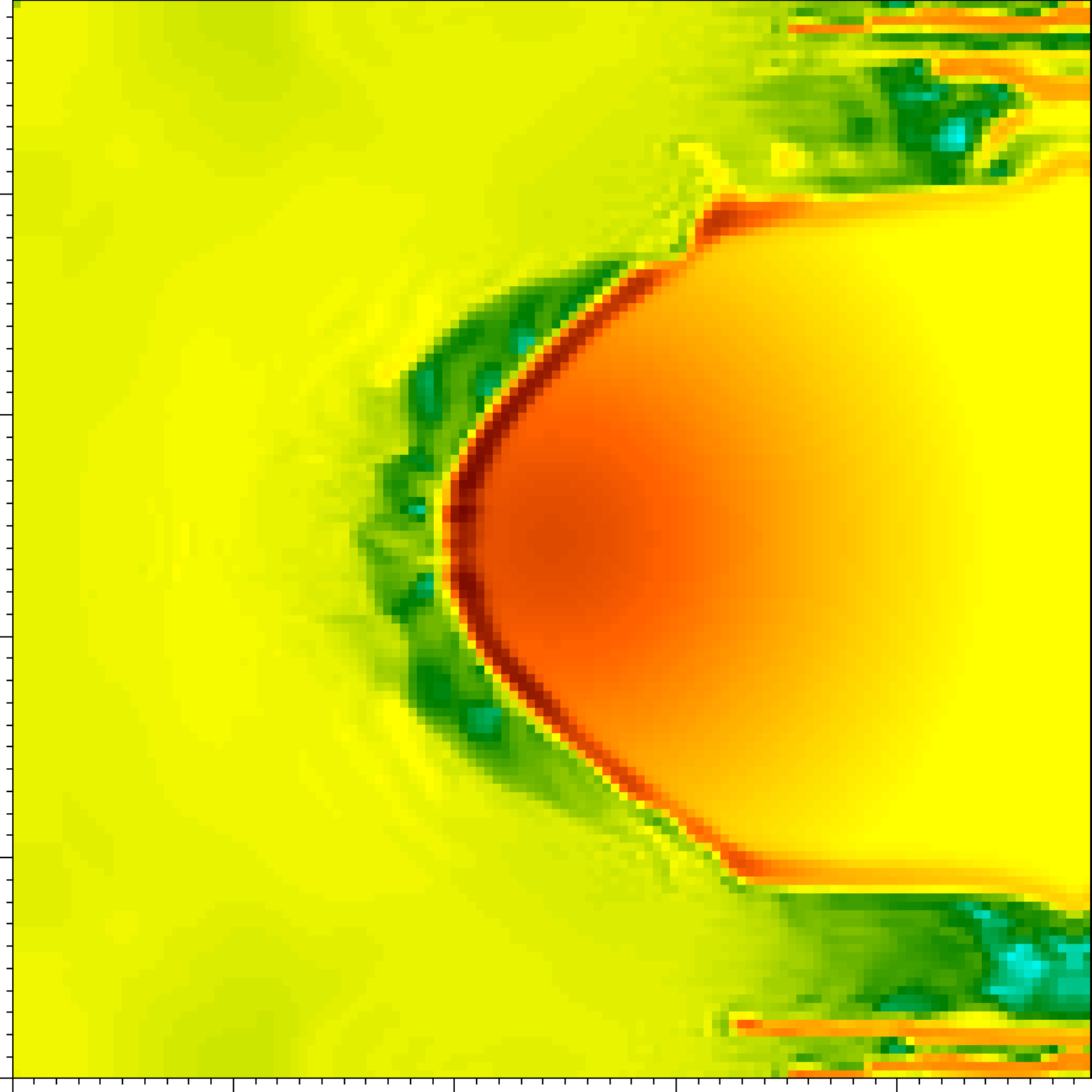}
		\hspace{3pt}
		\includegraphics[width=51mm, height=51mm]{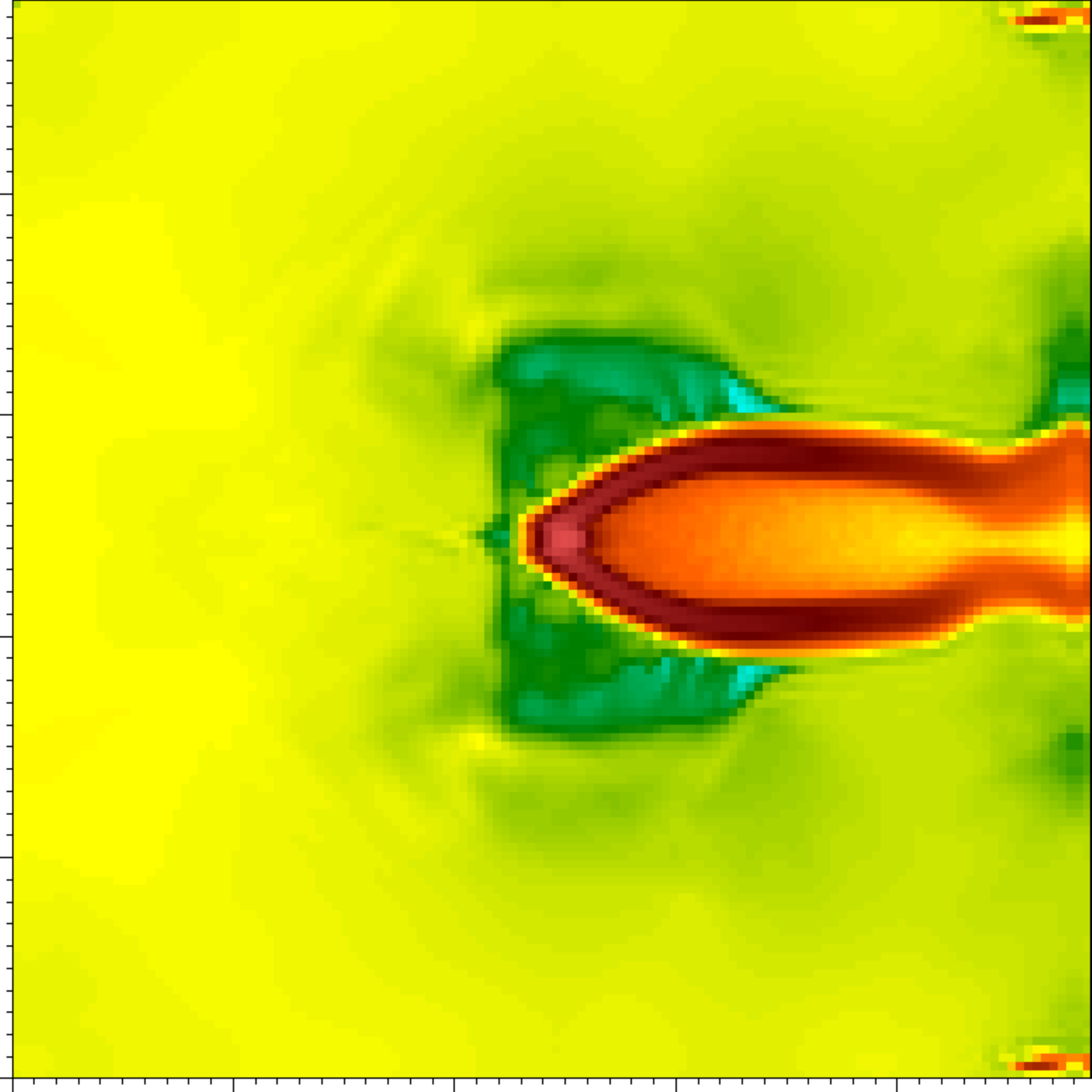}

  \caption{The medium flux model logarithmic density distributions (cgs). Left column) Monochromatic models. Middle column) Polychromatic models. Right column) Polychromatic-diffuse models. Time is increasing from top to bottom, with snapshots at 50, 100, 150 and 200\,kyr. Each frame is a slice through the computational grid, which is a cube with sides 4.87\,pc long. Major ticks are separated by 1\, pc.}
		\label{RDI_MED}
\end{figure*}

\begin{figure*}
%
%
%
		\hspace{10pt}
		\includegraphics[width=51.mm, height=51mm]{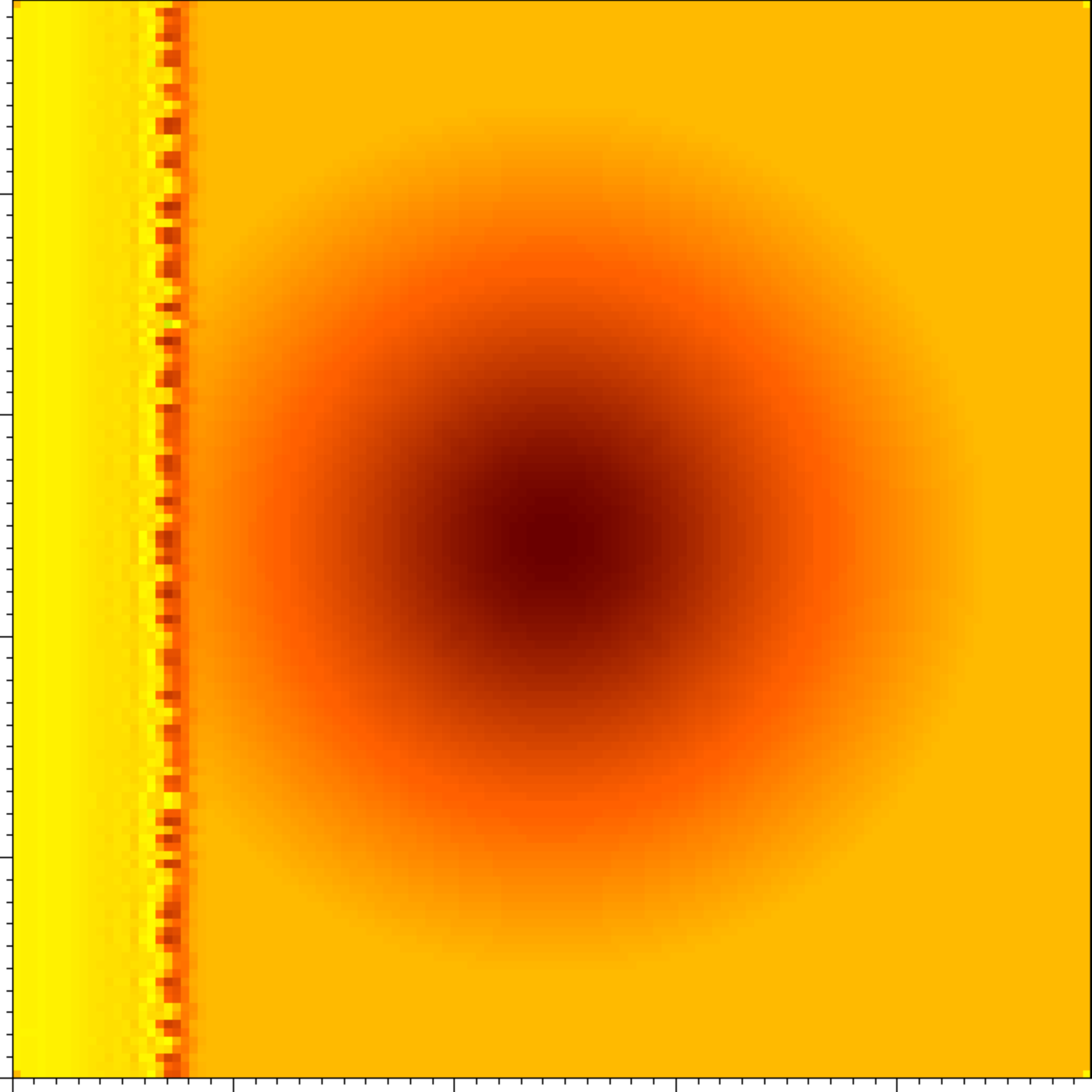}
		\hspace{3pt}
		\vspace{10pt}
		\includegraphics[width=51mm, height=51mm]{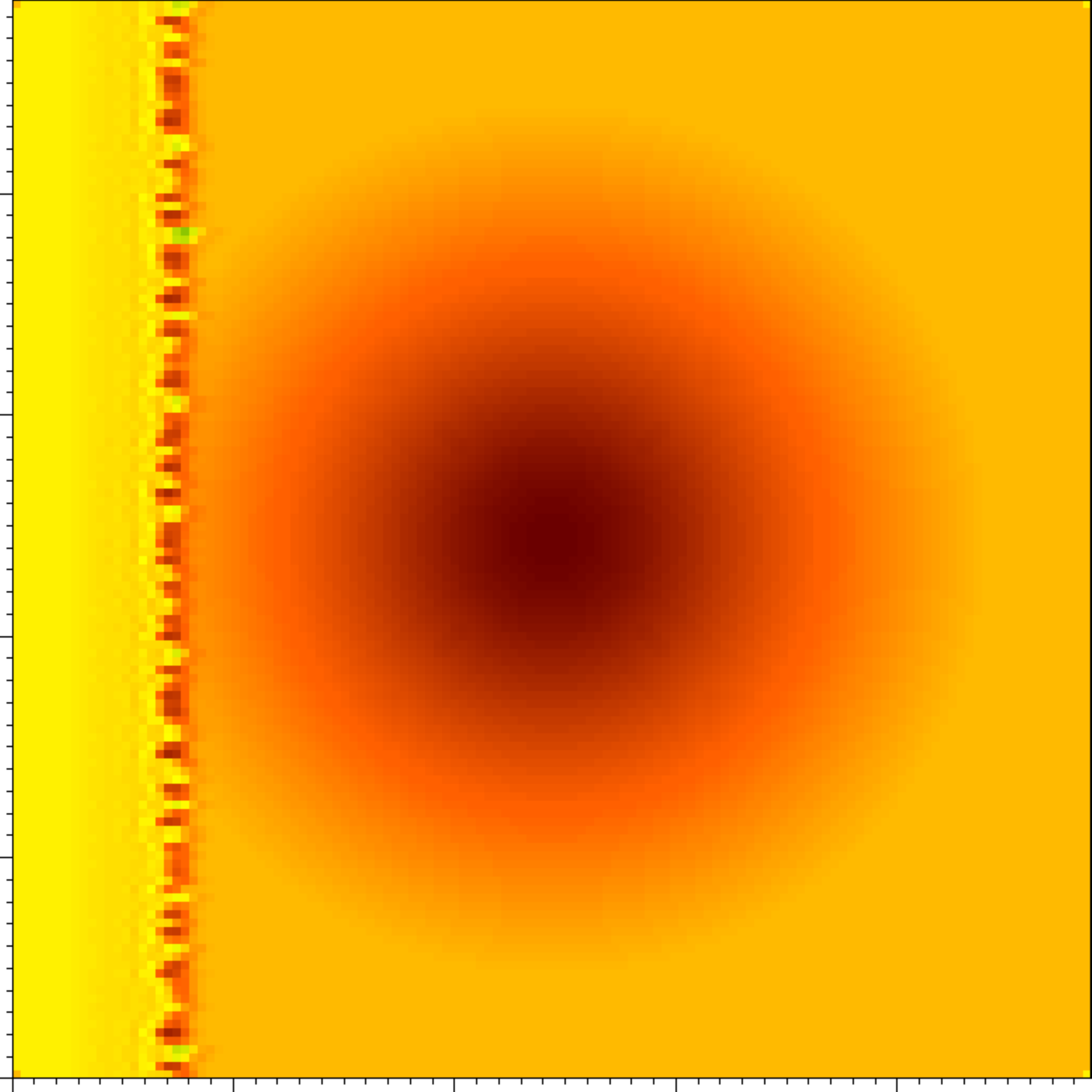} 
		\hspace{3pt}
		\includegraphics[width=61.7mm, height=51.2mm]{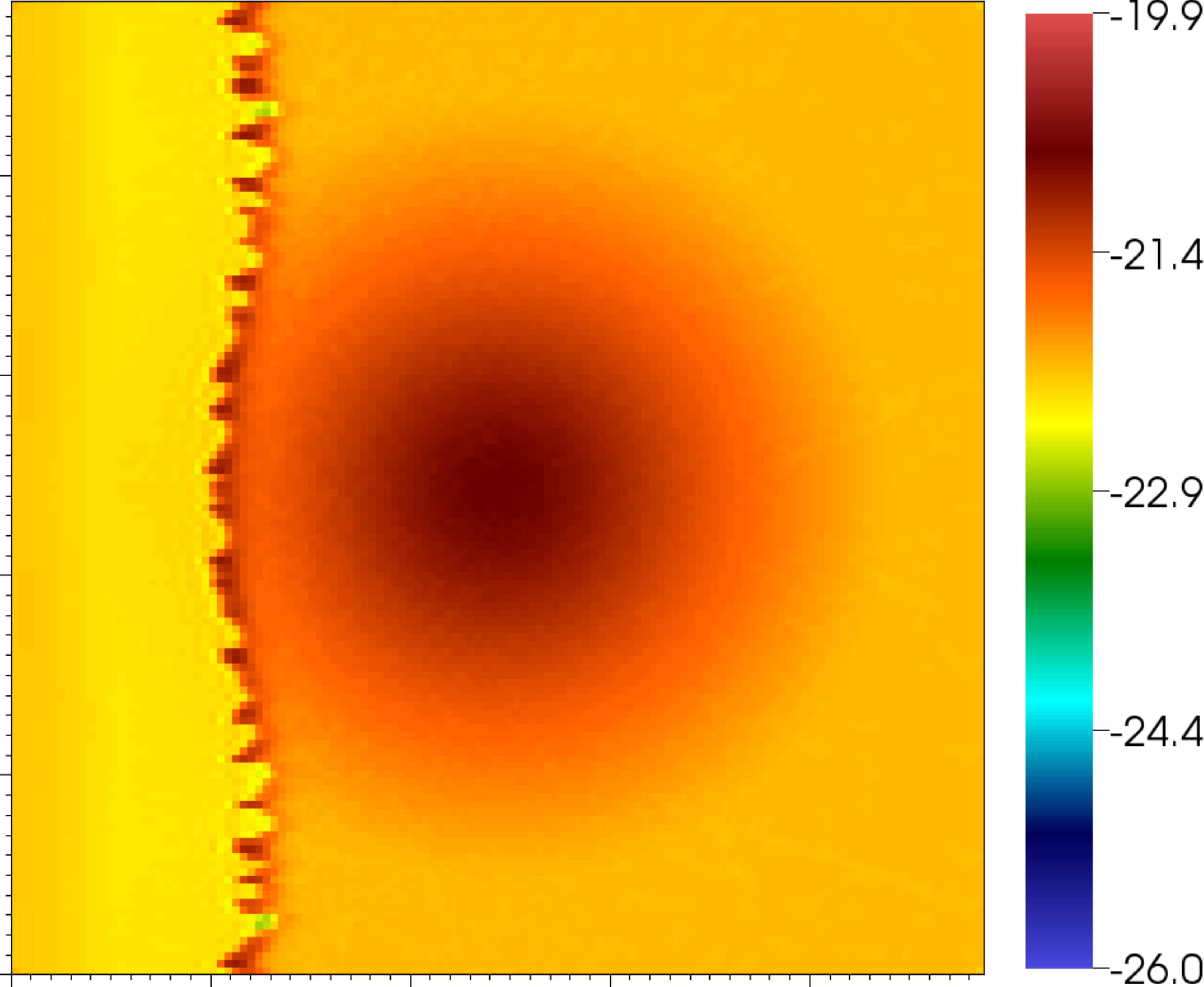}

		\hspace{-20pt}
		\vspace{10pt}
		\includegraphics[width=51mm, height=51mm]{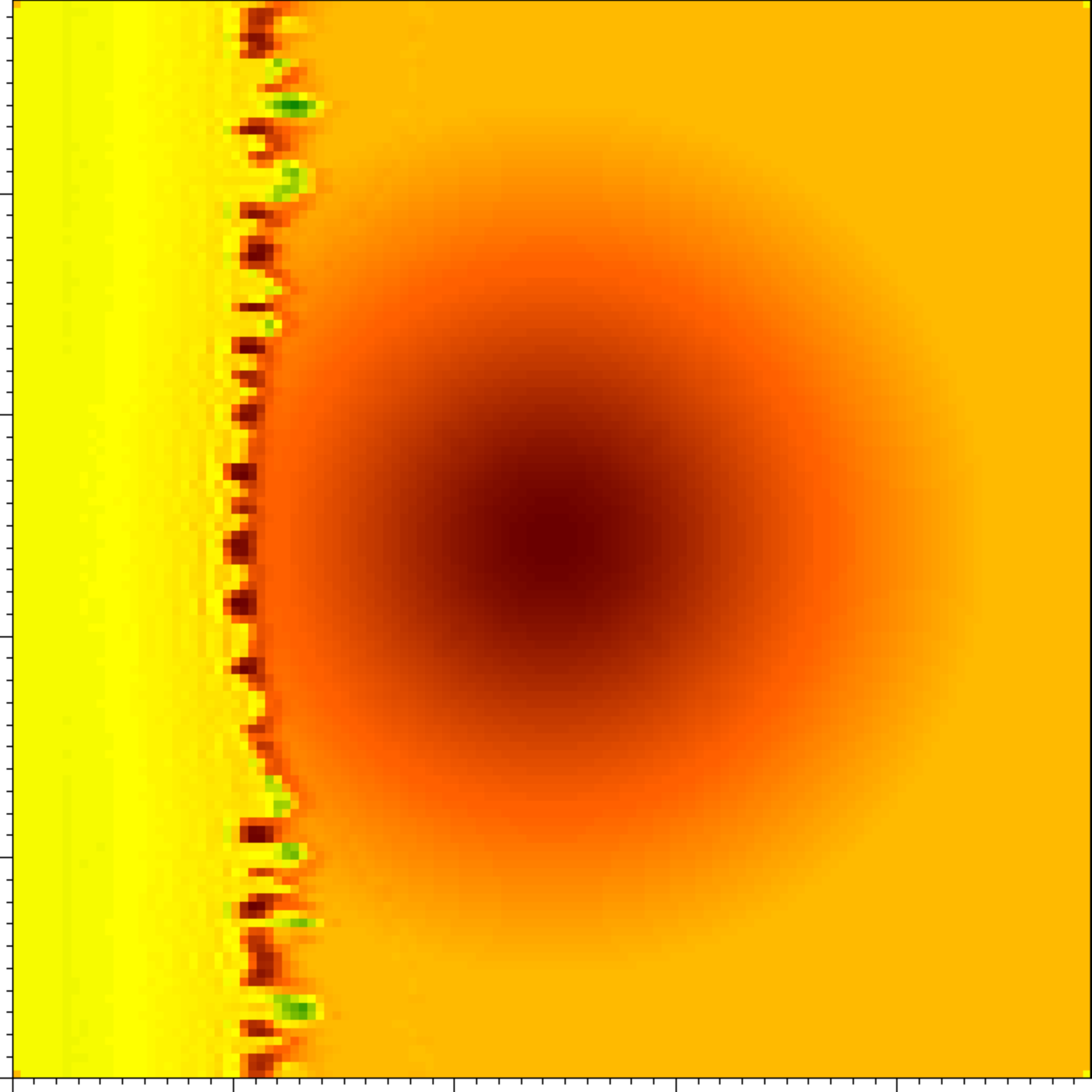}
		\hspace{3pt}
		\includegraphics[width=51mm, height=51mm]{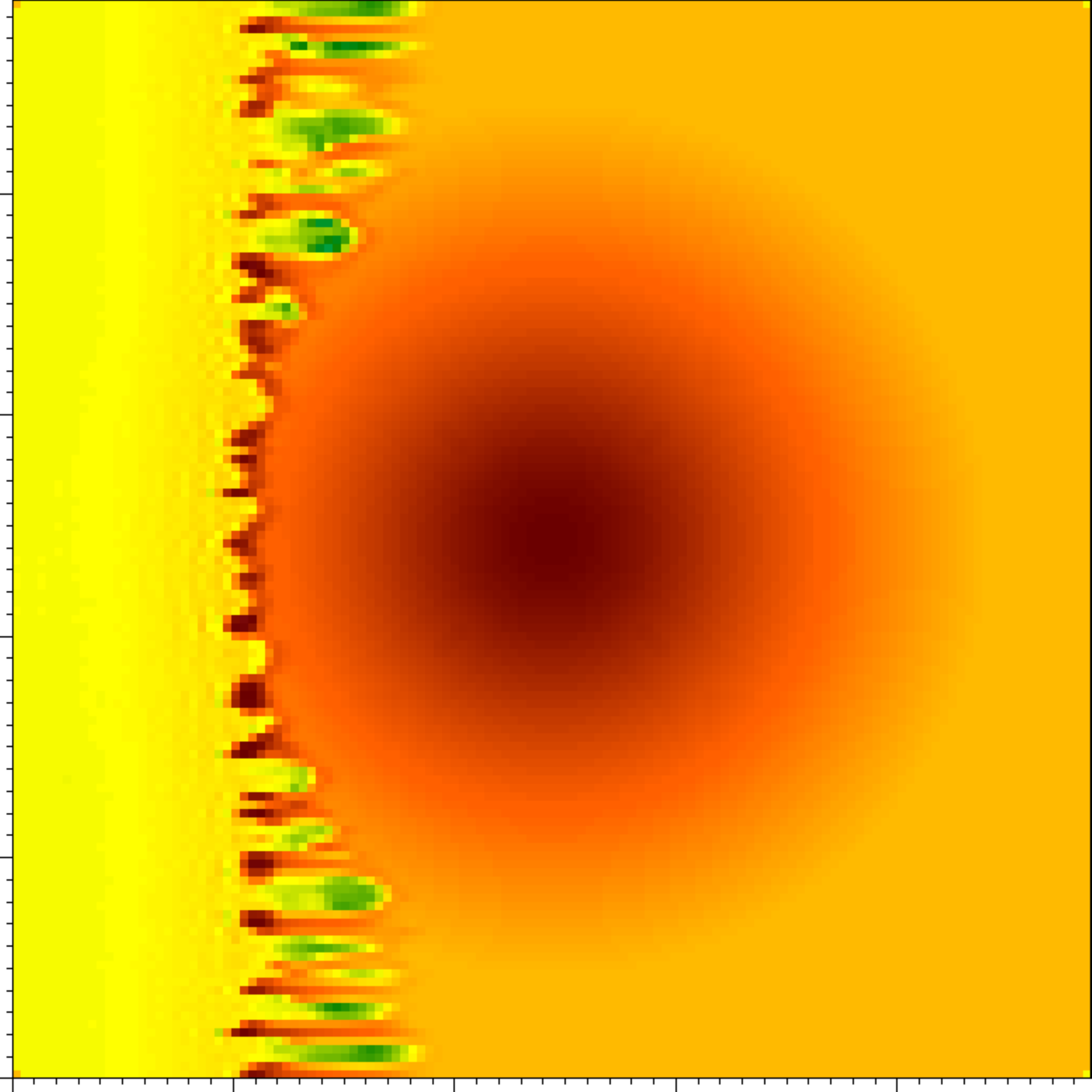}
		\hspace{3pt}
		\includegraphics[width=51mm, height=51mm]{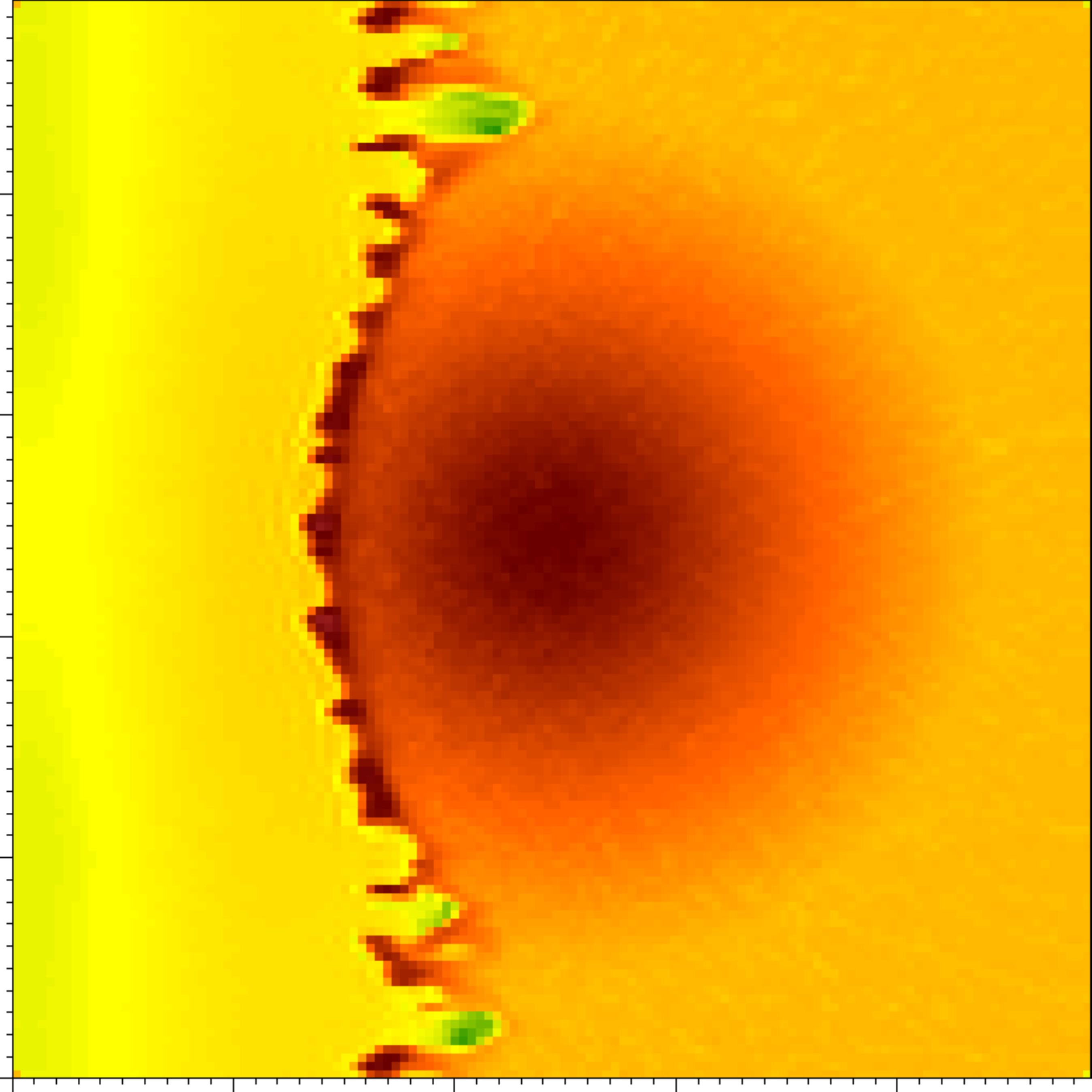}

		\hspace{-20pt}
		\vspace{10pt}
		\includegraphics[width=51mm, height=51mm]{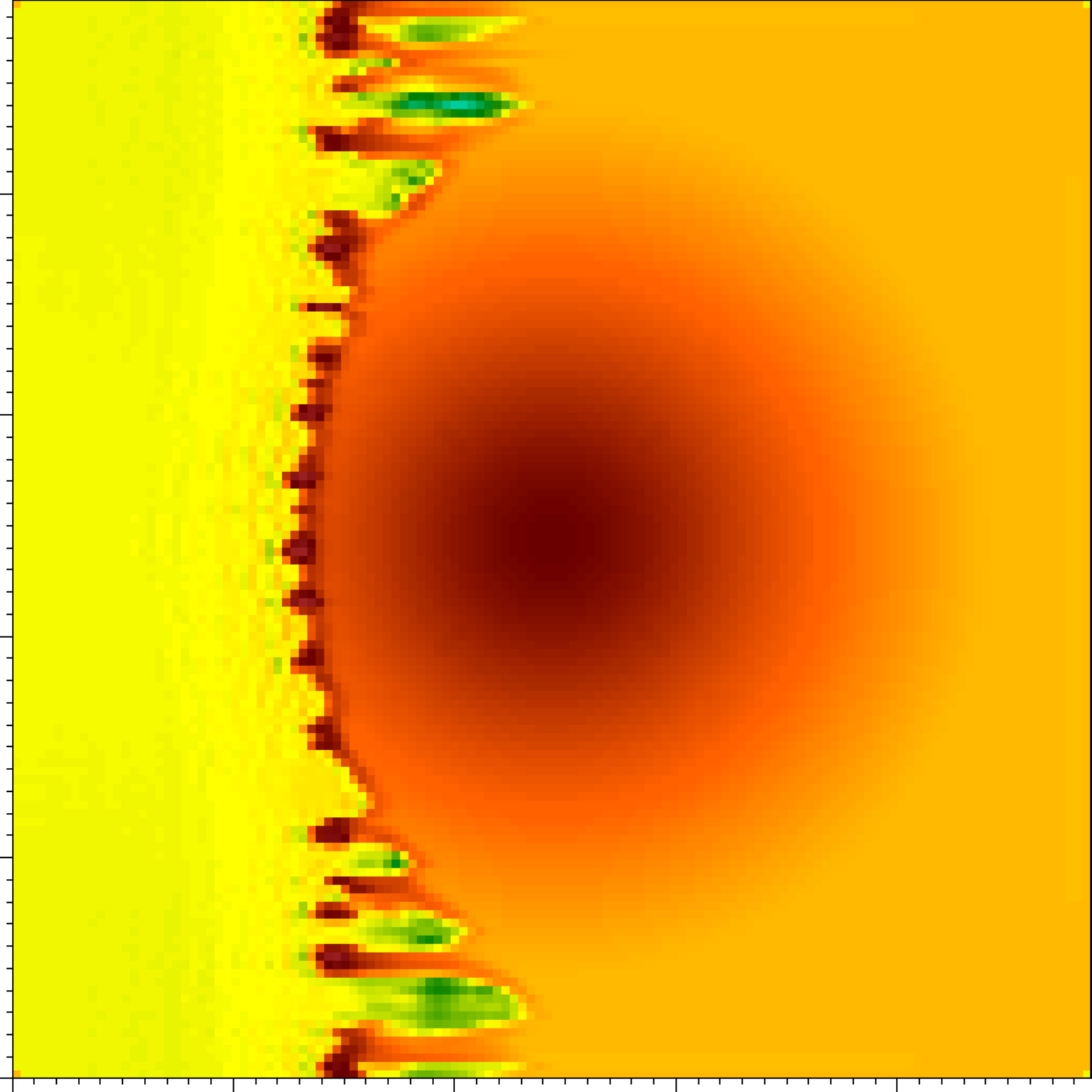}
		\hspace{3pt}
		\includegraphics[width=51mm, height=51mm]{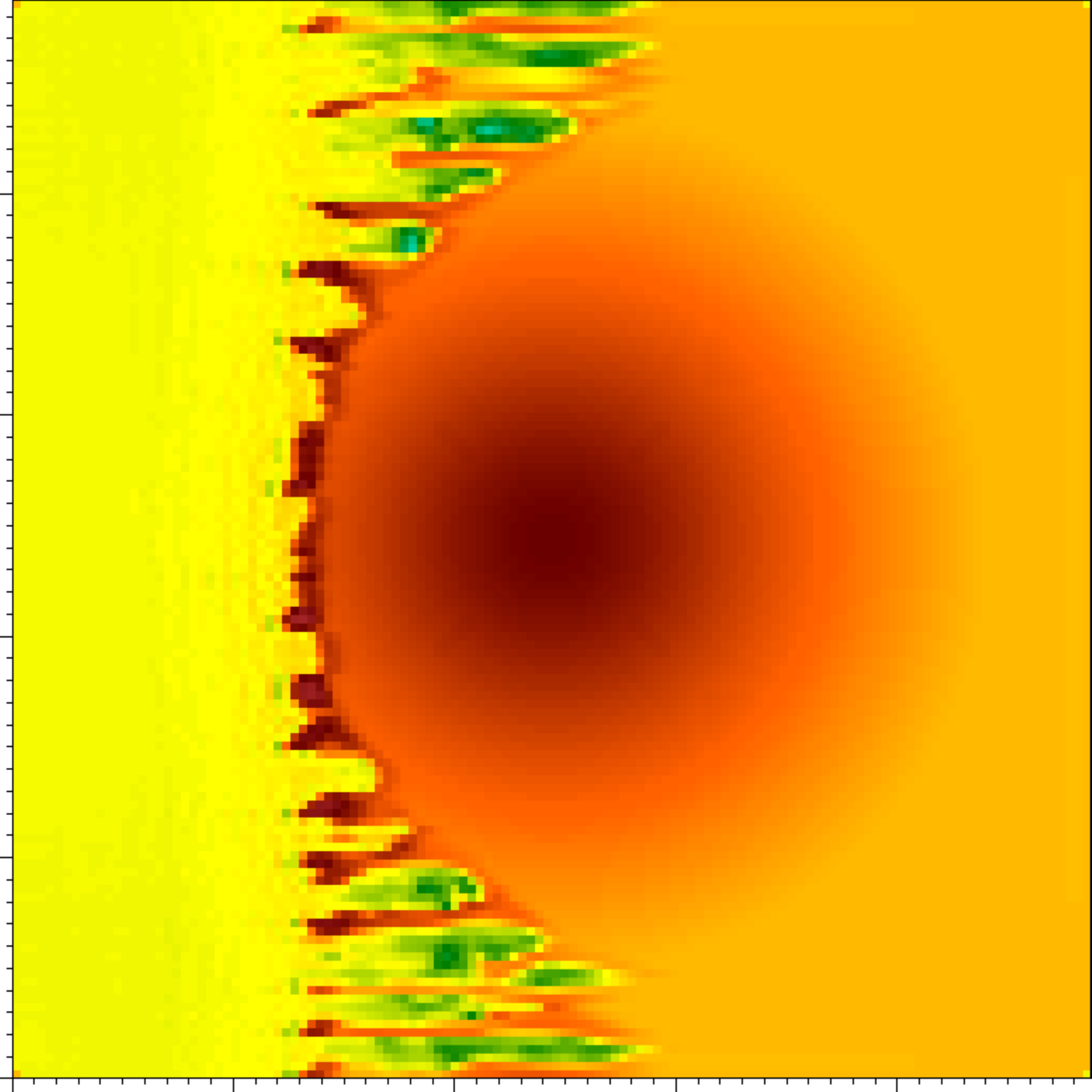}
		\hspace{3pt}
		\includegraphics[width=51mm, height=51mm]{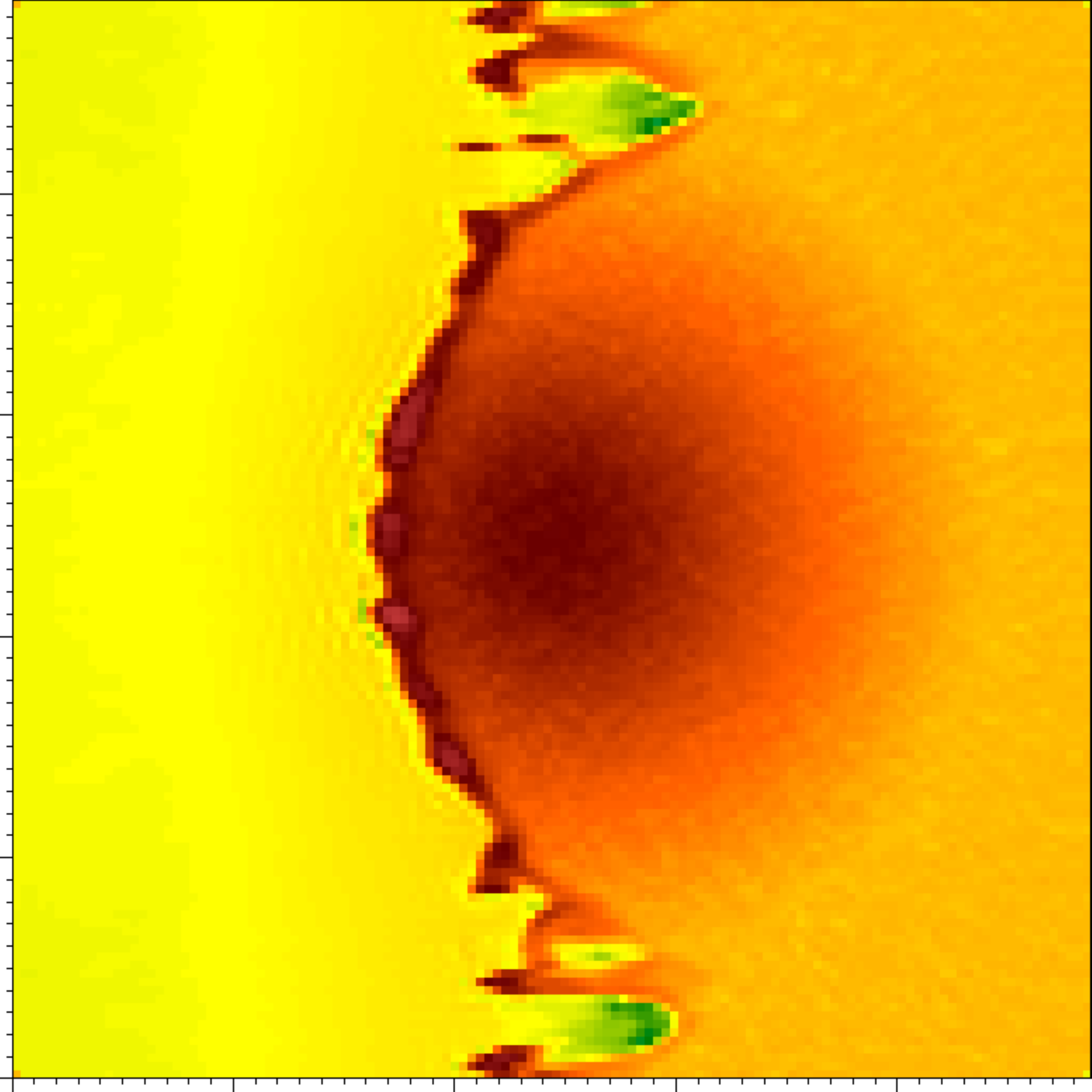}

		\hspace{-20pt}
		\vspace{10pt}
		\includegraphics[width=51mm, height=51mm]{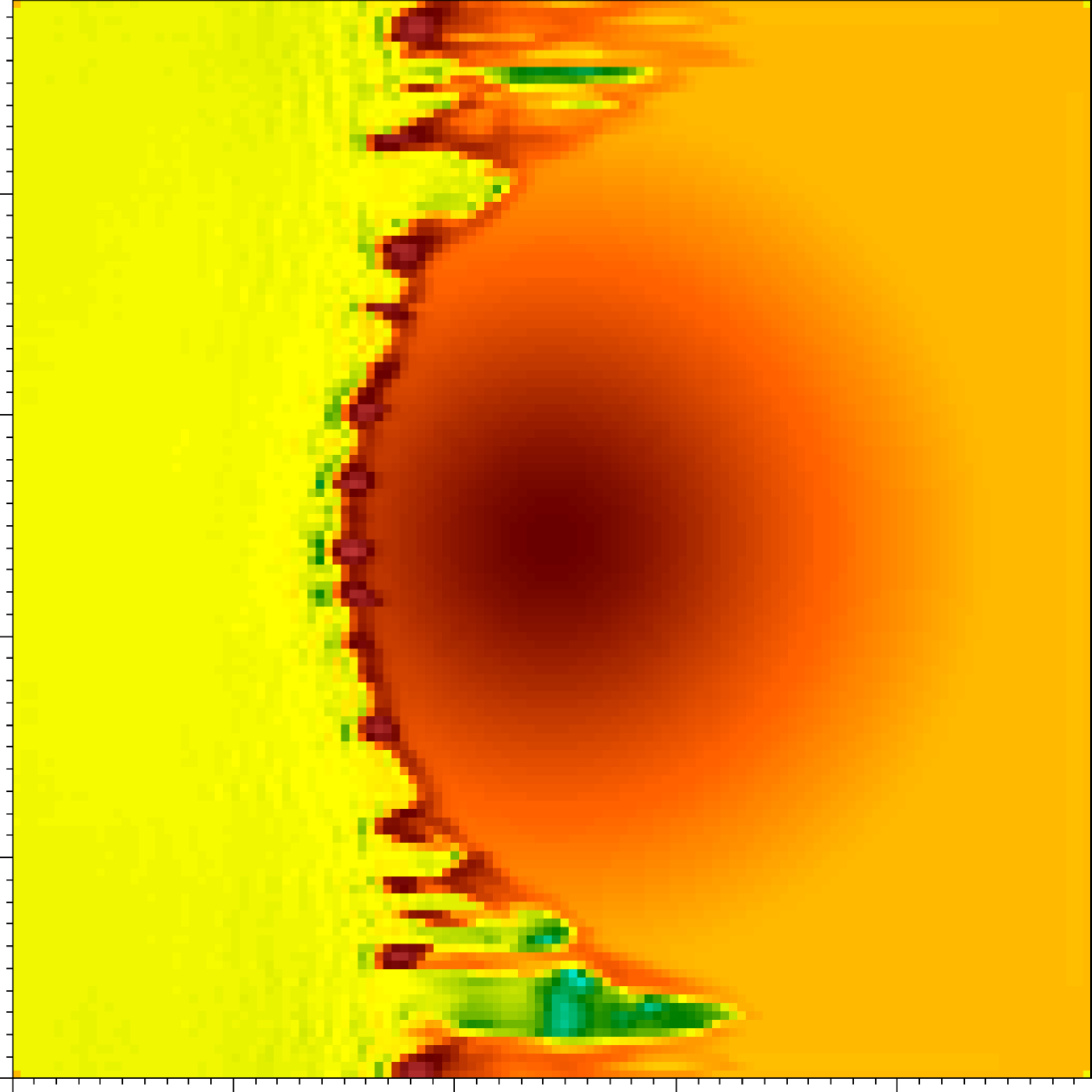}
		\hspace{3pt}
		\includegraphics[width=51mm, height=51mm]{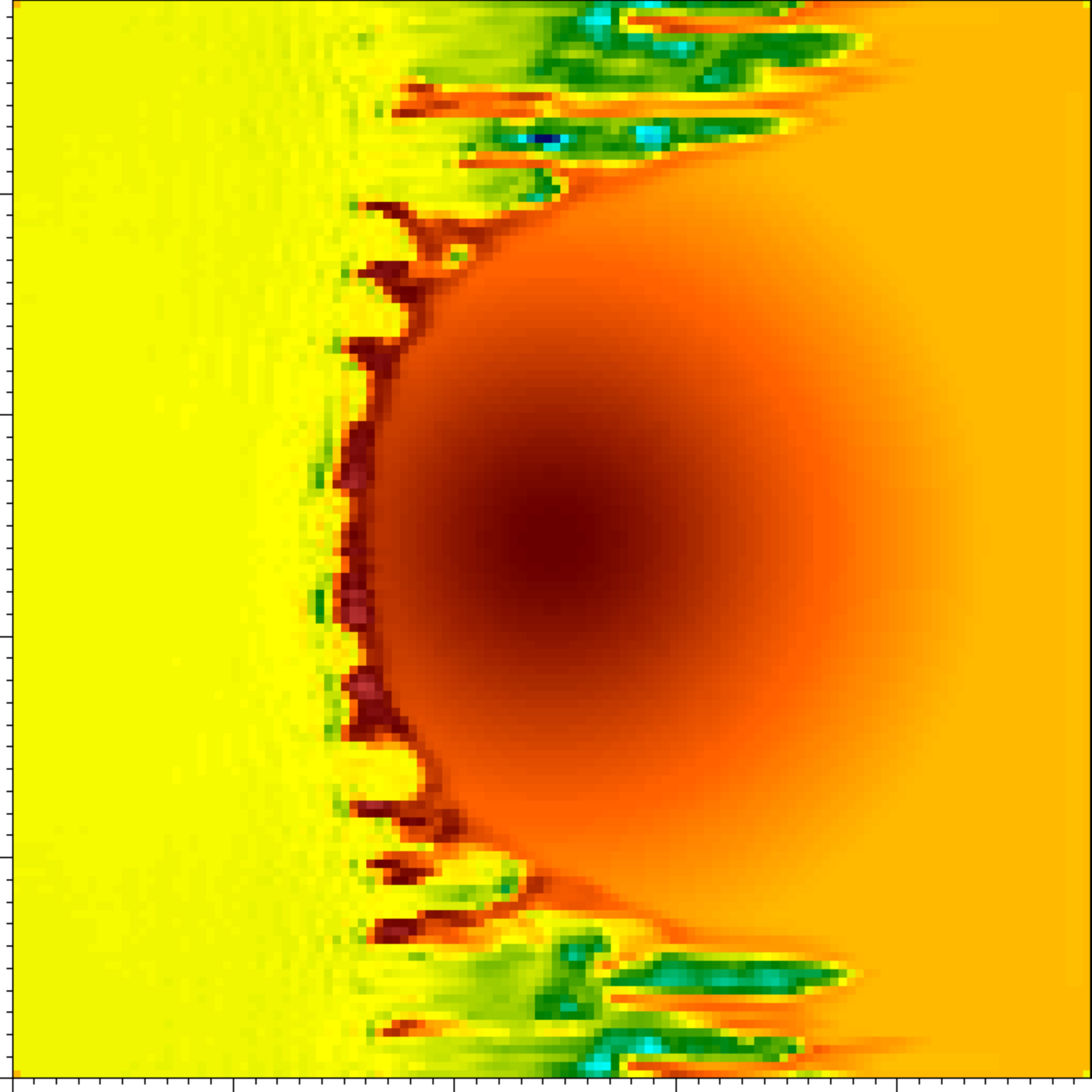}
		\hspace{3pt}
		\includegraphics[width=51mm, height=51mm]{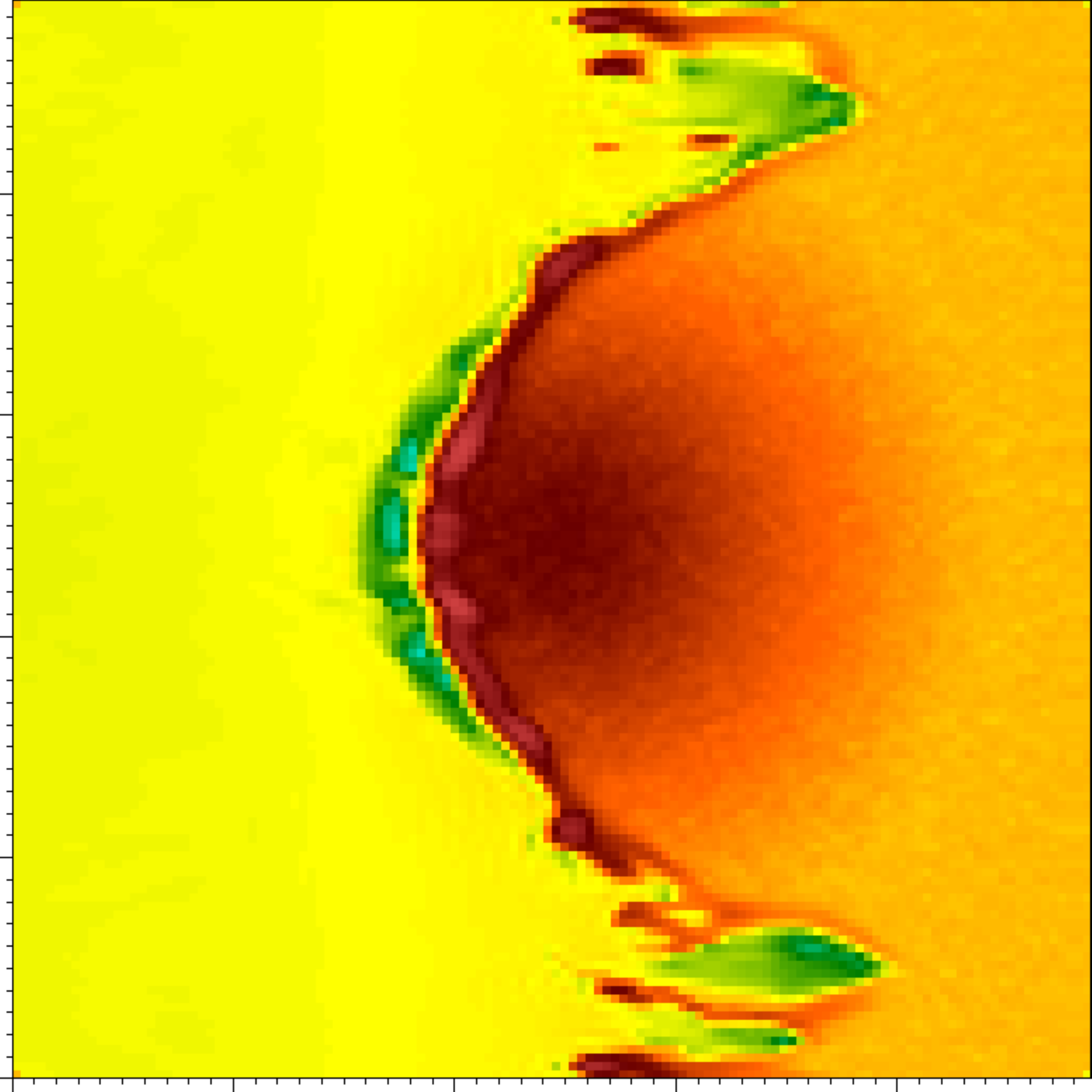}

  \caption{The low flux model logarithmic density distributions (cgs). Left column) Monochromatic models. Middle column) Polychromatic models. Right column) Polychromatic-diffuse models. Time is increasing from top to bottom, with snapshots at 50, 100, 150 and 200\,kyr. Each frame is a slice through the computational grid, which is a cube with sides 4.87\,pc long. Major ticks are separated by 1\, pc.}
		\label{RDI_LOW}
\end{figure*}

The logarithmic density distribution for the high, medium and low flux
models over all three treatments of the radiation field, are shown
side by side at 50, 100, 150 and 200\,kyr in Figures \ref{RDI_HI},
\ref{RDI_MED} and \ref{RDI_LOW} respectively.  In each of these
figures the monochromatic models are represented by the left hand
column, the polychromatic models by the central column and the
polychromatic-diffuse models by the right hand column. It is clear
that there are some marked differences between the evolutions of the
system under the different radiation treatments. A case-by-case study
of the evolution of the separate models is given below in sections
\ref{hiAnalysis}, \ref{medAnalysis} and \ref{lowAnalysis}.

\subsubsection{The formation and evolution of instabilities}
\label{sec:ins}
A number of the models presented here are subject to the formation of
instabilities. These arise when the thin shell swept up by the
ionization front propagates though the lower density regions of the
computational domain such as in the wings or (in the low flux regime)
prior to driving into the BES.

There are two main sources of perturbation that may seed these
instabilities. The first is `angle of incidence'
\cite{2002MNRAS.331..693W} in which regions of the I-front that are
not entirely perpendicular to the incoming radiation field are subject
to varying ionizing flux and therefore differential acceleration. This
may occur as the I-front wraps around the BES. The second is
numerically induced perturbation via noise in the calculated
ionization fraction. As mentioned in section \ref{photoionSection} the
induced temperature, and therefore pressure, is directly proportional
to the calculated ionization fraction (equation
\ref{temperature}). The resulting pressure gradient then determines
the induced advecting velocity. If a thin shell of material does not
encounter any disruption in its propagation (like encountering a high
density component of a BES) then even a small amount of numerical
spread in the ionization fraction along the I-front will eventually
lead to it bending on small scales and therefore induce thin shell
instabilities \citep{1983ApJ...274..152V, 1996ApJ...469..171G}. Once
the I-front structure is disrupted, the faster propagating components
will lose mass by transfer to the cool neutral neighboring material
perpendicular to the direction of the I-front propagation. This leads
to an accumulation of material ahead of the slower moving components
of the I-front, which further brakes the expansion in these
regions. This transport of material also reduces the density in the
faster moving components of the I-front, allowing photoionizing
radiation to propagate more deeply (c.f. equation \ref{stromgren}) and
accelerate these components of the front further.  Improving the
accuracy of the ionization fractions in the front will delay the onset
of numerically induced instabilities, however they should eventually
arise for any non-analytical radiation hydrodynamics code if the
I-front is allowed to propagate for long enough without being
disrupted by some other means.  \\

It is clear that other radiation-hydrodynamical methods should also
seed these instabilities. For example, combining SPH hydrodynamics
with ray-tracing will lead to a `noisy' I-front due to the
random-variation of the SPH representation of the density
field. Grid-based codes with ray-tracing radiation-transfer will also
be susceptible to instabilities as the angular sampling of the
radiation field may not coincide perfectly with the axes of the
underlying hydrodynamical grid. Of course within star forming regions
themselves density perturbations will inevitably lead to the
growth of instabilities.

Regardless of the seed of these instabilities in the simulations,
their evolution occurs in a manner consistent with the instability
studies referenced above and result in `elephant trunk'
structures. Their evolution also highlights interesting differences
between the different treatments of the radiation field, the details
of which are discussed in the following sections.

\subsubsection{High flux models}
\label{hiAnalysis}
The high flux density evolutions are given in Figure \ref{RDI_HI} and
broadly exhibit two different behaviours.  In the polychromatic OTS
and polychromatic-diffuse models the system is initially ionized to
the extent that a strong shock cannot form quickly enough to
effectively drive into the BES. What material is accumulated
propagates for a short time until the higher density BES brakes it and
a photo-evaporative flow is established
\citep{1990ApJ...354..529B}. The evolution of the resulting structure
is then a consequence of rocket motion as heated, dense, material is
evaporated away from the surface of the cloud into the low density
external material \citep{1955ApJ...121....6O}.  The tunnelling of
material at the tip and along the length of these models into the
cometary structure occurs where there are differences between
rocket-motion velocities due to either variations in the accumulated
density or the density internal to the shell. An interesting
difference between the OTS and diffuse field models is also revealed
in the rocket-driven phase, with the OTS model being accelerated only
along components facing the ionizing source and the diffuse field
model being accelerated across the entire cometary surface due
to diffuse-driven photo-evaporation.

\begin{figure}
	\hspace{8pt}
	\includegraphics[width=78mm, height=65.2mm]{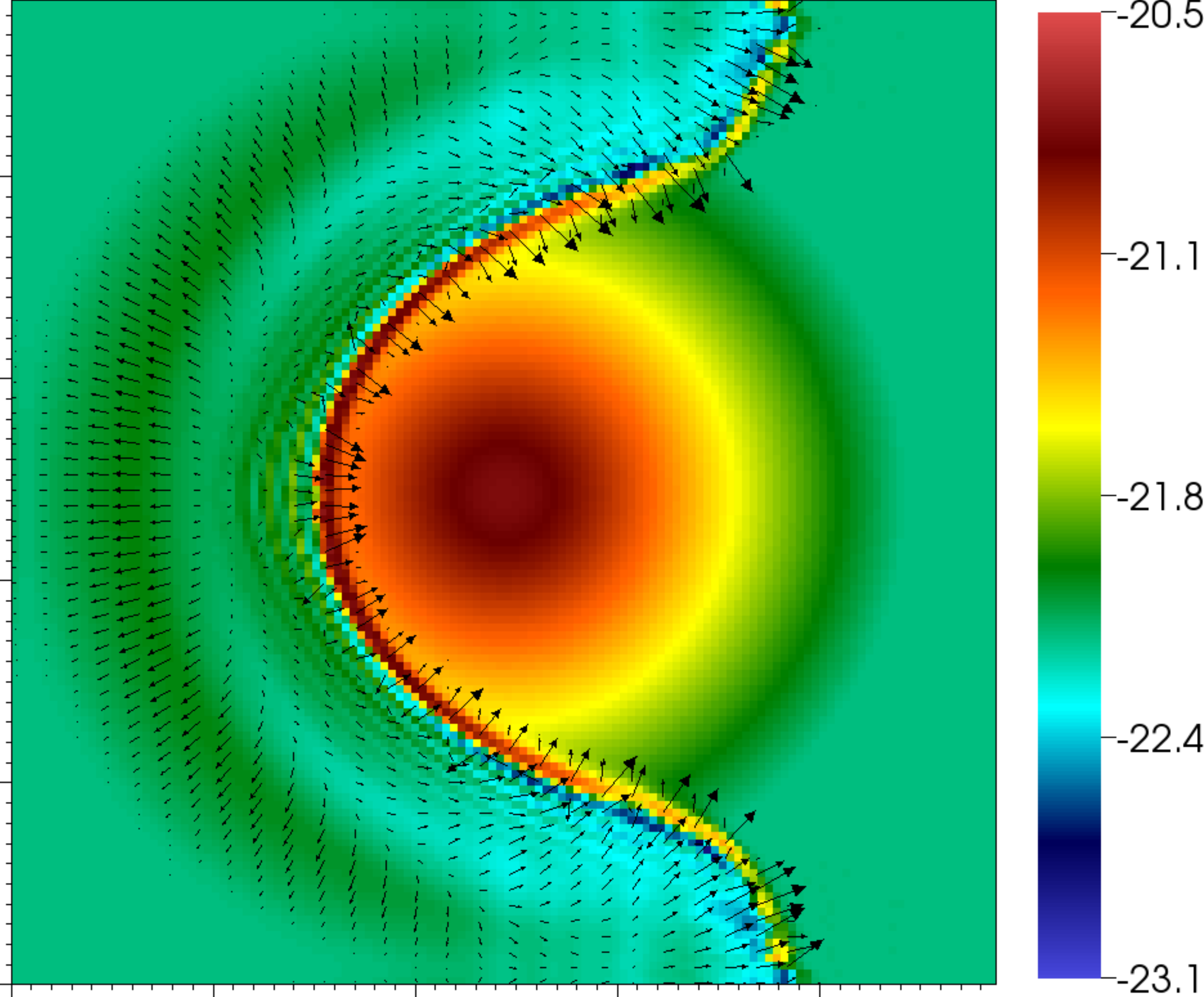}

	\vspace{6pt}
	\hspace{8pt}
	\includegraphics[width=65mm, height=65mm]{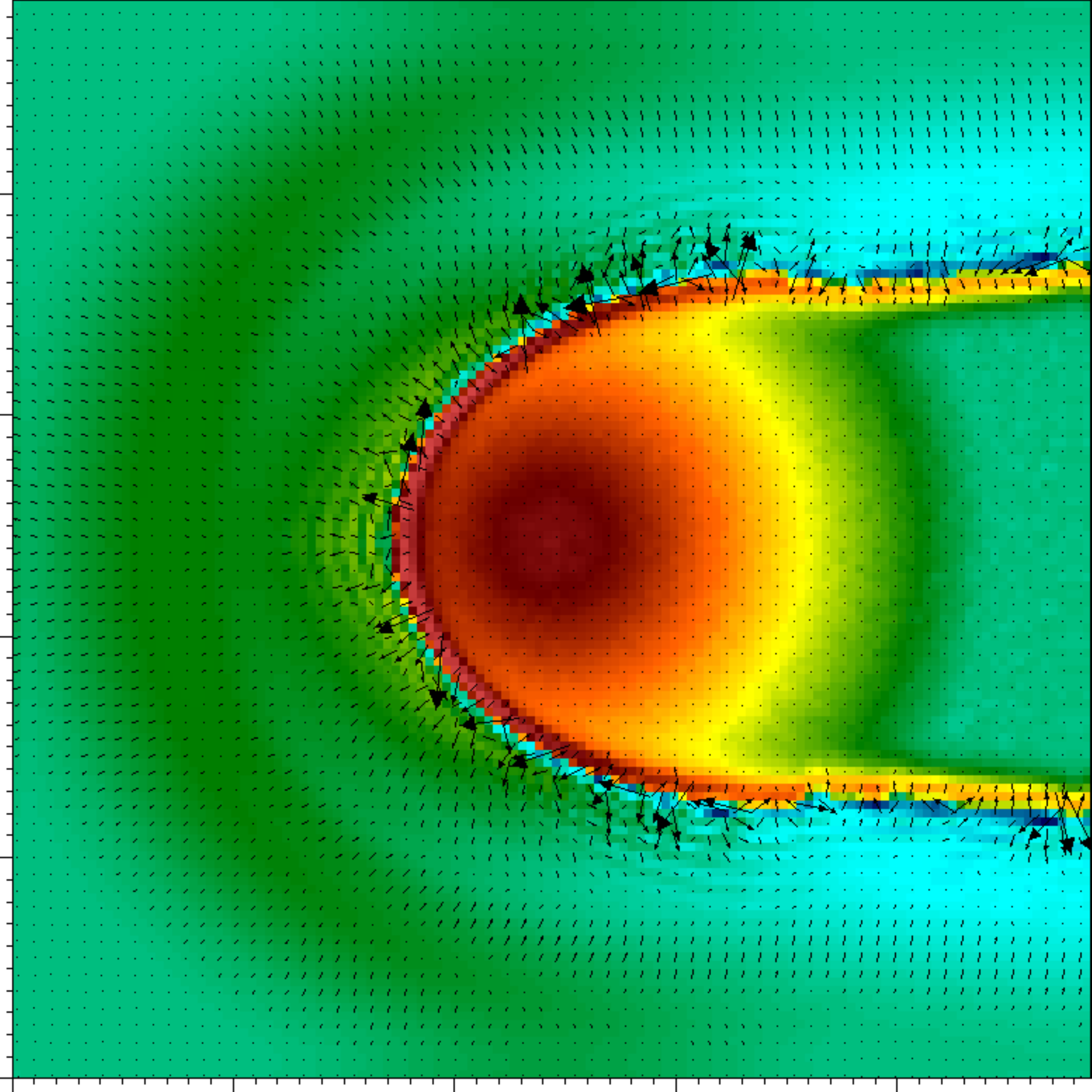}
	\caption{The logarithmic density field in g\,cm$^{-3}$, with velocity vectors, for the high flux models at 50\,kyr. This demonstrates the difference between the driving shock of the monochromatic OTS model (top frame) and the photo-evaporative flow of the polychromatic-diffuse model (bottom frame). Each frame is a slice through the computational grid, which is a cube with sides 4.87\,pc long. Major ticks are separated by 1\, pc. Typical shock (upper frame) and outflow (lower frame) velocities are 5-7\, km\,s$^{-1}$ and 1-4\,km\,s$^{-1}$ respectively.}
	\label{vels}
\end{figure}

On the other hand, the monochromatic OTS model does form a strong shock sufficiently rapidly to effectively drive into the BES, this leads to greater compression and accumulation of material into a relatively thick shell around the edge of the resulting bow structure. 
To illustrate the early braking of the polychromatic models, the difference in the velocity field between the monochromatic OTS and diffuse models at 50\,kyr is illustrated in Figure \ref{vels}. At this point the monochromatic model is still driving a shock into the BES and accumulating material, whereas the polychromatic-diffuse model is beginning a photo-evaporative flow.
 \\

\begin{figure}
	\hspace{8pt}
	\includegraphics[width=78mm, height=65.2mm]{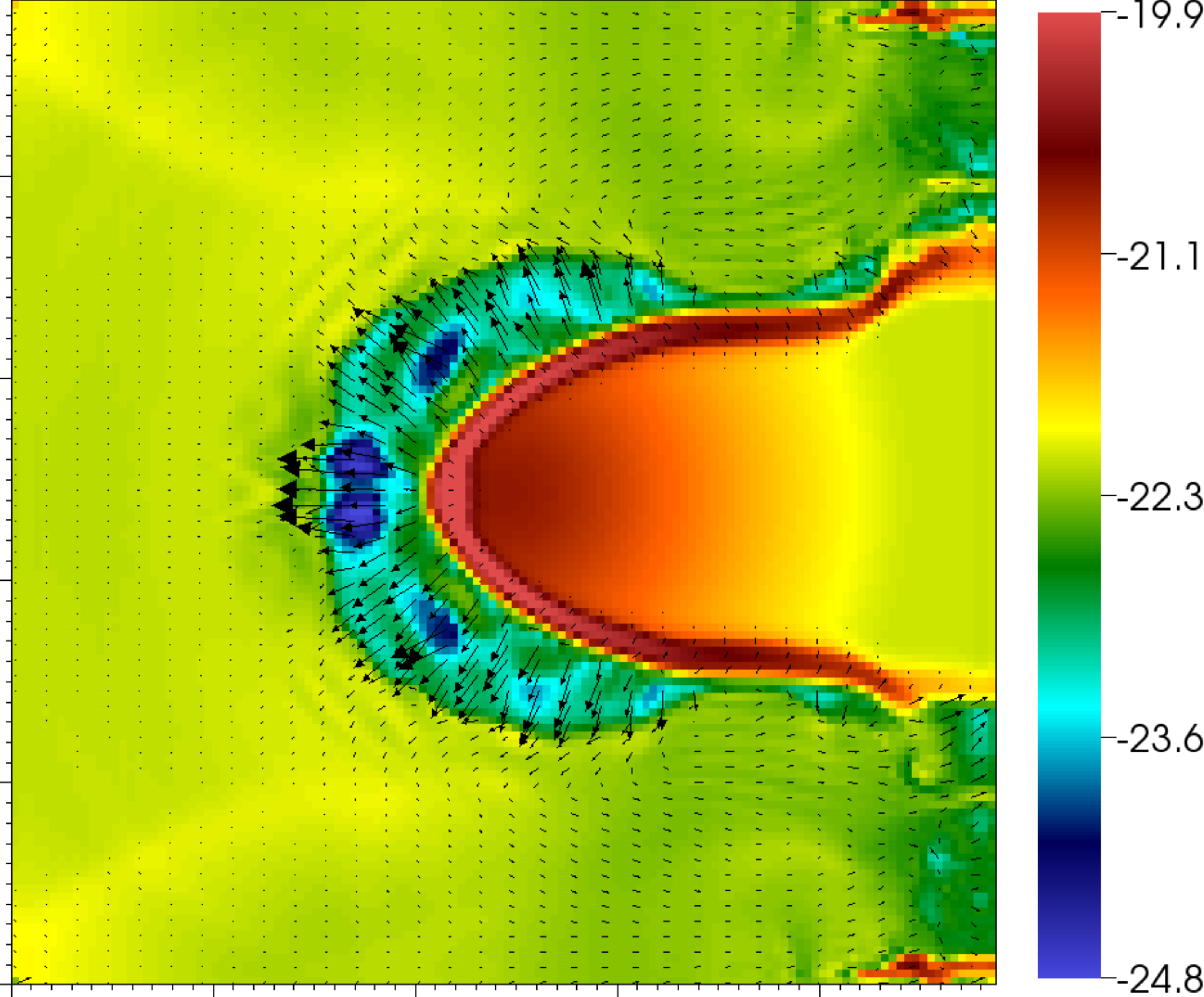}

	\vspace{6pt}
	\hspace{8pt}
	\includegraphics[width=65mm, height=65mm]{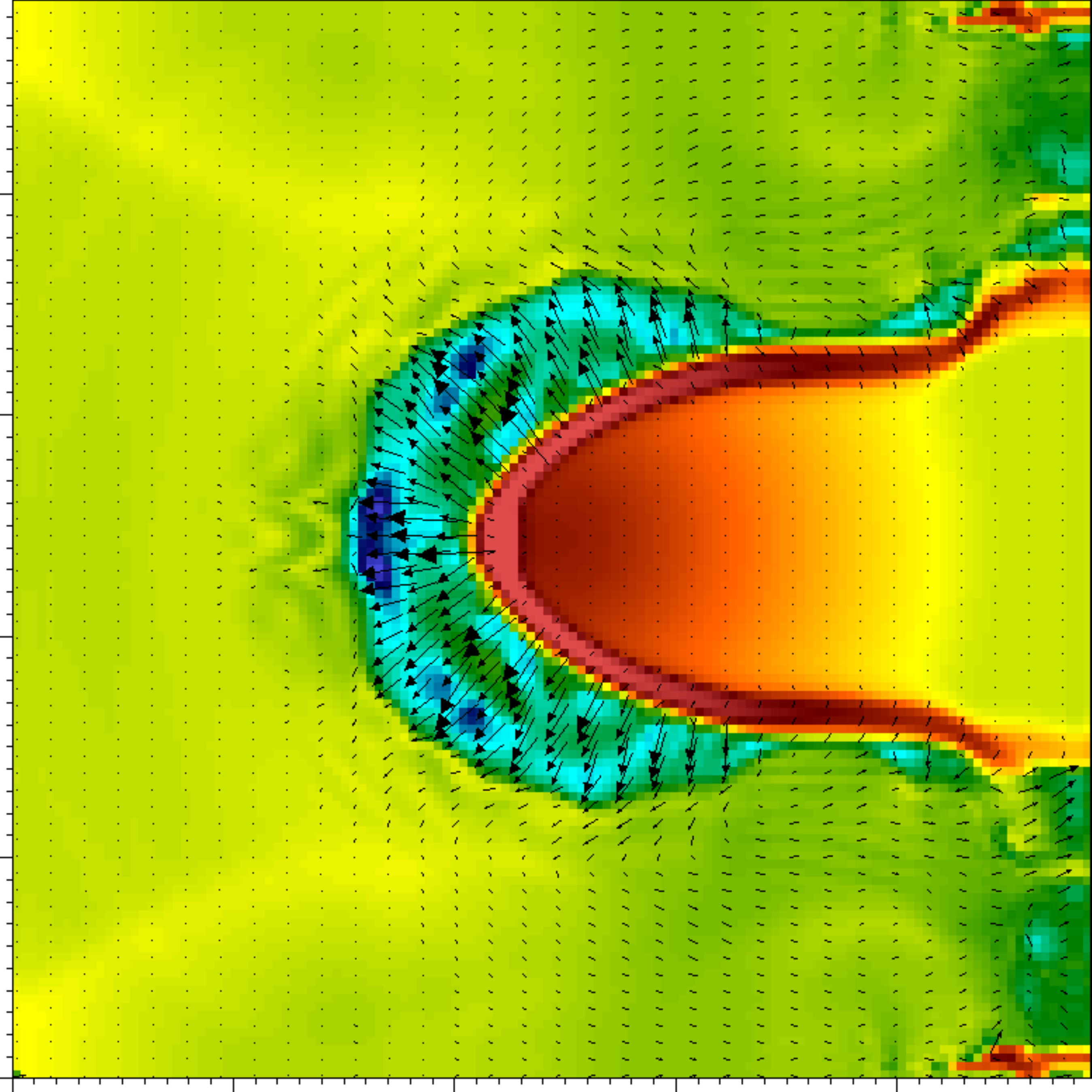}

	\vspace{6pt}
	\hspace{8pt}
	\includegraphics[width=65mm, height=65mm]{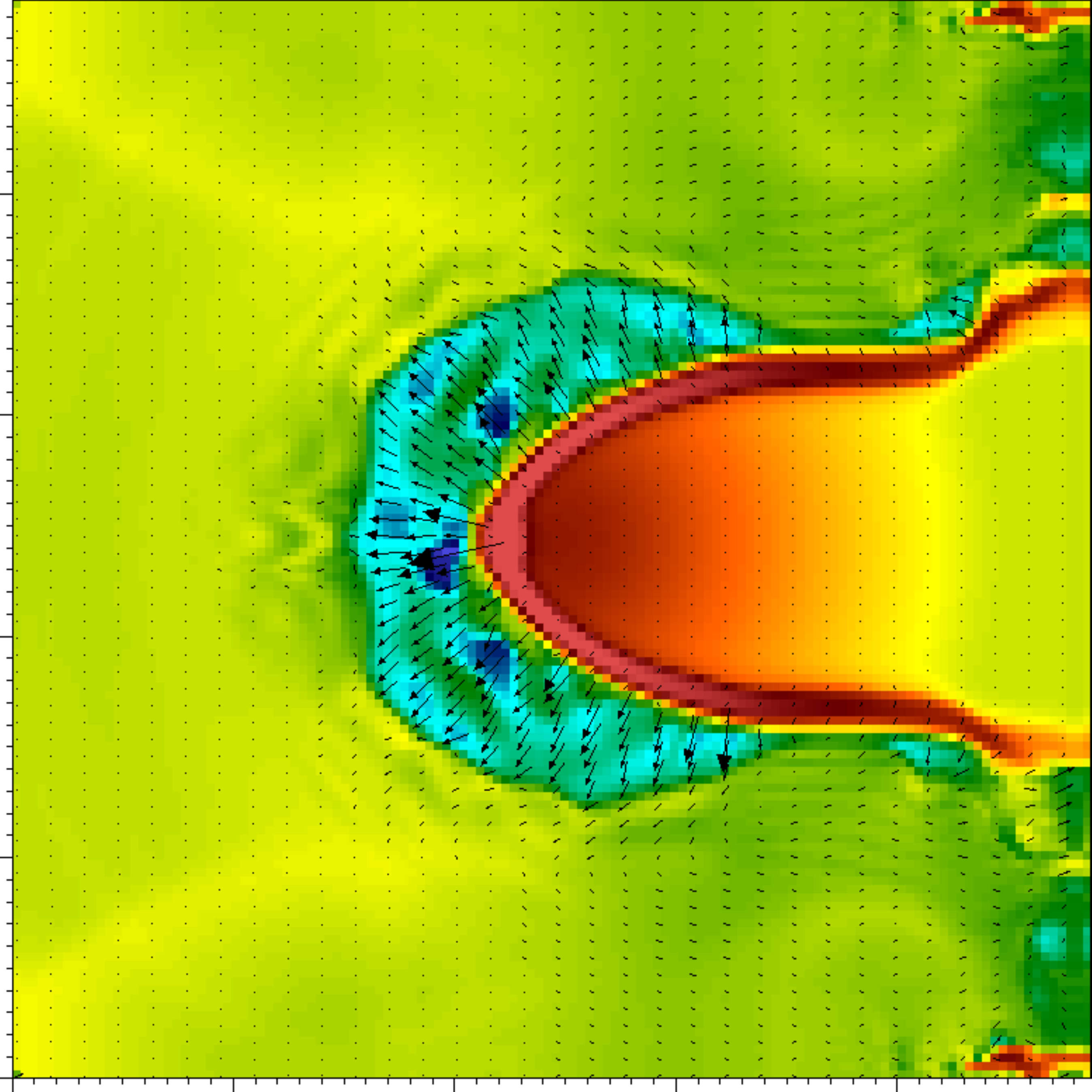}
	\caption{Logarithmic density snapshots in g\,cm$^{-3}$ with velocity vectors for the high flux monochromatic model at 180, 185 and 190\,kyr.  This illustrates the ejection of distinct knots of material and the evolution of the disrupted region around the tip of the bow. Each frame is a slice through the computational grid, which is a cube with sides 4.87\,pc long. Major ticks are separated by 1\, pc. Typical velocities in the outflow region of each frame are 25-35\,km\,s$^{-1}$.}
	\label{EPE}

\end{figure}

In the OTS model, sufficient material is rapidly accumulated for a
short, strong, photoevaporative flow to occur prior to substantial
braking of the bow.  The resulting rocket-motion is therefore much
stronger than normal photo-evaporative flow, giving rise to the
ejection of a significant amount of material that accelerates the
existing shock and carves out a low density wake.  Subsequent
ejections also occur episodically along the length of the bow that is
exposed to ionizing radiation. The result is a disrupted region
surrounding the tip of the bow structure in which densities can be
excavated to levels lower than the ambient surroundings. A possible
cause of the episodic nature of this process is that the outer shell
density oscillates about some critical value as the shell sequentially
accumulates and ejects.  This `episodic photo-evaporative ejection'
further drives the collapse very effectively and contracts the bow
perpendicularly to the ejection direction, tapering the head of the
cometary structure. Figure \ref{EPE} shows the disrupted region around
the tip of the bow of the monochromatic model at 180, 185 and 190\,kyr
and illustrates the motion of discrete knots of material away from the
surface, rather than a continuous stream.

\begin{figure}
		\vspace{-20pt}
		\hspace{-25pt}
		\includegraphics[scale = 0.3361]{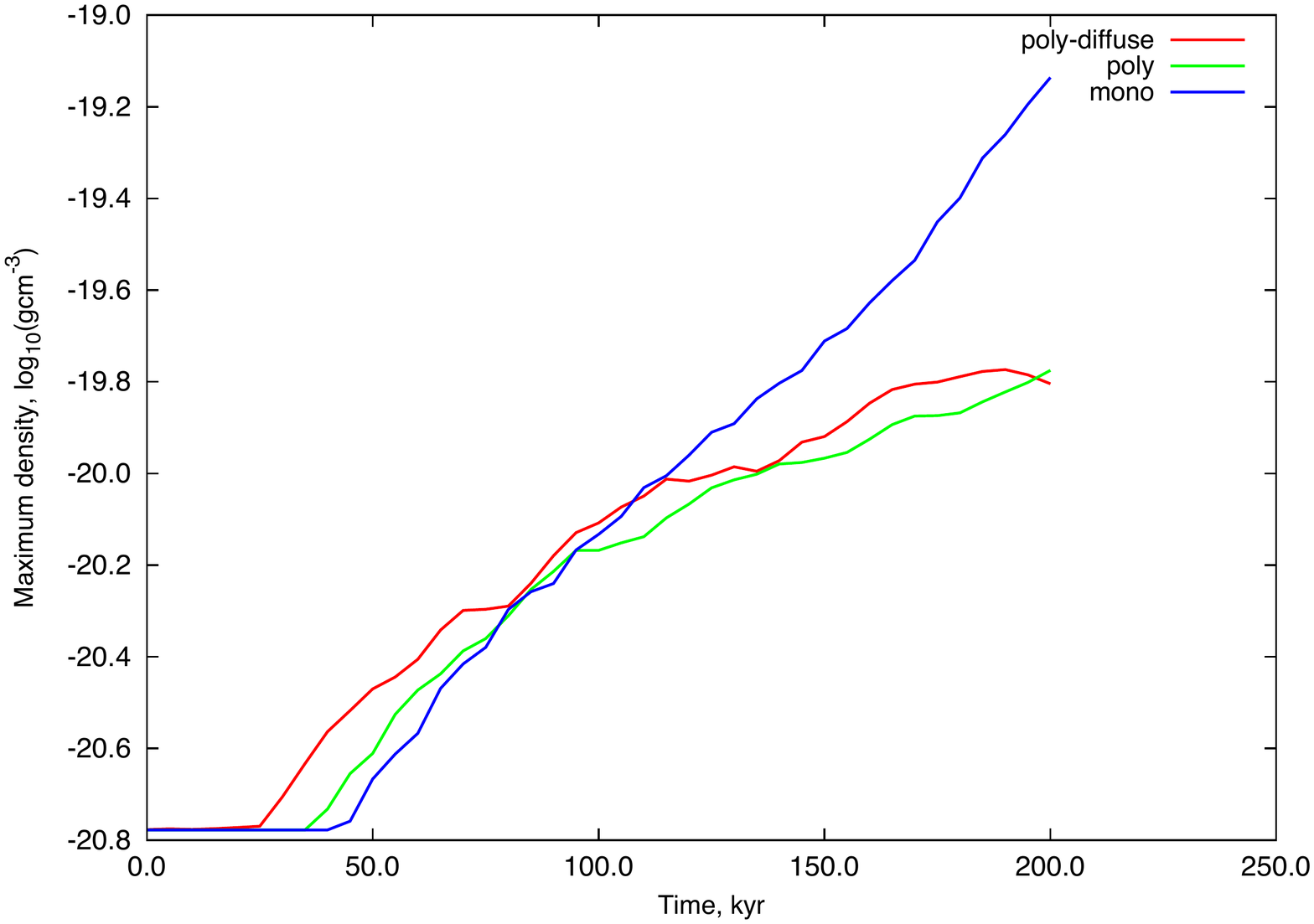}
		\vspace{-30pt}

		\hspace{-25pt}
		\vspace{-30pt}
		\includegraphics[scale = 0.3361]{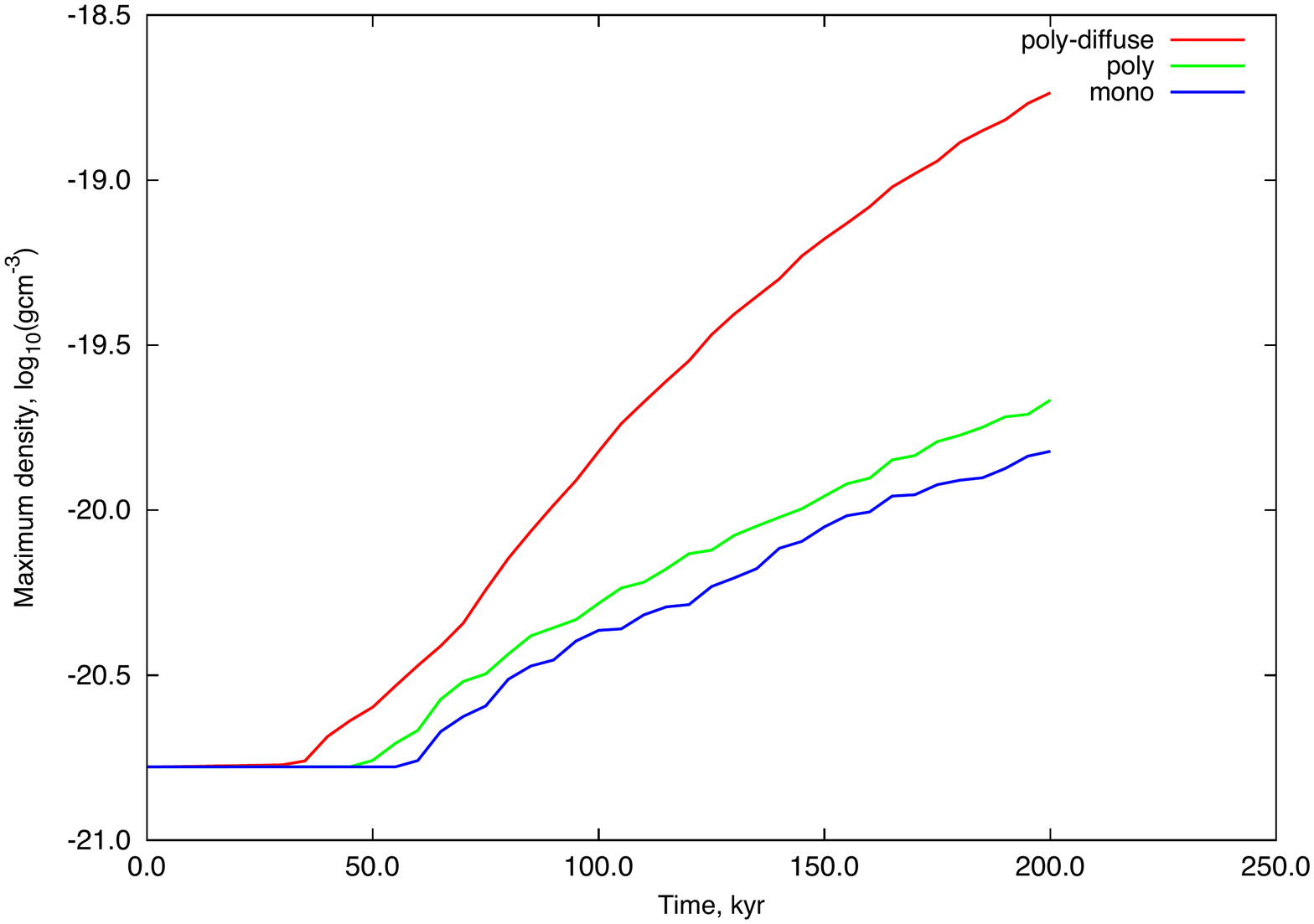}

		\hspace{-25pt}
		\includegraphics[scale = 0.3361]{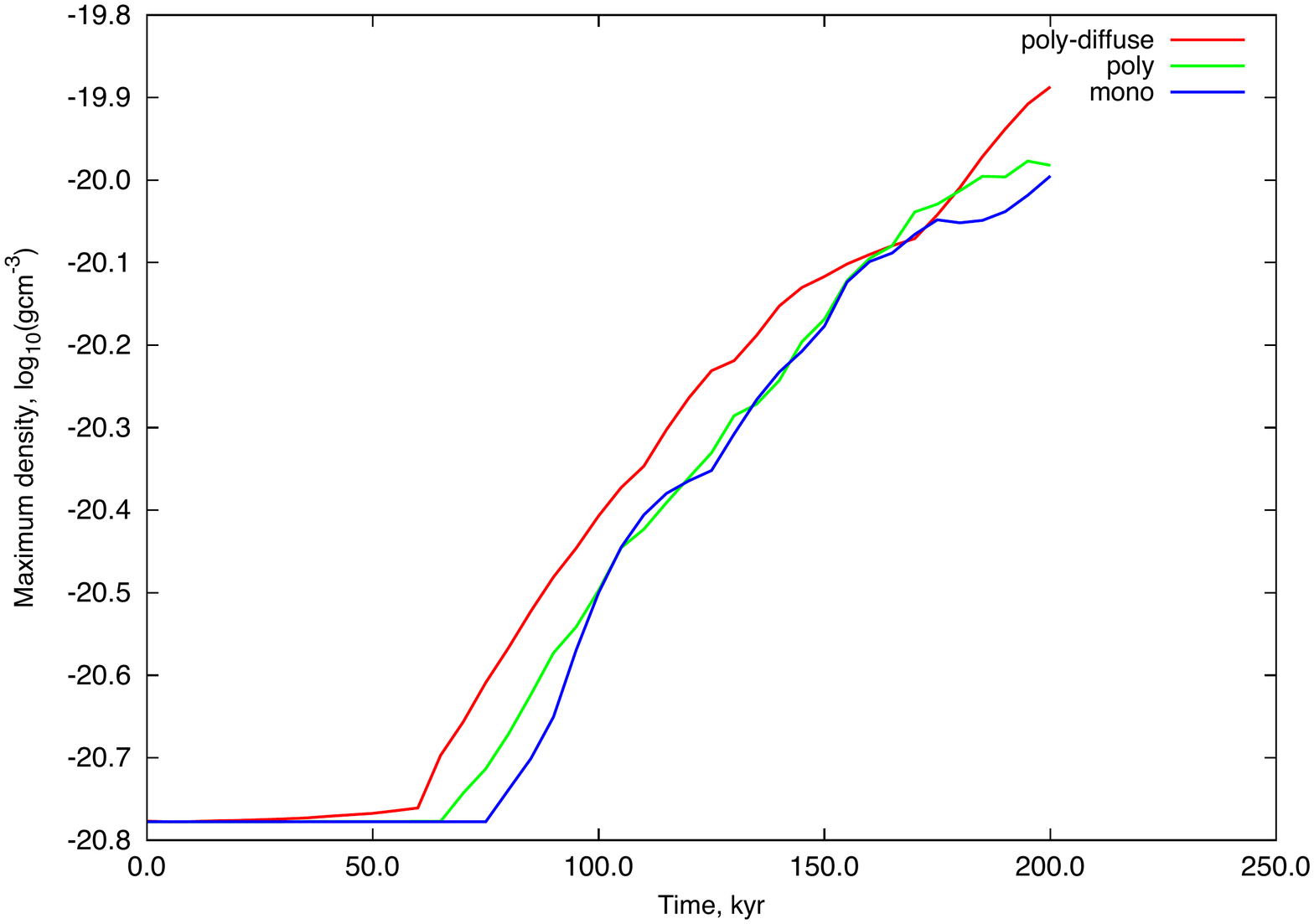}
  \caption{The evolution of maximum logarithmic densities
    respectively. Flux regimes are high to low from top to bottom.}
		\label{RDI_EVOLUTION}
\end{figure}

The evolution of the maximum cell densities for each high flux model
are shown in the top frame of Figure \ref{RDI_EVOLUTION}. This can be
used in conjunction with the appropriate density map to 
illustrate the rate at which material is collected.  Here the
maximum density evolution highlights the differences between the two
different behaviours noted above. In the polychromatic and
polychromatic-diffuse models the maximum density increases at a
declining rate and eventually plateaus. The monochromatic model
continues accumulating material as it is effectively rocket-driven
towards the core of the BES, finishing with a maximum density
approximately 4.5 times that of the other models.  Star formation at
this flux regime will actually occur more slowly when polychromatic
radiation or the diffuse field is accounted for than in the simplified
calculation.

The formation of thin shell instabilities has little to no impact in
the evolution of these models, appearing only in the wings of the
polychromatic and monochromatic OTS models and are simply  swept
off the grid.

\subsubsection{Medium flux models}
\label{medAnalysis}
The medium flux density evolution is presented in Figure \ref{RDI_MED} and exhibits the largest difference between the final states of the OTS and diffuse field models. The OTS models both develop shocks that drive effectively into the BES, resulting in rapidly formed high density bow structures. The accumulated density is high enough to give rise to a scaled down version of the episodic photo-evaporatively driven collapse exhibited by the high flux monochromatic model and leads to the same rocket-motion that continues driving material towards the centre of the BES and tapers the bow head.
In the wings of the OTS models instabilities form via the mechanism described in section \ref{sec:ins} and propagate linearly, having no effect of the rate of collapse and eventually being swept off of the grid. Note that the low density regions in the wake of these instabilities are not due to photo-evaporative flow, rather they have been excavated by the high velocity instability shocks, with the material funnelled laterally to form knots and elephant trunks.

The diffuse field model behaves slightly differently, resembling a more effectively collapsed version of the monochromatic OTS high flux model. Again a strong shock is developed which drives into the BES and transitions to episodic photo-evaporatively driven collapse. The major difference lies in the wings of the model, which effectively drive into the side of the BES resulting in a final bullet-like structure with a very high density at what was the core of the BES. The reason for this is that a photo-evaporative flow can establish itself along a greater extent of the bow, into regions that would otherwise be shadowed from heating, because of the diffuse radiation field. This results in a whip-like progression of photo-evaporation along the trunk as the shell sequentially becomes dense enough to readily eject material and ceases in the tail regions when the diffuse ionizing flux becomes too low to cause heating. 
This will most likely result in much more rapid star formation in the model that includes the diffuse field. 
The evolution of the maximum  cell density for these medium flux models is shown in the middle frame of Figure \ref{RDI_EVOLUTION}. This clearly highlights the fairly extreme difference between the OTS and diffuse field models, with a final difference in maximum density at 200\,kyr of well over a factor of 10.

\subsubsection{Low flux models}
\label{lowAnalysis}
The low flux density evolutions are given in Figure \ref{RDI_LOW}. These models exhibit the largest susceptibility to thin shell instability as instabilities can begin developing across the entirety of the I-front before it impacts the BES. Instabilities that collide with the BES are braked and result in high density knots along the rim of the resulting bow structure. Those that continue in the wings of the model are elongated into elephant trunks with high density tips via the mechanism described in section \ref{sec:ins}.
The evolution of these elephant trunks varies with each treatment of the radiation field. The polychromatic OTS trunks are more elongated than the monochromatic ones because hard radiation carves out a path more rapidly. Any compression of the trunks in the OTS models is due to thermal pressure.
In the polychromatic-diffuse model, diffuse field radiation effectively drives into the material perpendicular to any displacement in the ionization front and therefore actually prevents the formation of a number of potential trunks by smoothing out dimples. Those trunks that do form will also continue to be both compressed thermally and exposed to diffuse ionizing radiation. 

The RDI process for the OTS models occurs very weakly, with more material being accumulated through instability than compression. The diffuse field model drives into the BES more effectively, forming a smoother high density bow compared to the knotted structures of the other models and manages to initiate some photo-evaporative flow.
The maximum cell density evolution for the low flux models is given in the bottom frame of Figure \ref{RDI_EVOLUTION}. At 200\,kyr the diffuse field model has accumulated only a slightly larger maximum density than the OTS models, though it is clear from Figure \ref{RDI_LOW} that it has achieved this order of density over a relatively large volume. The diffuse field model will most likely form stars first and on a larger scale than the models that use the OTS approximation. 

These low flux results, comprising a bright rimmed cloud with pillars along its wings, bear resemblance to observed systems such as IC 1848E, as shown in Figure 1 of \cite{2011PASJ...63..795C}. In the aforementioned work, instability was suggested as the formation mechanism of the elephant trunks in this region of IC 1848E, our unstable radiatively driven shock driving into a pre-existing density structure supports this hypothesis.

\subsubsection{Comparison with iVINE}
There are some differences between the results obtained here and those of \cite{2009MNRAS.393...21G}, the work on which our model parameters are based. In particular the maximum density evolutions derived by TORUS are much weaker, to the extent that no clear gravitational collapse has occurred by 200\,kyr. This discrepancy is attributed to the difference in grid size, combined with the use of periodic boundary conditions. A significant proportion of the compression of the BES in the iVINE models arises because hot gas is advected off the edge of the periodic boundary and impacts the cloud on the opposite side of the grid \citep[see section 4.1.2 of][]{2009MNRAS.393...21G}. This effect is justified by the authors as they assume that the molecular cloud completely surrounds the triggering star, however the effect will still give rise to differences in the results obtained by our models and theirs since motion of the hot gas has longer to decay over our larger grid and is also impeded by the elephant trunks that have formed in the wings of our models. In our models, lateral compression of the BES due to motion in the hot gas through the boundaries has negligible effect and as such the BES can be considered to be isolated.

With regard to the effect of the diffuse field, a comparison with the results found using iVINE/DiVINE \citep{2011MNRAS.413..401E} is not straightforward as the systems being modelled are very different. However we broadly agree that treating the diffuse field can lead to higher density resulting structures, particularly in our medium flux model (section \ref{medAnalysis}). In our low flux model (section \ref{lowAnalysis}), where elephant trunks of a similar form to those created in the iVINE/DiVINE models arise, we also find that inclusion of  the diffuse field increases the density and decreases the number of elephant trunks that form. We also agree that these trunks are narrowed, sometimes to the extent that the head of the trunk can become completely detached.

\section{Conclusions}
It is clear that a direct treatment of the diffuse field has a
significant impact upon the evolution of radiation hydrodynamics
calculations on parsec scales. Not only is the effectiveness of RDI
sensitive to the way in which the radiation field is treated, the
formation and evolution of elephant trunk structures via instability
also varies.

At low and intermediate flux regimes inclusion of the diffuse field results in more efficient RDI, with both more widespread accumulation of material and, particularly in the medium flux case, a higher final maximum density. In the high flux regime however, the material around the BES is ionized so rapidly in the diffuse and polychromatic OTS models that only a weak shock forms and RDI does not occur effectively. Generally, it is found that the extent to which RDI occurs depends strongly on the strength of the shock that is accumulated prior to driving into the BES. Regardless of the ionizing flux, if only a weak driving shock has formed collapse will occur slowly. 
When sufficient material is accumulated a photo-evaporative flow is found to occur before the driving shock is braked, reinforcing the shock with the resulting rocket motion. This occurs in short, sharp, bursts with the reason for this episodic behaviour hypothesised as being due to the refilling time for material in the dense shell. The high velocity ejecta carve out material around the head of the cometary structure resulting in a low density disrupted bubble. This rocket motion has a significant effect on the collapse of the BES: with the inclusion of the diffuse field, it can accur across a large part of the otherwise shadowed region and significantly narrow the resulting cometary structure.

Elephant trunks that arise due to thin shell instability are harder to
form in the presence of the diffuse field as the dimples that seed
them are smoothed out, and those that do form are then subject to the
expected combination of thermal compression and diffuse field
photoionizing radiation. Despite this the formation of elephant trunks,
as the result of instability in a radiatively driven shock, still
occurs and provides a mechanism for the formation of systems such as
that discussed in \cite{2011PASJ...63..795C}.

The radiation-hydrodynamical effect of the diffuse field is
cumulative and significant, and is strongly coupled to the
hydrodynamical evolution of the system. This implies that the
treatment of the diffuse field should be accurate throughout a
simulation, as consistent deficiencies could lead to a systematic
change in the overall radiation hydrodynamical evolution of the
system.  \\

We next intend to compute synthetic images to determine theoretical
observables for systems undergoing star formation as a result of
radiative feedback. We will also include the use of an adaptive mesh
so that we can follow collapse until the onset of star formation.

\section{Summary}
We have used the radiation hydrodynamics module of the TORUS code to
investigate the effects of using a monochromatic OTS, polychromatic
OTS and polychromatic-diffuse field on the radiatively driven
implosion of a Bonnor-Ebert sphere. We have found that incorporating
the diffuse field into this model over three flux regimes leads to
significantly different results to those obtained using the OTS
approximation. At intermediate and low flux regimes the rate of
compression is higher than that without inclusion of the diffuse field,
whereas at high flux compression actually occurs more slowly when the
diffuse or polychromatic OTS field is considered because there is
little opportunity for a material shock to drive into the BES. In the
event of accumulation of sufficient material, photo-evaporative flow
or ejection has been identified as a mechanism which can drive
collapse very effectively. This photo-evaporative flow is particularly effective at
driving and compressing the tail of the cometary structure  when the diffuse field is
treated. The formation of elephant trunk structures via instability
also occurs much less readily with the inclusion of the diffuse field
as perturbations to the ionization front are smeared out. We conclude
that in order to properly address quantitative questions regarding
triggered star formation thorough treatment of the diffuse field is
necessary in radiation hydrodynamics models.

\section*{Acknowledgments}

The calculations presented here were performed using the University of
Exeter Supercomputer, part of the DiRAC Facility jointly funded by
STFC, the Large Facilities Capital Fund of BIS, and the University of
Exeter. T. J. Haworth is funded by an STFC studentship.  We thank
David Acreman for useful discussions and support and the anonymous referee for 
their insightful comments.

\bibliographystyle{mn2e}
\bibliography{RDIBES}

\appendix

\bsp

\label{lastpage}

\end{document}